\documentclass[12pt,a4paper]{article}

\usepackage{ifthen} 
\usepackage{rotating,tabularx,booktabs}
\newboolean{pdflatex}
\setboolean{pdflatex}{true} 

\newboolean{articletitles}
\setboolean{articletitles}{true} 

\newboolean{uprightparticles}
\setboolean{uprightparticles}{false} 

\newboolean{inbibliography}
\setboolean{inbibliography}{false} 

\def\paperauthors{LHCb collaboration} 
\def\paperasciititle{Updated measurement of time-dependent CP-violating observables in BsJpsiKK decays} 
\def\papertitle{Updated measurement of time-dependent \CP-violating observables in $\Bs\to\jpsi K^+ K^-$ decays} 
\def\paperkeywords{{High Energy Physics}, {LHCb}} 
\def\papercopyright{\the\year\ CERN for the benefit of the LHCb collaboration} 
\def\paperlicence{CC-BY-4.0 licence}
\def\paperlicenceurl{https://creativecommons.org/licenses/by/4.0/}

\usepackage[top=1in, bottom=1.25in, left=1in, right=1in]{geometry}

\usepackage{microtype}
\usepackage{lineno}  
\usepackage{xspace} 
\usepackage{caption} 

\usepackage{graphicx}  
\usepackage{color}
\usepackage{colortbl}
\graphicspath{{./figs/}} 

\usepackage{amsmath} 
\usepackage{amssymb}
\usepackage{amsfonts}
\usepackage{upgreek} 

\newcommand*\patchAmsMathEnvironmentForLineno[1]{%
\expandafter\let\csname old#1\expandafter\endcsname\csname #1\endcsname
\expandafter\let\csname oldend#1\expandafter\endcsname\csname
end#1\endcsname
 \renewenvironment{#1}%
   {\linenomath\csname old#1\endcsname}%
   {\csname oldend#1\endcsname\endlinenomath}%
}
\newcommand*\patchBothAmsMathEnvironmentsForLineno[1]{%
  \patchAmsMathEnvironmentForLineno{#1}%
  \patchAmsMathEnvironmentForLineno{#1*}%
}
\AtBeginDocument{%
\patchBothAmsMathEnvironmentsForLineno{equation}%
\patchBothAmsMathEnvironmentsForLineno{align}%
\patchBothAmsMathEnvironmentsForLineno{flalign}%
\patchBothAmsMathEnvironmentsForLineno{alignat}%
\patchBothAmsMathEnvironmentsForLineno{gather}%
\patchBothAmsMathEnvironmentsForLineno{multline}%
\patchBothAmsMathEnvironmentsForLineno{eqnarray}%
}

\usepackage{hyperxmp}

\usepackage[pdftex,
            pdfauthor={\paperauthors},
            pdftitle={\paperasciititle},
            pdfkeywords={\paperkeywords},
            pdfcopyright={Copyright (C) \papercopyright},
            pdflicenseurl={\paperlicenceurl}]{hyperref}

\usepackage[all]{hypcap} 

\usepackage{xspace} 
\usepackage{upgreek}

\def\lhcb {\mbox{LHCb}\xspace}

\def\MagUp {\mbox{\em Mag\kern -0.05em Up}\xspace}

\ifthenelse{\boolean{uprightparticles}}%
{

 \def\Pmu         {\ensuremath{\upmu}\xspace}

 \def\Ppi         {\ensuremath{\uppi}\xspace}

 \def\Ppsi        {\ensuremath{\uppsi}\xspace}

 \def\PDelta      {\ensuremath{\Delta}\xspace}                 
 \def\PXi      {\ensuremath{\Xi}\xspace}                 
 \def\PLambda      {\ensuremath{\Lambda}\xspace}                 
 \def\PSigma      {\ensuremath{\Sigma}\xspace}                 
 \def\POmega      {\ensuremath{\Omega}\xspace}                 
 \def\PUpsilon      {\ensuremath{\Upsilon}\xspace}                 
 
 \def\PB      {\ensuremath{\mathrm{B}}\xspace}                 
                  
 \def\PD      {\ensuremath{\mathrm{D}}\xspace}

 \def\PJ      {\ensuremath{\mathrm{J}}\xspace}                 
 \def\PK      {\ensuremath{\mathrm{K}}\xspace}

 \def\Pb      {\ensuremath{\mathrm{b}}\xspace}                 
 \def\Pc      {\ensuremath{\mathrm{c}}\xspace}                 
 \def\Pd      {\ensuremath{\mathrm{d}}\xspace}

 \def\Pi      {\ensuremath{\mathrm{i}}\xspace}

 \def\Ps      {\ensuremath{\mathrm{s}}\xspace}

}
{

 \def\Pmu         {\ensuremath{\mu}\xspace}

 \def\Ppi         {\ensuremath{\pi}\xspace}

 \def\Ppsi        {\ensuremath{\psi}\xspace}                 
                  
 \mathchardef\PDelta="7101
 \mathchardef\PXi="7104
 \mathchardef\PLambda="7103
 \mathchardef\PSigma="7106
 \mathchardef\POmega="710A
 \mathchardef\PUpsilon="7107
                  
 \def\PB      {\ensuremath{B}\xspace}                 
                  
 \def\PD      {\ensuremath{D}\xspace}

 \def\PJ      {\ensuremath{J}\xspace}                 
 \def\PK      {\ensuremath{K}\xspace}

 \def\Pb      {\ensuremath{b}\xspace}                 
 \def\Pc      {\ensuremath{c}\xspace}                 
 \def\Pd      {\ensuremath{d}\xspace}

 \def\Pi      {\ensuremath{i}\xspace}

 \def\Ps      {\ensuremath{s}\xspace}

}

\makeatletter
\ifcase \@ptsize \relax
  \newcommand{\miniscule}{\@setfontsize\miniscule{4}{5}}
\or
  \newcommand{\miniscule}{\@setfontsize\miniscule{5}{6}}
\or
  \newcommand{\miniscule}{\@setfontsize\miniscule{5}{6}}
\fi
\makeatother

\DeclareRobustCommand{\optbar}[1]{\shortstack{{\miniscule (\rule[.5ex]{1.25em}{.18mm})}
  \\ [-.7ex] $#1$}}

\def\mup        {{\ensuremath{\Pmu^+}}\xspace}

\def\dquark    {{\ensuremath{\Pd}}\xspace}

\def\squark    {{\ensuremath{\Ps}}\xspace}

\def\cquark    {{\ensuremath{\Pc}}\xspace}

\def\bquark    {{\ensuremath{\Pb}}\xspace}
\def\bquarkbar {{\ensuremath{\overline \bquark}}\xspace}

\def\pion   {{\ensuremath{\Ppi}}\xspace}

\def\pip    {{\ensuremath{\pion^+}}\xspace}
\def\pim    {{\ensuremath{\pion^-}}\xspace}

\def\kaon    {{\ensuremath{\PK}}\xspace}
  \def\Kbar    {{\kern 0.2em\overline{\kern -0.2em \PK}{}}\xspace}

\def\KorKbar    {\kern 0.18em\optbar{\kern -0.18em K}{}\xspace}

\def\Kp      {{\ensuremath{\kaon^+}}\xspace}
\def\Km      {{\ensuremath{\kaon^-}}\xspace}

  \def\Dbar    {{\kern 0.2em\overline{\kern -0.2em \PD}{}}\xspace}
\def\D       {{\ensuremath{\PD}}\xspace}

\def\DorDbar    {\kern 0.18em\optbar{\kern -0.18em D}{}\xspace}

\def\Dsm     {{\ensuremath{\D^-_\squark}}\xspace}
\def\Dspm    {{\ensuremath{\D^{\pm}_\squark}}\xspace}

\def\Dssm    {{\ensuremath{\D^{*-}_\squark}}\xspace}

\def\B       {{\ensuremath{\PB}}\xspace}
\def\Bbar    {{\ensuremath{\kern 0.18em\overline{\kern -0.18em \PB}{}}}\xspace}

\def\BorBbar    {\kern 0.18em\optbar{\kern -0.18em B}{}\xspace}
\def\Bz      {{\ensuremath{\B^0}}\xspace}

\def\Bu      {{\ensuremath{\B^+}}\xspace}

\def\Bd      {{\ensuremath{\B^0}}\xspace}
\def\Bs      {{\ensuremath{\B^0_\squark}}\xspace}
\def\Bsb     {{\ensuremath{\Bbar{}^0_\squark}}\xspace}

\def\Bc      {{\ensuremath{\B_\cquark^+}}\xspace}

\def\jpsi     {{\ensuremath{{\PJ\mskip -3mu/\mskip -2mu\Ppsi\mskip 2mu}}}\xspace}

  \def\Y#1S{\ensuremath{\PUpsilon{(#1S)}}\xspace}

\def\Lz          {{\ensuremath{\PLambda}}\xspace}
\def\Lbar        {{\ensuremath{\kern 0.1em\overline{\kern -0.1em\PLambda}}}\xspace}
\def\LorLbar    {\kern 0.18em\optbar{\kern -0.18em \PLambda}{}\xspace}

\def\Lb      {{\ensuremath{\Lz^0_\bquark}}\xspace}
\def\Lbbar   {{\ensuremath{\Lbar{}^0_\bquark}}\xspace}

\def\Lcbar   {{\ensuremath{\Lbar{}^-_\cquark}}\xspace}

\def\to                 {\ensuremath{\rightarrow}\xspace}

\def\CP                {{\ensuremath{C\!P}}\xspace}

\newcommand{\dms}{{\ensuremath{\Delta m_{\squark}}}\xspace}

\newcommand{\DGs}{{\ensuremath{\Delta\Gamma_{\squark}}}\xspace}

\newcommand{\Gs}{{\ensuremath{\Gamma_{\squark}}}\xspace}
\newcommand{\Gd}{{\ensuremath{\Gamma_{\dquark}}}\xspace}

\newcommand{\phis}{{\ensuremath{\phi_{\squark}}}\xspace}

\def\AT#1     {\ensuremath{A_{\mathrm{T}}^{#1}}\xspace}           

\def\C#1      {\ensuremath{\mathcal{C}_{#1}}\xspace}                       
\def\Cp#1     {\ensuremath{\mathcal{C}_{#1}^{'}}\xspace}                    
\def\Ceff#1   {\ensuremath{\mathcal{C}_{#1}^{\mathrm{(eff)}}}\xspace}        
\def\Cpeff#1  {\ensuremath{\mathcal{C}_{#1}^{'\mathrm{(eff)}}}\xspace}       
\def\Ope#1    {\ensuremath{\mathcal{O}_{#1}}\xspace}                       
\def\Opep#1   {\ensuremath{\mathcal{O}_{#1}^{'}}\xspace}                    

\newcommand{\braket}[2]{\ensuremath{\langle #1|#2\rangle}} 

\newcommand{\tev}{\ifthenelse{\boolean{inbibliography}}{\ensuremath{~T\kern -0.05em eV}}{\ensuremath{\mathrm{\,Te\kern -0.1em V}}}\xspace}
\newcommand{\gev}{\ensuremath{\mathrm{\,Ge\kern -0.1em V}}\xspace}
\newcommand{\mev}{\ensuremath{\mathrm{\,Me\kern -0.1em V}}\xspace}
\newcommand{\kev}{\ensuremath{\mathrm{\,ke\kern -0.1em V}}\xspace}
\newcommand{\ev}{\ensuremath{\mathrm{\,e\kern -0.1em V}}\xspace}
\newcommand{\gevc}{\ensuremath{{\mathrm{\,Ge\kern -0.1em V\!/}c}}\xspace}
\newcommand{\mevc}{\ensuremath{{\mathrm{\,Me\kern -0.1em V\!/}c}}\xspace}
\newcommand{\gevcc}{\ensuremath{{\mathrm{\,Ge\kern -0.1em V\!/}c^2}}\xspace}
\newcommand{\gevgevcccc}{\ensuremath{{\mathrm{\,Ge\kern -0.1em V^2\!/}c^4}}\xspace}
\newcommand{\mevcc}{\ensuremath{{\mathrm{\,Me\kern -0.1em V\!/}c^2}}\xspace}

\def\mum  {\ensuremath{{\,\upmu\mathrm{m}}}\xspace}

\def\invfb   {\ensuremath{\mbox{\,fb}^{-1}}\xspace}

\def\ps   {\ensuremath{{\mathrm{ \,ps}}}\xspace}
\def\fs   {\ensuremath{\mathrm{ \,fs}}\xspace}

\def\invps{\ensuremath{{\mathrm{ \,ps^{-1}}}}\xspace}

\def\deriv {\ensuremath{\mathrm{d}}}

\def\gsim{{~\raise.15em\hbox{$>$}\kern-.85em
          \lower.35em\hbox{$\sim$}~}\xspace}
\def\lsim{{~\raise.15em\hbox{$<$}\kern-.85em
          \lower.35em\hbox{$\sim$}~}\xspace}

\def\sPlot{\mbox{\em sPlot}\xspace}

\def\ptot       {\mbox{$p$}\xspace}
\def\pt         {\mbox{$p_{\mathrm{ T}}$}\xspace}

\def\mrad{\ensuremath{\mathrm{ \,mrad}}\xspace}
\def\rad{\ensuremath{\mathrm{ \,rad}}\xspace}

\def\evtgen     {\mbox{\textsc{EvtGen}}\xspace}

\def\geant      {\mbox{\textsc{Geant4}}\xspace}

\def\photos     {\mbox{\textsc{Photos}}\xspace}

\def\pythia     {\mbox{\textsc{Pythia}}\xspace}

\def\tell1  {TELL1\xspace}
\def\ukl1   {UKL1\xspace}

\def\maglambda{|\lambda|}

\usepackage{cite} 
\usepackage{mciteplus}

\usepackage{longtable} 

\begin{document}

\renewcommand{\thefootnote}{\fnsymbol{footnote}}
\setcounter{footnote}{1}


\begin{titlepage}
\pagenumbering{roman}

\vspace*{-1.5cm}
\centerline{\large EUROPEAN ORGANIZATION FOR NUCLEAR RESEARCH (CERN)}
\vspace*{1.5cm}
\noindent
\begin{tabular*}{\linewidth}{lc@{\extracolsep{\fill}}r@{\extracolsep{0pt}}}
\ifthenelse{\boolean{pdflatex}}
{\vspace*{-1.5cm}\mbox{\!\!\!\includegraphics[width=.14\textwidth]{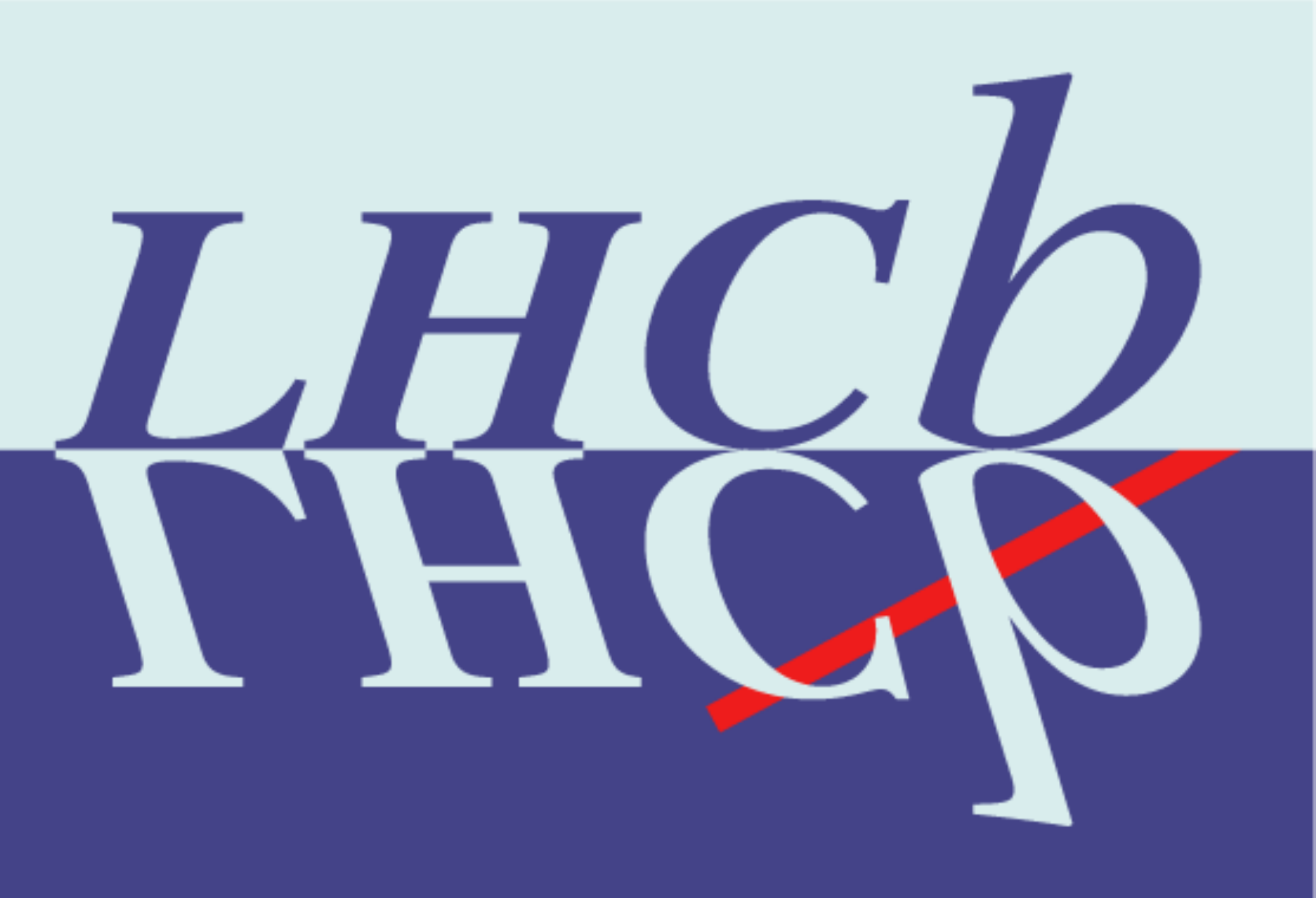}} & &}%
{\vspace*{-1.2cm}\mbox{\!\!\!\includegraphics[width=.12\textwidth]{lhcb-logo.eps}} & &}%
\\
 & & CERN-EP-2019-108\\  
 & & LHCb-PAPER-2019-013 \\  
 & & 19 March 2020 \\
 & & \\
\end{tabular*}

\vspace*{2.0cm}

{\normalfont\bfseries\boldmath\huge
\begin{center}
  \papertitle
\end{center}
}

\vspace*{0.5cm}

\begin{center}
\paperauthors\footnote{Authors are listed at the end of this paper.}
\end{center}

\vspace{0.5cm}

\begin{abstract}
  \noindent
The decay-time-dependent \CP asymmetry in $\Bs\to\jpsi \Kp \Km$ decays is measured
    using proton-proton collision data, corresponding to an integrated luminosity of
    $1.9\invfb$, collected with the LHCb detector at a centre-of-mass energy of $13\,\mathrm{TeV}$ in 2015 and 2016. Using a sample of approximately 117\,000 signal decays with an invariant $\Kp \Km$ mass in the vicinity of the $\phi(1020)$ resonance, the \CP-violating
    phase $\phi_s$ is measured, along with the difference in decay widths of the
    light and heavy mass eigenstates of the $\Bs$-$\Bsb$ system, $\Delta\Gamma_s$.
    The difference of the average $\Bs$ and $\Bd$ meson decay widths, $\Gamma_s-\Gamma_d$, is determined using in addition a sample of \mbox{$\Bz \to \jpsi \Kp \pim$} decays. 
    The values obtained are
    \mbox{$\phi_s = -0.083\pm0.041\pm0.006\rad$}, \mbox{$\Delta\Gamma_s = 0.077 \pm 0.008 \pm 0.003 \invps$}
    and \mbox{$\Gamma_s-\Gamma_d = -0.0041 \pm 0.0024  \pm 0.0015\invps$},
    where the first
    uncertainty is statistical and the second systematic. These are the most
    precise single measurements of these quantities to date and are consistent
    with expectations based on the Standard Model and with a previous LHCb analysis of this decay using
    data recorded at centre-of-mass energies 7 and 8 TeV. Finally, the results are combined with recent results from $\Bs\to\jpsi\pip\pim$ decays obtained using the same dataset as this analysis, and with previous independent LHCb results.
\end{abstract}

\vspace*{0.5cm}

\begin{center}
  Submitted to EPJC
\end{center}

\vspace{\fill}

{\footnotesize
\centerline{\copyright~\papercopyright. \href{\paperlicenceurl}{\paperlicence}.}}
\vspace*{2mm}

\end{titlepage}


\newpage
\setcounter{page}{2}
\mbox{~}

\cleardoublepage


\renewcommand{\thefootnote}{\arabic{footnote}}
\setcounter{footnote}{0}



\pagestyle{plain} 
\setcounter{page}{1}
\pagenumbering{arabic}


%

\section{Introduction}
\label{sec:Introduction}


The existence of new phenomena beyond those predicted by the Standard Model
(SM), hereafter referred to as New Physics (NP), could introduce sizeable
effects on \CP-violating observables. In the SM, \CP violation originates from an irreducible complex phase in the 
Cabibbo--Kobayashi--Maskawa (CKM) matrix that describes the mixing of the mass and weak interaction eigenstates of the quarks~\cite{Cabibbo:1963yz,Kobayashi:1973fv}.
In decays of a $\Bs$ meson to a \CP eigenstate, \CP violation can originate from the 
interference of the amplitude of the decay and that of the adjoint decay preceded by $\Bs$-$\Bsb$ oscillation.
It manifests itself through a nonzero value of the phase $\phi_s = - \mathrm{arg} \left( \lambda \right)$, where
the parameter $\lambda \equiv \left({q}/{p}\right) \left({\overline{A}}/{A}\right)$ describes \CP violation in the interference between mixing and decay.
Here, $A$ and $\overline{A}$ are the amplitudes for a $B_s^{0}$
or a $\Bsb$ meson to decay to the same final state and the complex parameters
$p = \braket{B_s^{0}}{B_{\rm L}}$ and $q=\braket{\Bsb}{B_{\rm L}}$ describe the
relation between the flavour and the mass eigenstates, light, L, and heavy, H. The two eigenstates have a decay width difference $\Delta \Gamma_s \equiv \Gamma_{\rm L} - \Gamma_{\rm H}$ and a mass difference $\Delta m_s \equiv m_{\rm H} - m_{\rm L}$. In the absence of $\CP$ violation in the decay and assuming negligible $\CP$ violation in $\Bs$-$\Bsb$ mixing~\cite{LHCb-PAPER-2013-033}, $\maglambda$ is expected to be unity.
In the SM, ignoring subleading contributions,
the phase $\phi_{s}$ can be related to the CKM matrix elements
$V_{ij}$, such that $\phi_s \approx
-2\beta_s$, where $\beta_s \equiv \arg [ - (V_{ts}V_{tb}^*)/(V_{cs}V_{cb}^*)]$. Global
fits to experimental data, assuming unitarity of the CKM matrix, give a precise prediction of a small value, 
namely $-2\beta_s = -0.0369 ^{+0.0010}_{-0.0007}\rad$ according to the
CKMfitter group~\cite{CKMfitter2015} and $-2\beta_s = -0.0370 \pm 0.0010\rad$
according to the UTfit collaboration~\cite{UTfit-UT}. 
However, many NP
models~\cite{Buras:2009if,Chiang:2009ev} predict larger values for this phase if
non-SM particles were to contribute to $\Bs$-$\Bsb$ oscillations, while
satisfying all existing constraints.  
Thus, a measurement of $\phi_s$ 
different from the SM prediction would provide clear evidence for NP.

Due to its high yield and clean experimental signature, the most sensitive decay channel to NP contributions is $\Bs \to \jpsi(\to \mu^+ \mu^-) \Kp \Km$~\cite{Faller:2008gt},
where the kaon pair predominantly originates from the decay of a $\phi(1020)$ resonance.\footnote{The inclusion of charge-conjugate processes
is implied throughout this paper, unless otherwise noted. For simplicity, the resonance $\phi(1020)$ is referred to as $\phi$ in the following.} Angular momentum
conservation in the decay
implies that the final state is an admixture of \CP-even and \CP-odd components,
with orbital angular momentum of 0 or 2, and 1, respectively. Moreover, along with the three polarisation states  
of the $\phi$ meson (P-wave states), there is also a \CP-odd $\Kp \Km$ component in an S-wave state~\cite{Stone:2008ak}. The data can therefore be described considering four polarisation amplitudes $A_g = |A_g|e^{-i\delta_g}$, 
where the indices $g \in \{0,\parallel,\perp,S\}$
refer to the
longitudinal, transverse-parallel and transverse-perpendicular relative orientations of the
linear polarisation vectors of the $J/\psi$ and $\phi$ mesons and $S$ to the single S-wave amplitude, respectively.
The \CP-even and \CP-odd components are disentangled by a decay-time-dependent angular analysis, where the angular observables $\cos\theta_{K}$, $\cos\theta_{\mu}$ and $\phi_{h}$ are defined in the helicity basis as described in Ref.~\cite{LHCb-PAPER-2013-002}.
The polar angle $\theta_{K}$ ($\theta_{\mu}$) is the angle between the $K^+$ ($\mu^{+}$) momentum and the
direction opposite to the \Bs momentum in the $K^{+}K^{-}$ ($\mu^{+}\mu^{-}$) centre-of-mass system and $\phi_{h}$ is
the azimuthal angle between the $K^{+}K^{-}$ and $\mu^{+}\mu^{-}$ decay planes. The $\phi_{h}$ angle is defined by a rotation from the $K^{-}$ side of the $K^{+}K^{-}$ plane to the $\mu^{+}$  side of the $\mu^{+}\mu^{-}$ plane. The rotation is positive in the $\mu^{+}\mu^{-}$ direction in the \Bs rest frame.

A decay-time-dependent angular analysis also allows the determination of $\Delta \Gamma_s$,
and of the average $\Bs$ decay width, $\Gamma_s \equiv \left( \Gamma_{\rm L} + \Gamma_{\rm H}\right)/2$.
In the SM, $\Gamma_s$ and $\Delta\Gamma_s$ can be calculated within the framework of the heavy quark expansion (HQE) 
theory~\cite{KhozeShifman,VoloskinShifman,VoloskinShifman1986,Bigi:1992su,Bigi:1995jr,Uraltsev:1998bk,Neubert:1997gu},
where a perturbative expansion of the amplitudes in inverse powers of the $b$-quark mass is used to calculate $b$-hadron observables. The ratio of the average decay width of $\Bs$ and $\Bd$ mesons, $\Gamma_s/\Gamma_d$, is usually the preferred observable to compare with experimental measurements as it allows the suppression of common uncertainties in the calculation. The predictions are $\Delta \Gamma_s=0.088\pm0.020\invps$~\cite{Artuso:2015swg} and
$\Gamma_s/\Gamma_d=1.0006\pm0.0025$~\cite{Kirk:2017juj}. 
The high precision of the ratio $\Gamma_s/\Gamma_d$ makes it an excellent testing ground for 
the validity of the HQE ~\cite{Jager:2017gal,Kirk:2017juj}. In addition, $\Delta\Gamma_s$ can provide bounds complementary to those from $\Gamma_s/\Gamma_d$ on
quark-hadron duality violation~\cite{Jubb:2016mvq}.

Measurements of $\phi_s$, $\Delta\Gamma_s$ and $\Gamma_s$
using $\Bs \to \jpsi \Kp \Km$ decays, with $\jpsi \to \mu^+ \mu^-$, have been previously reported by
the D0~\cite{Abazov:2011ry}, CDF~\cite{Aaltonen:2012ie},
ATLAS~\cite{Aad:2014cqa,*ATLAS-CONF-2019-009}, CMS~\cite{Khachatryan:2015nza} and
LHCb~\cite{LHCb-PAPER-2014-059} collaborations. The LHCb collaboration has also
exploited different decay channels, namely $\Bs \to \jpsi \pip
\pim$~\cite{LHCb-PAPER-2014-019}, $\Bs \to \psi(2S)
\phi$~\cite{LHCb-PAPER-2016-027}, $\Bs \to D_s^+
D_s^-$~\cite{LHCb-PAPER-2014-051} and $\Bs \to \jpsi \Kp \Km$ for the $\Kp\Km$
invariant-mass region above 1.05\gevcc~\cite{LHCb-PAPER-2017-008}. The world-average values, including all of the above mentioned results, are
$\phi_s=-0.021 \pm 0.031\rad$, $\Delta
\Gamma_s = 0.085 \pm 0.006\invps$ and
$\Gamma_s/\Gamma_d= 1.006 \pm 0.004$~\cite{HFLAV16}. They are in agreement with the abovementioned predictions.

The main parameters of interest in this paper are $\phis$, $\maglambda$, $\Gs-\Gd$, $\DGs$ and $\Delta m_{s}$ measured in $\Bs \to \jpsi \Kp\Km$ decays, in the $\Kp\Km$ mass region \mbox{0.99--1.05\gevcc}. The new measurement reported is based on a data sample of proton-proton collisions recorded at a centre-of-mass energy of $\sqrt{s}=
13\tev$ in 2015 and 2016 during Run 2 of LHC operation, corresponding to an integrated luminosity
of $0.3\invfb$ and $1.6\invfb$, respectively. 
The decay width difference $\Gamma_s-\Gamma_d$ is determined using $\Bz \to \jpsi \Kp \pim$ decays as a reference, reconstructed in the same data set as the signal. The $\Kp \pim$ in the final state originates predominantly from the decay of a $K^*(892)^0$ resonance. The analysis procedure gives access to $\Gamma_s-\Gamma_d$ rather than $\Gamma_s$ due to the dependence of the time efficiency parametrisation on $\Gamma_d$. This allows the determination of $\Gamma_s-\Gamma_d$ with a significant reduction of the systematic uncertainty associated with lifetime-biasing selection requirements compared to the previous measurement. Taking as an input the 
precisely known value of $\Gamma_d$~\cite{HFLAV16}, the ratio $\Gamma_s/\Gamma_d$ may be determined with higher precision with respect to measuring the two lifetimes independently.

In this analysis, the polarisation-independent \CP{}-violating parameter $\lambda_r$, associated with
each polarisation state $r$, is defined such that $\lambda_r= \eta_r \lambda$, where $\eta_r =  +1   \:\  \text{for $r\in\{0,\parallel\}$}$ and $\eta_r =-1 \: \text{for $r\in\{\perp,{\rm S}\}$}$. As a consequence, $\phi_s = -\arg{\lambda}$.
However, this assumption can be relaxed such that the values of
$\phi_{s}^{r}$ and $|\lambda_r|$ are measured separately for each polarisation state. In addition, the following quantities are
measured: the $\phi$ polarization fractions $|A_{0}|^{2}$ and $|A_{\perp}|^{2}$; the strong-phase differences $\delta_{\perp} - \delta_{0}$ and $\delta_{\parallel} - \delta_{0}$; 
the fraction of S-wave, $F_{\rm S}$, and the phase difference $\delta_{S}-\delta_{\perp}$. The S-wave parameters are measured in bins of $m(\Kp\Km)$. The sum
$|A_{0}|^{2}+|A_{\perp}|^{2}+|A_{\parallel}|^{2}$
equals unity and by convention $\delta_{0}$ is zero.

After a brief description of the LHCb detector in Sec.~\ref{sec:Detector}, the
candidate selection and the background subtraction using the $sPlot$ technique~\cite{Pivk:2004ty} are outlined in Sec.~\ref{sec:Selection}. The relevant inputs
to the analysis, namely the decay-time resolution, the decay-time efficiency, the angular efficiency and the flavour-tagging calibration, are described in
Secs.~\ref{sec:TimeResolution}, \ref{sec:TimeAcc}, \ref{sec:AngAcc} and \ref{sec:Tagging}, respectively. The $sFit$ procedure~\cite{sFit}, 
the evaluation of the systematic uncertainties and the results are discussed in Secs.~\ref{sec:Fit}, \ref{sec:Systematics} and \ref{sec:Results}, respectively. 
The combination of the results obtained in this analysis with those measured by the LHCb collaboration using data collected in 2011 and 2012 and determined using 2015 and 2016 $\Bs\to\jpsi\pip\pim$ data is presented in Sec.~\ref{sec:Combination}.
Finally, conclusions are drawn in Sec.~\ref{sec:Conclusions}.

\section{Detector and simulation}
\label{sec:Detector}

The \lhcb detector~\cite{LHCb-DP-2012-002,LHCb-DP-2014-002} is a single-arm forward
spectrometer covering the \mbox{pseudorapidity} range $2<\eta <5$, designed for
the study of particles containing \bquark or \cquark quarks. The detector
includes a high-precision tracking system consisting of a silicon-strip vertex
detector surrounding the $pp$ interaction
region, a large-area silicon-strip detector
located upstream of a dipole magnet with a bending power of about
$4{\mathrm{\,Tm}}$, and three stations of silicon-strip detectors and straw
drift tubes placed downstream of the magnet.
The tracking system provides a measurement of the momentum, \ptot, of charged
particles with a relative uncertainty that varies from 0.5\% at low momentum to
1.0\% at 200\gevc.  The minimum distance of a track to a primary vertex (PV),
the impact parameter (IP), is measured with a resolution of $(15+29/\pt)\mum$,
where \pt is the component of the momentum transverse to the beam, in\,\gevc.
Different types of charged hadrons are distinguished using information from two
ring-imaging Cherenkov detectors.  Photons,
electrons and hadrons are identified by a calorimeter system consisting of
scintillating-pad and preshower detectors, an electromagnetic and a
hadronic calorimeter. Muons are identified by a system composed of alternating
layers of iron and multiwire proportional
chambers.

Samples of simulated events are used to optimise the signal selection, 
to derive the angular efficiency and to correct the decay-time efficiency. In simulations, $pp$ collisions are generated using
\pythia~\cite{Sjostrand:2007gs, *Sjostrand:2006za} with a specific
\lhcb configuration~\cite{LHCb-PROC-2010-056}.  Decays of hadronic particles are
described by \evtgen~\cite{Lange:2001uf}, in which final-state radiation is
generated using \photos~\cite{Golonka:2005pn}. The interaction of the generated
particles with the detector, and its response, are implemented using the \geant
toolkit~\cite{Allison:2006ve, *Agostinelli:2002hh} as described in
Ref.~\cite{LHCb-PROC-2011-006}. The $\Bs \to J/\psi \phi$ simulated sample used in this analysis
is generated taking into account the three possible polarization states of the $\phi$ meson while S-wave contributions
are not included.

\section{Selection and mass fit}
\label{sec:Selection}

Events are first required to pass an online event selection performed
by a trigger~\cite{LHCb-DP-2012-004}, which consists of a hardware stage, based
on information from calorimeters and muon systems, followed by a software stage,
which applies a full event reconstruction. At the hardware stage, events are
required to have a muon with high $\pt$ or a hadron, photon or electron with
high transverse-energy deposit in the calorimeters. A difference with respect to the previous analysis is that all the events passing any of the hardware
trigger requirements are accepted. This increases the signal yield by 13\% in 2015 and by 7\% in 2016 with respect to using the muon system information
only. The different signal gain in the two data taking years is due to tighter L0 trigger thresholds employed in the 2015 data. The subsequent software trigger consists of two separate stages. In the first stage, the events can be divided into two categories. In the first
category, they are required to have two well-identified oppositely charged
muons with invariant mass larger than \mbox{2700\mevcc}.  This trigger has an
almost uniform efficiency as a function of $\Bs$ decay time and will be referred to as
\textit{unbiased}. In the second category, events are retained if there is at least
one muon with transverse momentum larger than about 1\gevc and with a large impact-parameter significance with respect to all PVs in the event. The
latter is defined as the difference in the vertex-fit $\chi^2$ of the PV fitted
with and without the considered track. Events are also included in the second category if they pass the selection by a multivariate algorithm that identifies a two-track
good-quality secondary vertex with a large scalar sum of the $\pt$ of the
associated charged particles and a significant displacement from the PVs. These
triggers, whose selection thresholds changed slightly between 2015 and 2016 data taking,
introduce a nontrivial dependence of the efficiency on the $\Bs$ decay time and will
be referred to as \textit{biased}. In the second stage of the trigger, events
containing a $\mup \mu^-$ pair with invariant mass within 120\mevcc of the
$\jpsi$ mass~\cite{PDG2018} and which form a vertex that is significantly displaced
from the PV are selected, introducing another small decay-time bias. 

In the offline selection, the $\jpsi$ meson  candidates  are  formed from  two  oppositely charged
particles,  originating  from  a  common  vertex,  which  are identified
as muons and which have $\pt$ larger than 500\mevc.  The invariant mass of the
$\mup \mu^-$ pair, $m(\mup \mu^-)$, must be in the range \mbox{3020--3170\mevcc}.
The $\jpsi$ meson  candidates are combined with $\Kp\Km$ candidates formed  from  two
oppositely charged  particles  that  are  identified  as  kaons  and
that  originate  from  a  common  vertex.   The $\Kp\Km$ pair is  required  to
have $\pt$ larger  than  500\mevc.   The  invariant  mass  of  the $\Kp\Km$
pair, $m(\Kp\Km)$, must be in the range 
\mbox{990--1050\mevcc}.  The $\Bs$
candidates are reconstructed by combining the $\jpsi$ candidate with the
$\Kp\Km$ pair, requiring that they form a good vertex and have an invariant mass, $m(\jpsi\Kp\Km)$, in the range \mbox{5200--5550\mevcc}. The \Bs origin vertex is defined as the PV in the interaction, or if multiple PVs are reconstructed the PV with the minimum value of the $\Bs$ impact parameter significance is associated with the candidate. The invariant mass is calculated from a kinematic fit that constrains the $\Bs$ candidate to originate from its
origin vertex and constrains $m(\mup \mu^-)$ to the known $\jpsi$ mass~\cite{PDG2018}.
When deriving the decay time, $t$, and the helicity angles of the $\Bs$ candidate the origin vertex constraint is also applied.
In addition, $t$ is required to be in the range \mbox{0.3--15.0\ps}, which 
suppresses a large fraction of prompt combinatorial
background whilst having a negligible effect on the sensitivity to $\phi_s$.
The kinematic fit also estimates a per-candidate decay-time uncertainty, $\delta_t$.

The selection is optimised with respect to the previous analysis \cite{LHCb-PAPER-2014-059}
by means of a gradient-boosted decision
tree~(BDT)~\cite{Breiman,AdaBoost}, which is used to further suppress combinatorial
background. To train the BDT, simulated $\Bs \to \jpsi \phi$ candidates are used as a
signal sample and data candidates with $m(\jpsi \Kp \Km)$ in the range \mbox{5450--5550\mevcc}
are used as a sample of combinatorial background. 
The simulation is corrected to match the distributions observed in data of
particle identification variables,
the $\Bs$ transverse momentum and pseudorapidity, the quality of the muon and kaon track fits and the number of tracks in an event with measurements both in the 
VELO and the tracking stations.
Various input quantities are used in the BDT to exploit the features of the signal decay in order to distinguish it from background, 
namely the track-fit $\chi^2$ of the final-state particles, the particle identification probability as
provided mainly from the RICH and muon systems, the quality of the candidate $\jpsi$
and $\Bs$ decay vertices, the $\pt$ of the $\Bs$ candidate and of the $\Kp\Km$ combination
and the $\Bs$ IP with respect to its origin vertex. The  selection  requirement on the BDT  output is
chosen to maximise the effective signal sample size approximated by the square of the sum of {\it sWeights} divided by sum of squared {\it sWeights}. 

\begin{figure}
    \centering
    \includegraphics[width=0.58\textwidth]{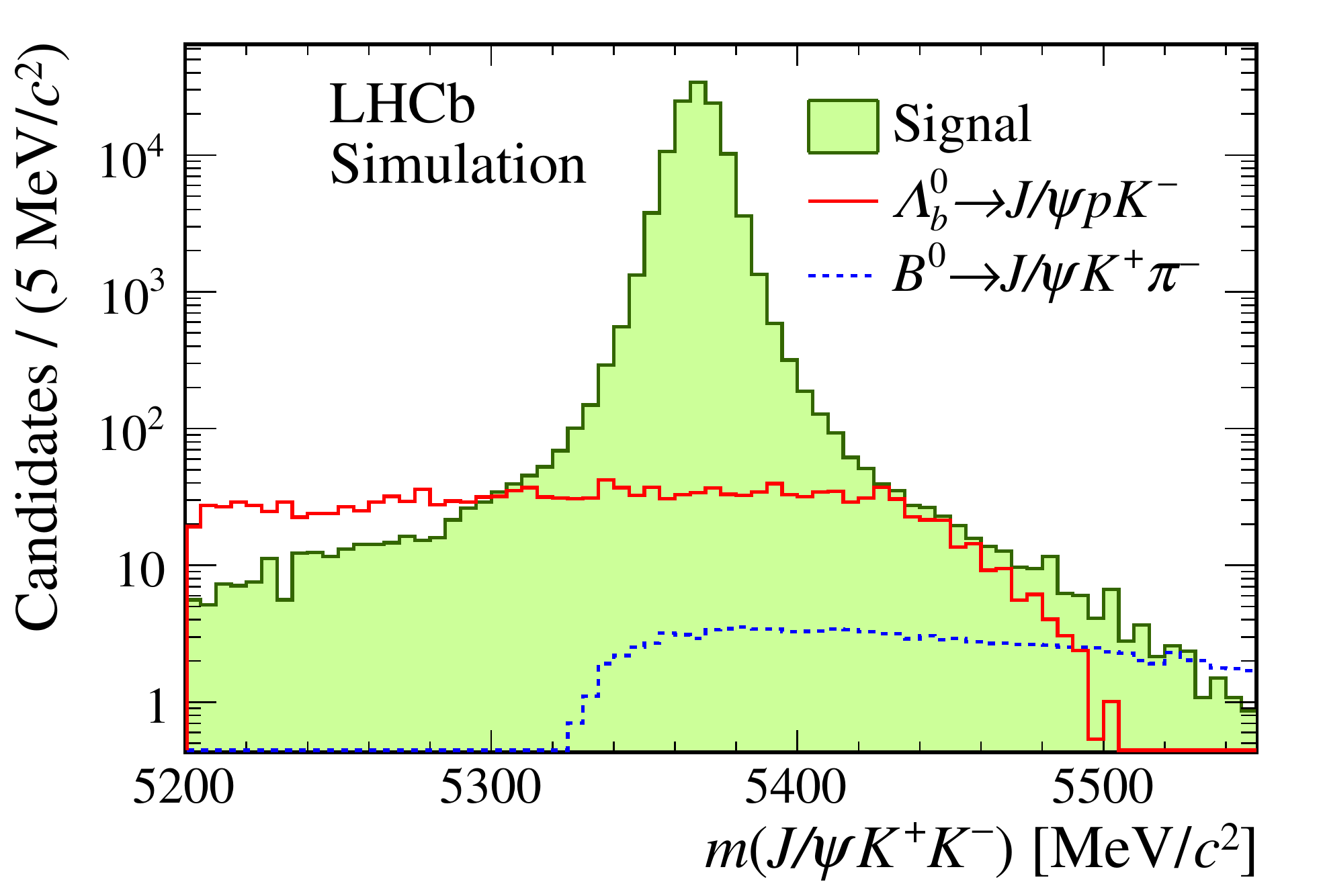}
    \caption{\label{fig:Bmass_peaking_bkg}\small Distribution of the invariant mass of \Bs candidates, selected from simulated
    \mbox{$\Bs\to\jpsi\Kp\Km$} (green filled area), \mbox{$\Lb \to \jpsi p \Km$}  (solid red line) and \mbox{$\Bd \to \jpsi \Kp \pim$} (dotted blue line) decays. The distributions are weighted to correct differences in the kinematics and the resonance content between simulation and data.}
\end{figure}

In addition to
combinatorial background, studies of the data in sidebands of the $m(\jpsi \Kp
\Km)$ spectrum show contributions from approximately 5200 $\Lb \to \jpsi p \Km$ (350 $\Bd \to \jpsi \Kp \pim$) 
decays where the proton (pion) is misidentified as a kaon. These backgrounds lie around the
$\Bs$ signal peak in the $m(\jpsi \Kp \Km)$ distribution, as shown in Fig.~\ref{fig:Bmass_peaking_bkg}.
These contributions are suppressed using more stringent kaon
identification requirements if the $m(\jpsi \Kp \Km)$
mass, with the kaon interpreted as a proton (pion), lies within $15\mevcc$
around the $\Lb$ ($\Bd$) known mass~\cite{PDG2018}. This reduces the $\Bd\to \jpsi \Kp \pim$
peaking background contribution to approximately 120 decays. 
This background is neglected and a systematic uncertainty is assigned to account for this approximation. 
The contribution due to the $\Lb$ background is $1600\pm160$, where the uncertainty includes statistical and systematic sources. The $\Lb$ background is statistically subtracted by inserting simulated $\Lb$ 
decays into the data sample with negative weights. This is done prior to the \sPlot procedure, in which the combinatorial background is subtracted in a fit to $m(\jpsi \Kp \Km)$. 
Correlations between the candidate mass and the angular variables are preserved and the simulated candidates are weighted such that the
distributions of the kinematic variables used in the fit, and their correlations,
match those of data.

\begin{figure}
    \centering
    \includegraphics[width=0.49\textwidth]{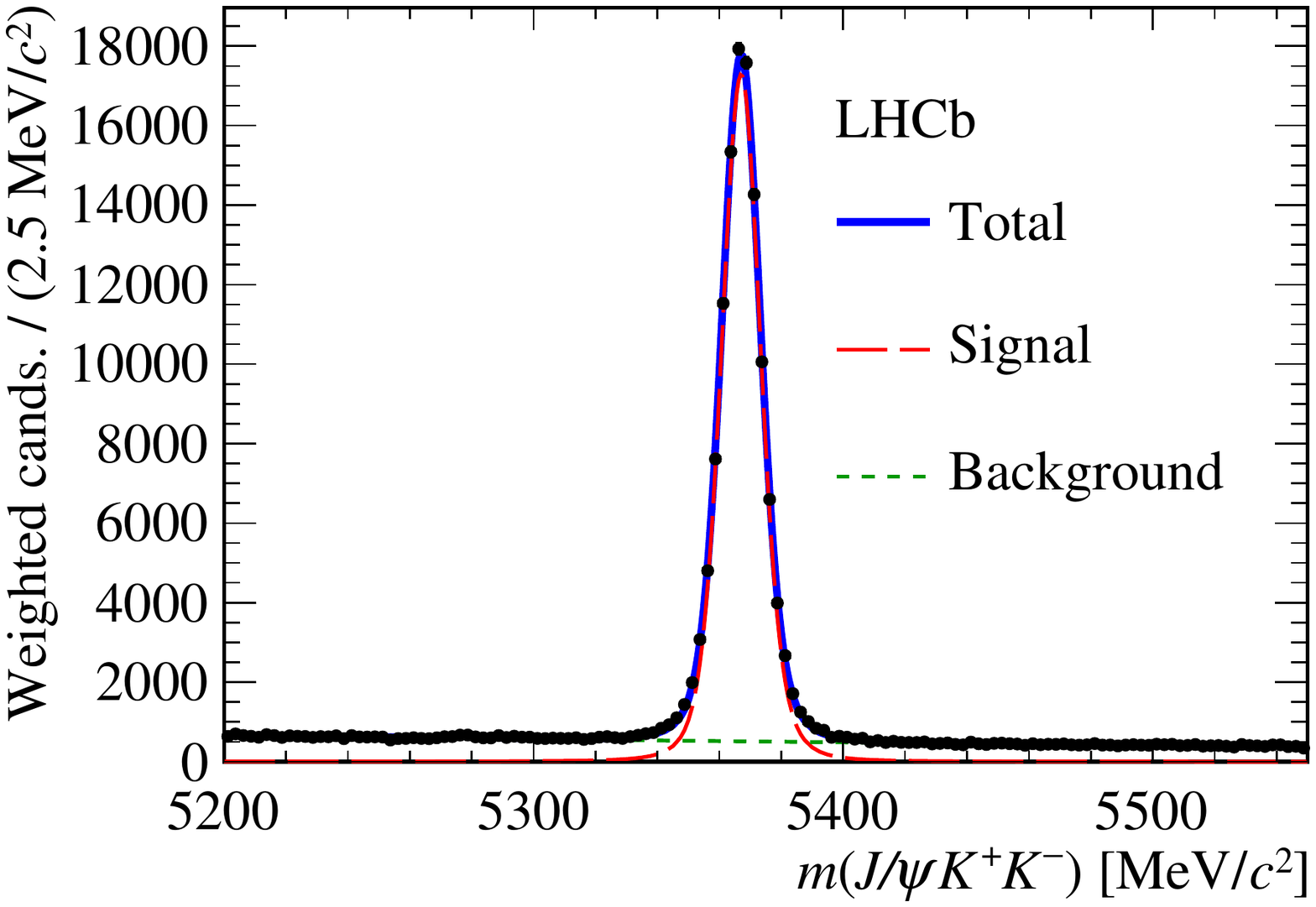}
    \put(-160,120){(a)}
    \includegraphics[width=0.49\textwidth]{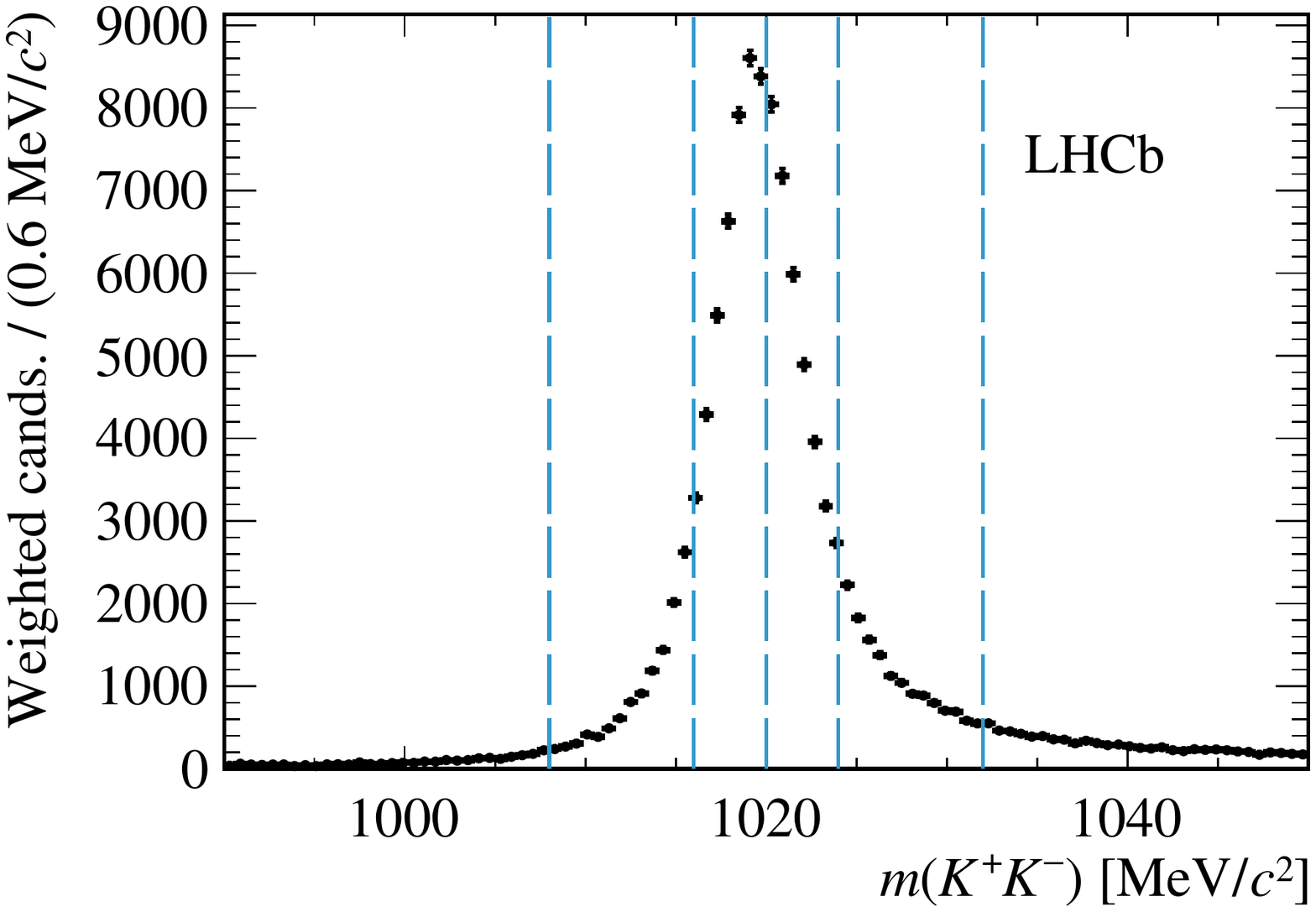}
    \put(-160,120){(b)}
    \caption{\label{fig:Bmass}\small (a) Distribution of the invariant mass of selected
    \mbox{$\Bs\to\jpsi\Kp\Km$} decays. The signal component is shown by the long-dashed red line, the background
    component by the dashed green line and the total fit function by the solid blue line. The background contribution due to \mbox{$\Lb \to \jpsi p \Km$} decays is statistically subtracted. The contribution from \mbox{$\Bz\to\jpsi\Kp\Km$}decays is not shown separately due to its small size. (b) Distribution of $\Kp\Km$
    invariant mass from selected
    \mbox{$\Bs\to\jpsi\Kp\Km$} decays. The background is subtracted using the $sPlot$ method. The dashed blue lines define the boundaries of the six $m(K^{+}K^{-})$ bins that are used in the analysis.}
\end{figure}

Figure~\ref{fig:Bmass}(a) shows the $m(\jpsi \Kp \Km)$ distribution and
the result of an unbinned maximum-likelihood fit to the
sample in the range 5200--5550\mevcc. The sample is divided into 24 independent subsamples, corresponding to six bins in $m(\Kp \Km)$ with boundaries at 990, 1008, 1016, 1020, 1024, 1032, 1050\mevcc,
to the biased and the unbiased trigger categories, and to the year of data taking.
The probability density function (PDF) used for the fit is independent for each of these subsamples and is 
composed of
a single double-sided Crystal Ball (CB)~\cite{Skwarnicki:1986xj} function for the signal and an
exponential function for the combinatorial background. The CB tail parameters are fixed to those obtained from simulation.

The \sPlot technique relies on the variable used for background subtraction to be uncorrelated with the variables to which the {\it sWeights} are applied. However, a correlation between the signal mass shape with $\cos\theta_{\mu}$ is observed, due to the dependence of the mass resolution on the transverse momentum of the muons.
The per-candidate mass uncertainty, $\sigma_m$, obtained in the vertex and kinematic fit used to
obtain $m(\jpsi \Kp \Km)$, is found to represent a good proxy of $\cos\theta_{\mu}$ due to its correlation with the \Bs candidate mass resolution. 
Therefore, the signal function uses $\sigma_m$ as a conditional observable. 
The width parameter $\sigma_{\mathrm{CB}}$ of the double-sided CB function is parametrised as a 
quadratic function of the per-candidate mass uncertainty such that
$\sigma_{\mathrm{CB}} = a_1\sigma_m + a_2\sigma_m^2$, $a_1$ and
$a_2$ are free parameters determined from the data. The quadratic dependence is motivated by simulation studies.

A small contribution from $\Bd \to \jpsi \Kp \Km$ background candidates is
observed at the known $\Bd$ mass~\cite{PDG2018}. This contribution is included in the PDF and is
modelled with a Gaussian distribution, where the mean is fixed to the
fitted $\Bs$ mass minus the difference between $\Bs$ and $\Bd$ masses~\cite{PDG2018}
and the resolution is fixed to 7\mevcc,
which is determined from a fit to the $\Bd \to \jpsi \Kp \pim$ data control channel. Figure~\ref{fig:Bmass}(b) shows the background-subtracted invariant-mass
distributions of the $\Kp\Km$ system in the selected
$\Bs\to\jpsi\Kp\Km$ candidates.
After the trigger and full offline selection requirements, the signal yield totals approximately 15\,000 and 102\,000 
$\Bs \to \jpsi \Kp \Km$ decays in the 2015 and 2016 data sets, respectively. 

The fraction of events containing more than
one \Bs candidate within the $m(J/\psi K^{+}K^{-})$ range \mbox{$5340$--$5400\mevcc$} is 0.3\%.
All candidates are retained in the subsequent stages of the
analysis and a systematic uncertainty on the impact of allowing multiple candidates per event to be present in the analysis is assigned.

\section{Decay-time resolution}
\label{sec:TimeResolution}
The value and the uncertainty of the decay-time resolution strongly affects the relative precision on $\phi_s$, 
thus the knowledge of the decay-time resolution calibration is pivotal. 
The resolution function is modelled with a Gaussian distribution with a mean of zero and a width $\sigma_{\rm eff}$, where $\sigma_{\rm eff}$ is determined using a 
sample of candidates constructed from combinations of $\jpsi$, $\Kp$ and $\Km$ candidates that originate predominantly in the primary interaction (prompt component).
This sample is referred to as the prompt $\jpsi\Kp\Km$ sample. It is selected as described in Sec.~\ref{sec:Selection} for
$\Bs\to\jpsi\Kp\Km$ decays except for the lower limit requirement for the decay time, by making use of a different trigger line which is heavily prescaled.

The prompt component has zero decay time and is used to
calibrate the detector resolution by studying the shape of the decay-time distribution around zero. This distribution is modelled by a delta function.
In addition to the prompt component, there is a contribution at later decay times originating from $\jpsi$ mesons produced in $b$-hadron decays, 
and a small fraction of a background due to candidates that have a decay time computed with respect to a wrong PV (wrong-PV component). The $b$-hadron component contributes to a tail at positive decay times and is described by two exponential functions. 
The shape of the wrong-PV component is determined from a data control sample in which the decay-time distribution of
candidates is constructed by computing their decay time with respect to an independent PV from the following event. This contribution is found to be approximately $0.5\%$ of the prompt sample.

The sum of the prompt and $b$-hadron components is
convolved with a triple-Gaussian resolution function
\begin{equation}
    {\cal R}(t) = \sum_{i=1}^{3} f_i \frac{1}{\sqrt{2\pi}\sigma_i}
            \exp\left[-\frac{(t-\mu)^2}{2\sigma_i^2}\right],
        \label{eqn:decay_time_res_model}
\end{equation}
where $\sum_i f_i = 1$, $\mu$ is a parameter that describes a bias in the decay time measurement
and $\sigma_i$ are the individual widths. 
The bias, $\mu$, is assumed to be zero and a systematic uncertainty is assigned studying a possible deviation from this value. The rest of the parameters are determined from the fit.

\begin{figure}[tb]
    \centering
    \includegraphics[width=0.46\textwidth]{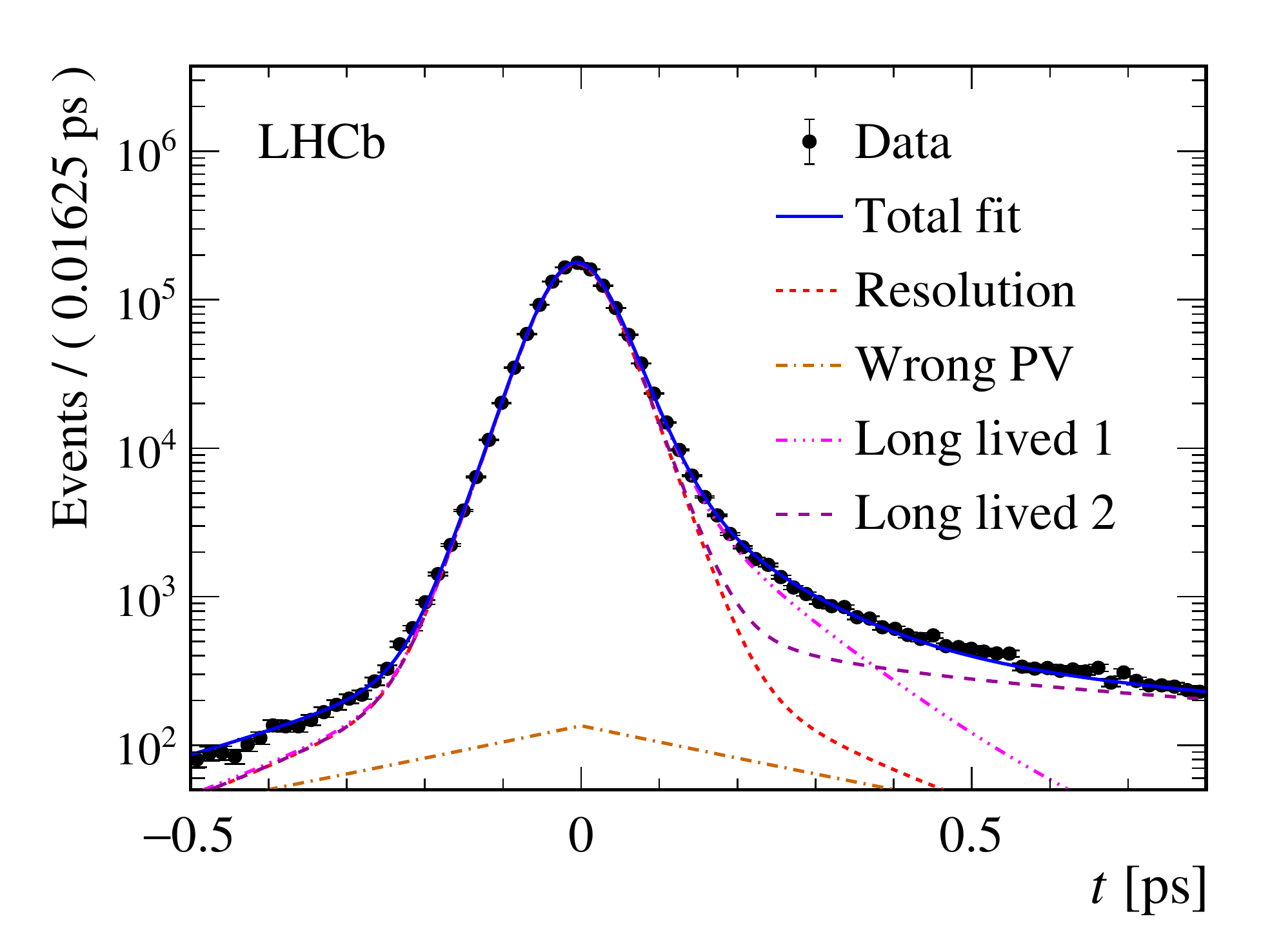}
    \put(-125,130){(a)}
    \includegraphics[width=0.52\textwidth]{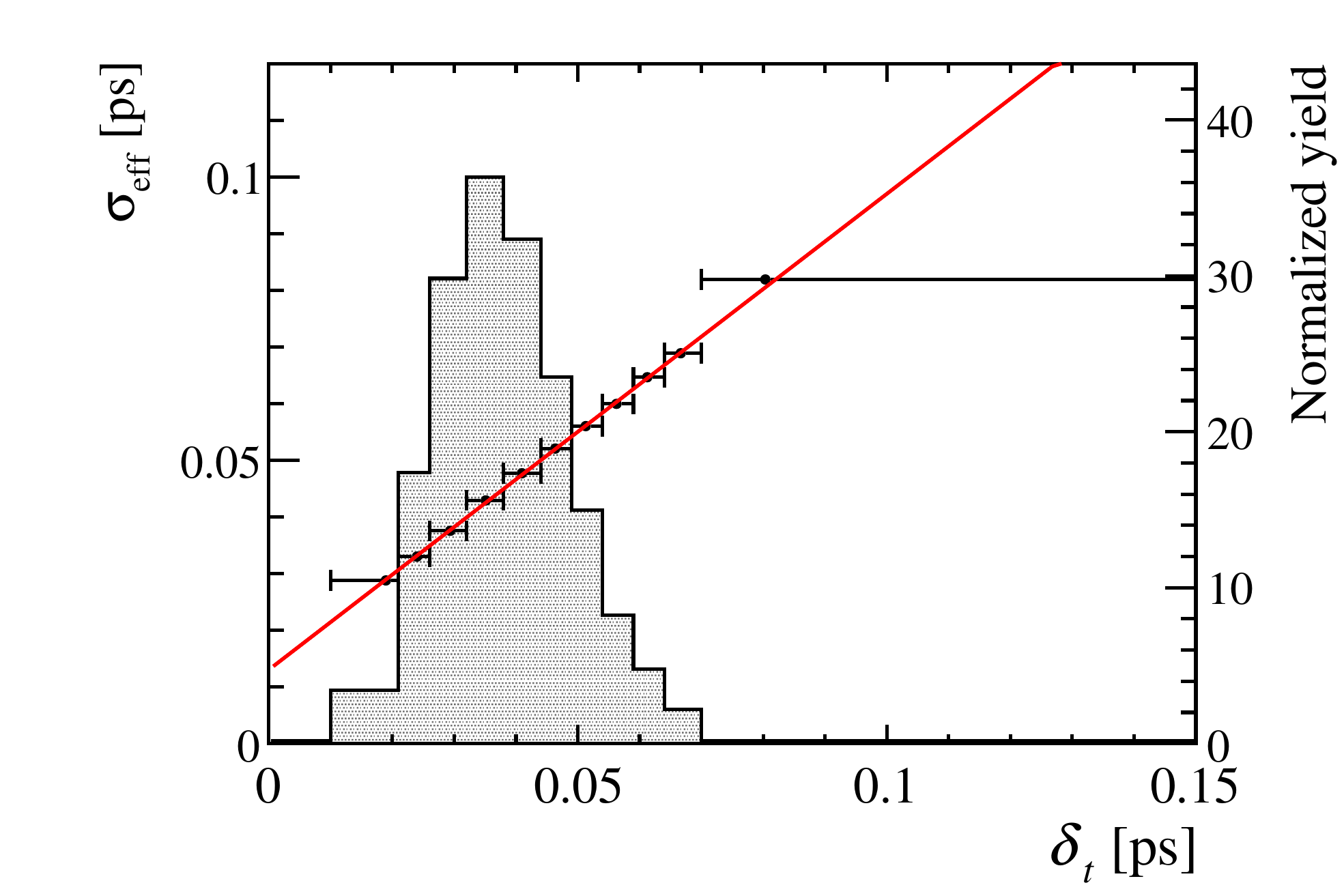}
    \put(-125,130){(b)}
    \caption{\label{fig:time_res_calib}\small (a) Decay-time distribution of the prompt $\jpsi\Kp\Km$ calibration sample with the result of an unbinned maximum-likelihood fit overlaid in blue. The overall triple-Gaussian resolution is represented by the dashed red line, while the two long-lived and the wrong-PV components are shown by the long-dashed-dotted and dashed-multiple-dotted brown and pink lines and the long-dashed purple line, respectively. (b) Variation of the effective single-Gaussian decay-time resolution,
    $\sigma_{\rm eff}$, as a
    function of the estimated per-candidate decay-time uncertainty, $\delta_t$, obtained from the prompt $\jpsi\Kp\Km$
    sample. The red line shows the result of a linear fit.  The data points are positioned at the barycentre of each $\delta_t$ bin. The shaded histogram (see right $y$ axis) shows the
    distribution of $\delta_t$ in the background-subtracted $\Bs\to\jpsi\Kp\Km$ sample.}
\end{figure}
The calibration sample is split into eleven subsets according to the per-candidate decay-time uncertainty, $\delta_t$. 
The model is fit to the decay-time distribution in order to extract the parameters governing the decay-time resolution of Eq.~\eqref{eqn:decay_time_res_model} as shown in Fig.~\ref{fig:time_res_calib}(a). 
The dilution of the amplitude of the $\Bs$-$\Bsb$ oscillation due to the calibrated resolution is determined in each bin of $\delta_t$ as 
\begin{equation}
    \label{eqn:time_dilution_tr}
    \mathcal{D} = \sum_{i=1}^{3}f_i \exp \left[-\sigma_i^2\Delta m_s^2/2\right],
\end{equation}
and is then used to evaluate an effective single-Gaussian width given by
\begin{equation}
    \label{eqn:time_eff_res_tr}
    \sigma_{\rm eff} = \sqrt{(-2/\Delta m_s^2)\ln \mathcal{D}}.
\end{equation}
This effective single-Gaussian resolution of width $\sigma_{\rm eff}$ gives the same
damping effect on the magnitude of the $\Bs$ meson oscillation as the triple-Gaussian model.
Figure~\ref{fig:time_res_calib}(b) shows the variation
of $\sigma_{\rm eff}$ as a function of $\delta_t$. The variation is
fit with a linear function
\mbox{$\sigma_{\mathrm{eff}}(\delta_t) = b_0 + b_1\delta_t$}
to determine the calibration parameters
\mbox{$b_0 = 12.97 \pm 0.22\fs$} and \mbox{$b_1 = 0.846 \pm 0.006$},
where the uncertainties are statistical only. A quadratic dependence is also evaluated and used as an alternate model to compute a systematic uncertainty. 
The calibration procedure is validated using simulated signal and prompt samples. The difference between the effective resolutions obtained in 
these simulated samples is approximately 0.8\fs and is treated as a source of systematic uncertainty.

The result of the calibration leads to an effective single-Gaussian resolution 
function averaged over the $\delta_t$ bins with \mbox{$\sigma_{\rm eff} = 45.54\pm0.04\pm0.05\fs$}, where the first uncertainty is statistical,
and the second contribution comes from the uncertainties on the calibration parameters. This corresponds to a dilution \mbox{${\cal D}=0.721\pm0.001$} assuming \mbox{$\Delta m_s = 17.757\pm0.021$\invps~\cite{HFLAV16}}.

\section{Decay-time efficiency}
\label{sec:TimeAcc}

The selection and reconstruction efficiency depends on the $B^0_s$ decay time due to
displacement requirements made on the signal tracks and a
decrease in reconstruction efficiency for tracks
with large impact parameter with respect to the beam line~\cite{LHCb-PAPER-2013-065}.
The efficiency as a function of the decay time is determined using a new technique with respect 
to Ref.~\cite{LHCb-PAPER-2014-059}, exploiting the $\Bd \to \jpsi \Kp \pim$ decay, with $\jpsi \to \mu^+ \mu^-$, as a control sample. 
This control mode is kinematically similar to the signal decay. Since
the decay-width difference between the two mass eigenstates in the $\Bz$ system is measured to be consistent with zero~\cite{PDG2018}, $\Bd \to \jpsi \Kp \pim$ candidates are
assumed to have a purely exponential decay-time distribution with lifetime $\tau^{\Bd}_{\mathrm{data}} = 1.520\ps$~\cite{HFLAV16}.
The \Bs efficiency is determined via a simultaneous fit to background-subtracted data and simulated samples through the relation
\begin{equation}
\varepsilon^{\Bs}_{\mathrm{data}}(t) = \varepsilon^{\Bd}_{\mathrm{data}}(t) \times \frac{\varepsilon^{\Bs}_{\mathrm{sim}}(t)}{\varepsilon^{\Bd}_{\mathrm{sim}}(t)} \,,
\label{eq:timeacc}
\end{equation}
where $\varepsilon^{\Bd}_{\mathrm{data}}(t)$ is the efficiency of the control channel and $\varepsilon^{\Bs}_{\mathrm{sim}}(t)/
\varepsilon^{\Bd}_{\mathrm{sim}}(t)$ is the ratio of
efficiencies of simulated signal and reference decays after reconstruction and selection.
Residual differences between either signal and control mode or data and simulation are automatically corrected for in the ratio of Eq.~\eqref{eq:timeacc}.
In order to correct first-order differences between the $\Bs$ and $\Bd$ data samples, the latter is weighted to match the $p$ and $\pt$ distribution of $\Bs$ data.
In addition, both $\Bs$ and $\Bd$ simulated samples are weighted to match the $\pt$ distribution of the $\Bs$ data sample.
The simulated samples are further corrected according to the ratio of the PDF used to generate them and the PDF obtained 
with the parameters measured in data~\cite{Aaij:2013cma,LHCb-PAPER-2014-059}.
Together with an additional weighting to match the $m(\Kp\pim)$ and $m(\Kp\Km)$ distributions in data, 
this procedure reproduces the correct mixture of P- and S-waves in the $\Kp\pim$ and $\Kp\Km$ final state.
The decay-time efficiency is obtained separately for the data-taking periods
2015 and 2016 and the two trigger categories.

The $\Bd \to \jpsi \Kp \pim$ candidates are selected using trigger and
preselection requirements similar to those of the
$\Bs \to \jpsi K^+ K^-$ channel. The main difference is an additional selection on the pion-identification requirement, in order to reduce the probability of reconstructing two different $\Bd$ candidates by swapping the kaon and pion mass hypotheses.
In addition, the \pt of the pion is required to be larger than
\mbox{$250\mevc$} to reduce the number of multiple candidates per event to 0.5$\%$ in the $m(\jpsi K^+ \pi^-)$ region \mbox{$5260$--$5300\mevcc$}. The invariant mass of the kaon-pion pair is required to be in the range \mbox{$826$--$966\mevcc$}. The BDT as trained and optimised on the signal channel is used, applying the same selection requirement. 
Several potential peaking backgrounds arising from the misidentification of particles are considered but they are all found to be negligible.
A small contribution from $B^0_s \to \jpsi K^+ \pi^-$ decays
is removed by selecting candidates with \mbox{$m(\jpsi K^+ \pi^-)<5350\mevcc$}.

\begin{figure}[tb]
        \centering
        \includegraphics[width=0.49\textwidth]{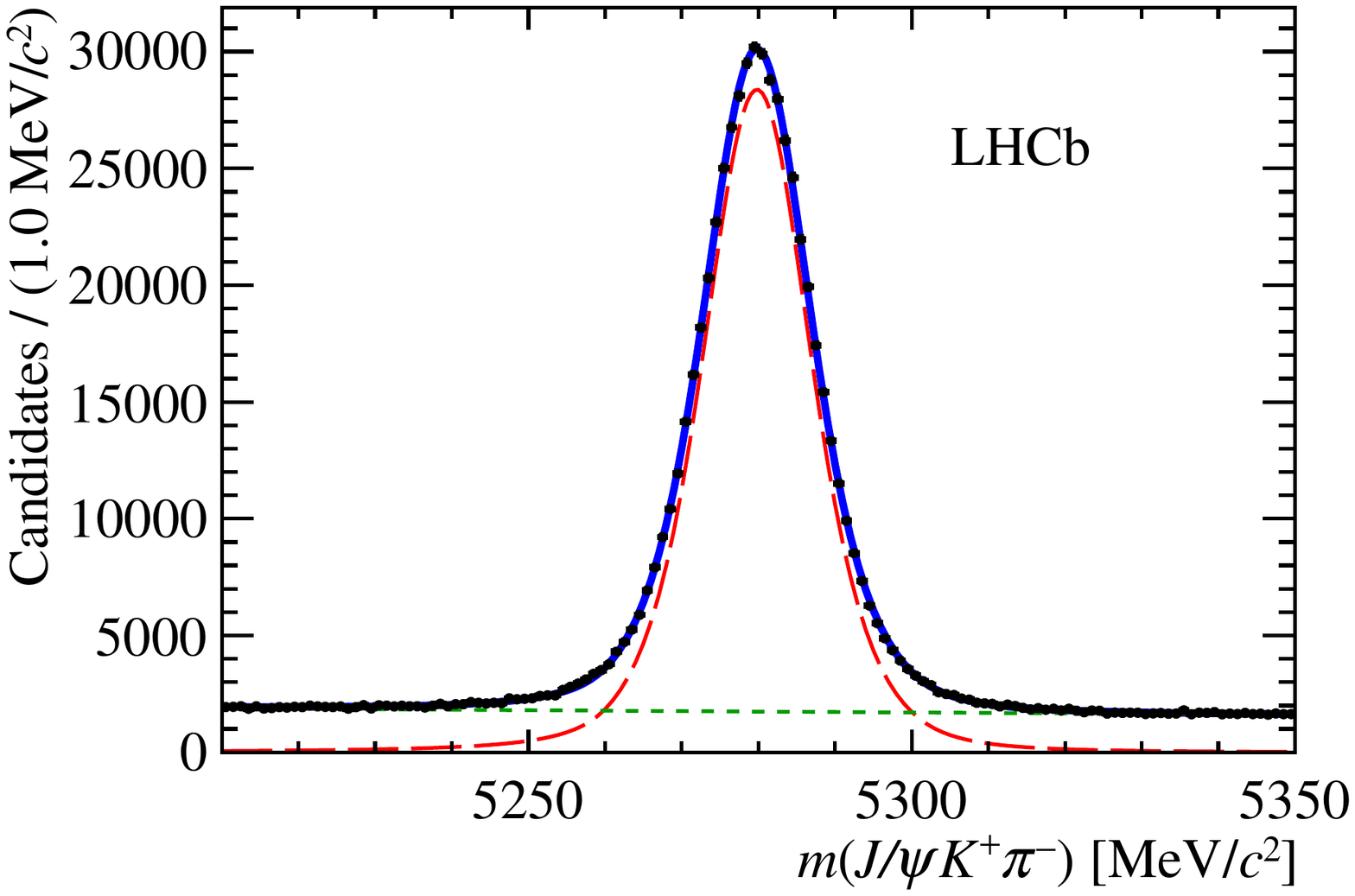}
    \put(-160,120){(a)}
        \includegraphics[width=0.49\textwidth]{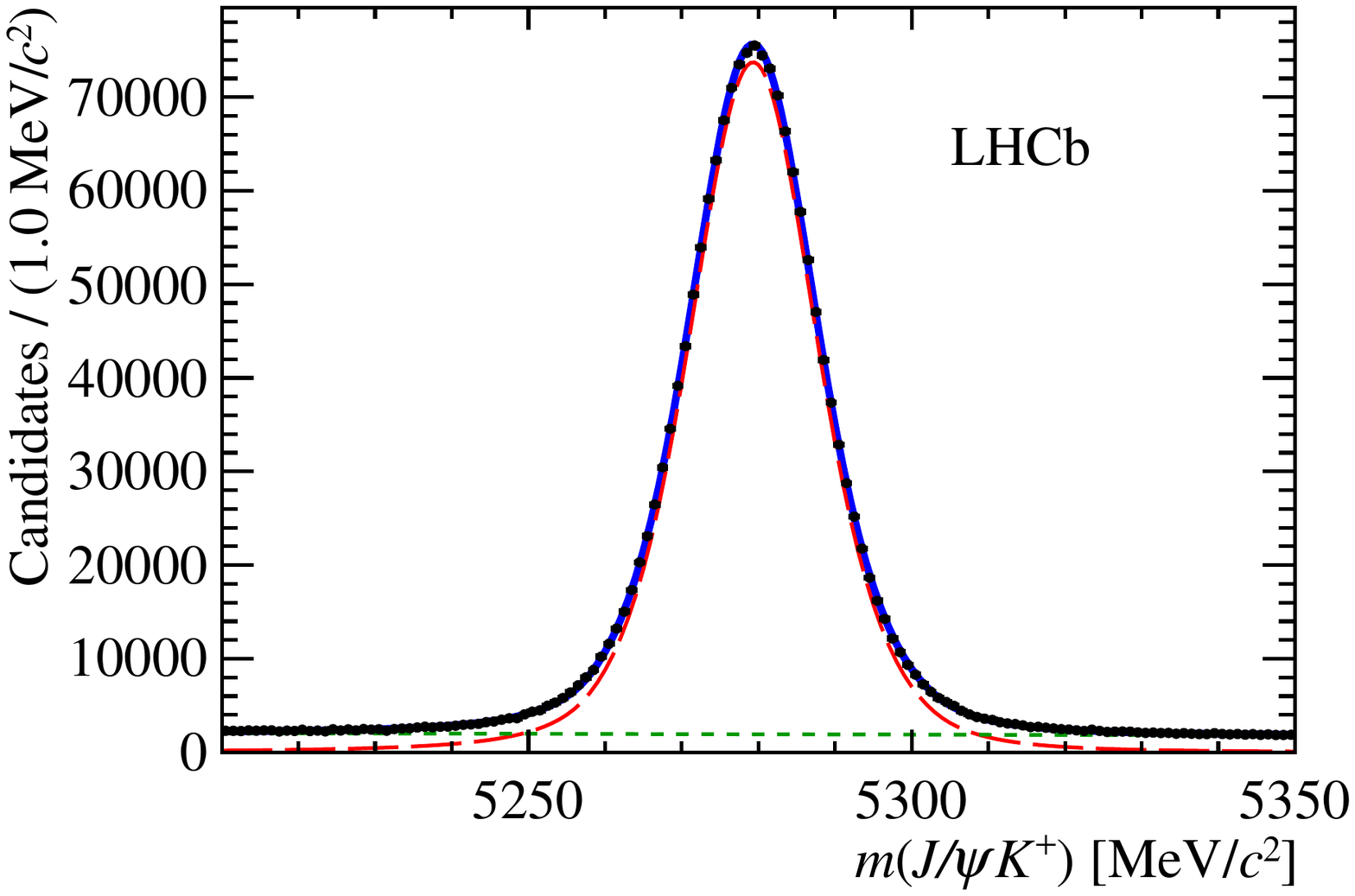}
    \put(-160,120){(b)}
    \caption{\label{fig:control_samples}\small Distribution of the invariant mass of selected
(a) \mbox{$\Bd \to \jpsi K^+ \pi^-$} and (b) \mbox{$B^+ \to \jpsi K^+$} decays used for the calibration and validation of the decay-time efficiency. The signal component is shown by the long-dashed red line, the background component by the dashed green line and the total fit function by the solid blue line.}
\end{figure}

Figure~\ref{fig:control_samples}(a) shows the $m(\jpsi K^+ \pi^-)$ distribution and corresponding result of an unbinned maximum-likelihood fit to the sample. The model used for the fit is the same for the 2015 and 2016 data-taking periods and the two trigger categories but with 
independently fitted parameters. It is composed of a
Hypatia~\cite{Santos:2013gra} function for the signal, where the parameters
describing the tails are fixed to the values obtained from simulation,  and an
exponential function for the combinatorial background. 
In total, $75\,000$ and $480\,000$ $\Bd$
mesons are found in 2015 and 2016, respectively.
The result of this fit is used to statistically subtract the background when determining
the decay-time efficiency in data, by using weights computed with the \sPlot technique. 

The PDF used to describe the decay-time distribution of the $\Bd$ data, and of the $\Bs$ and $\Bd$ simulated samples
is composed of the product of the efficiency function and a single exponential function,
convolved with a single Gaussian resolution function centred at zero. For the $\Bd$ candidates, the width of the resolution function is set to 39\fs and 42\fs for 
the simulated and data samples, respectively. The first value is obtained from simulation, and the second value is obtained by scaling 
the $\Bs$ resolution obtained in data, as described in Sec.~\ref{sec:TimeResolution},
by the ratio seen between the $\Bd$ and $\Bs$ resolutions in simulated samples.
A $\Bs$ simulated sample is generated with $\Delta \Gamma_s=0\invps$ and thus a
single exponential function is used to determine $\varepsilon_{\mathrm{sim}}^{\Bs}$.
As a cross-check, the decay-time efficiency is also derived from the nominal
$\Bs \to \jpsi \phi$ simulated sample, weighted to have $\Delta \Gamma_s = 0\invps$ such that the same fitting strategy can be used
as defined above. The difference between these two strategies is considered as a source of systematic uncertainty.

\begin{figure}[tb]
        \centering
         \includegraphics[width=0.49\textwidth]{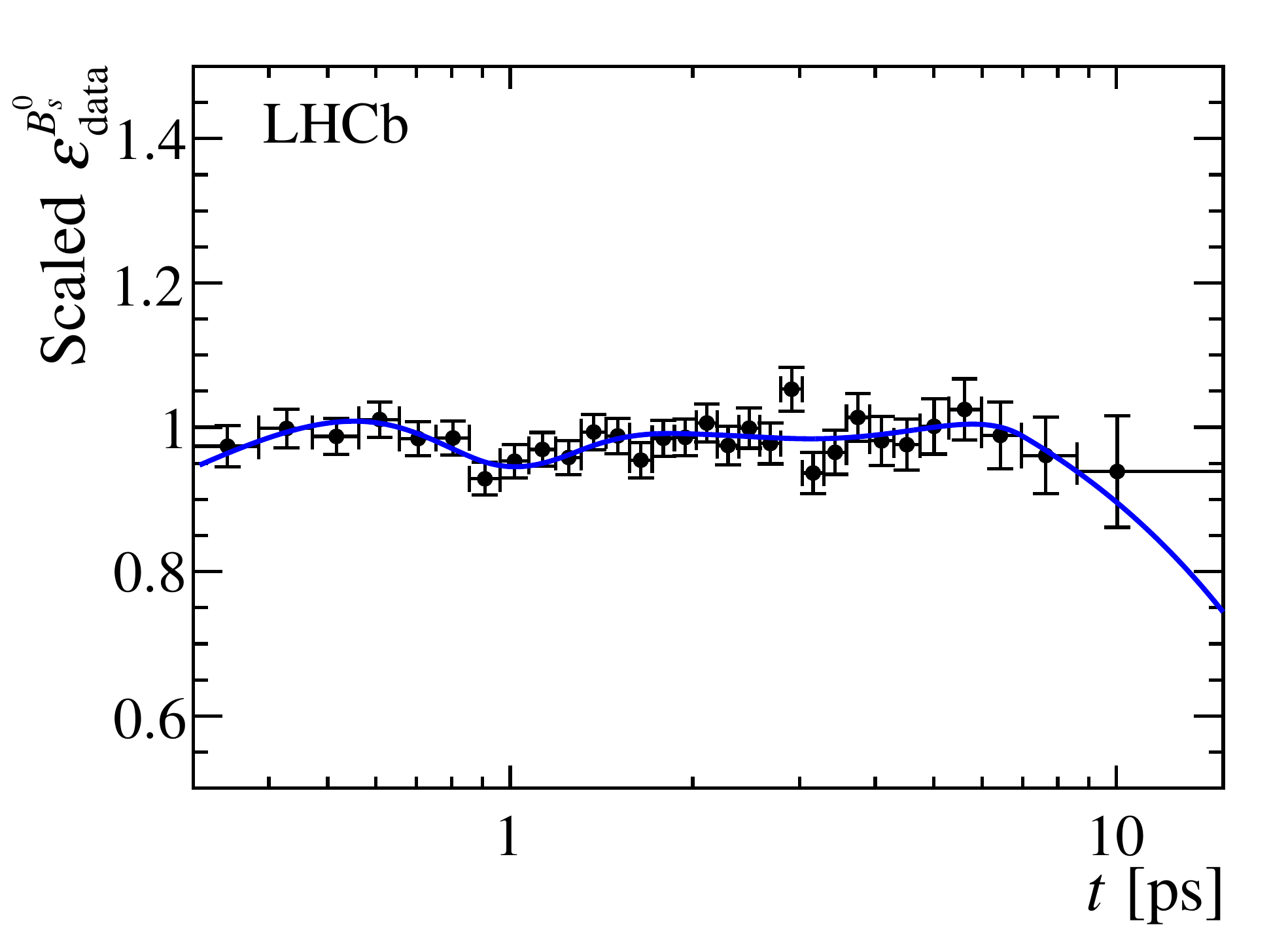}
        \put(-140,140){(a)}
         \includegraphics[width=0.49\textwidth]{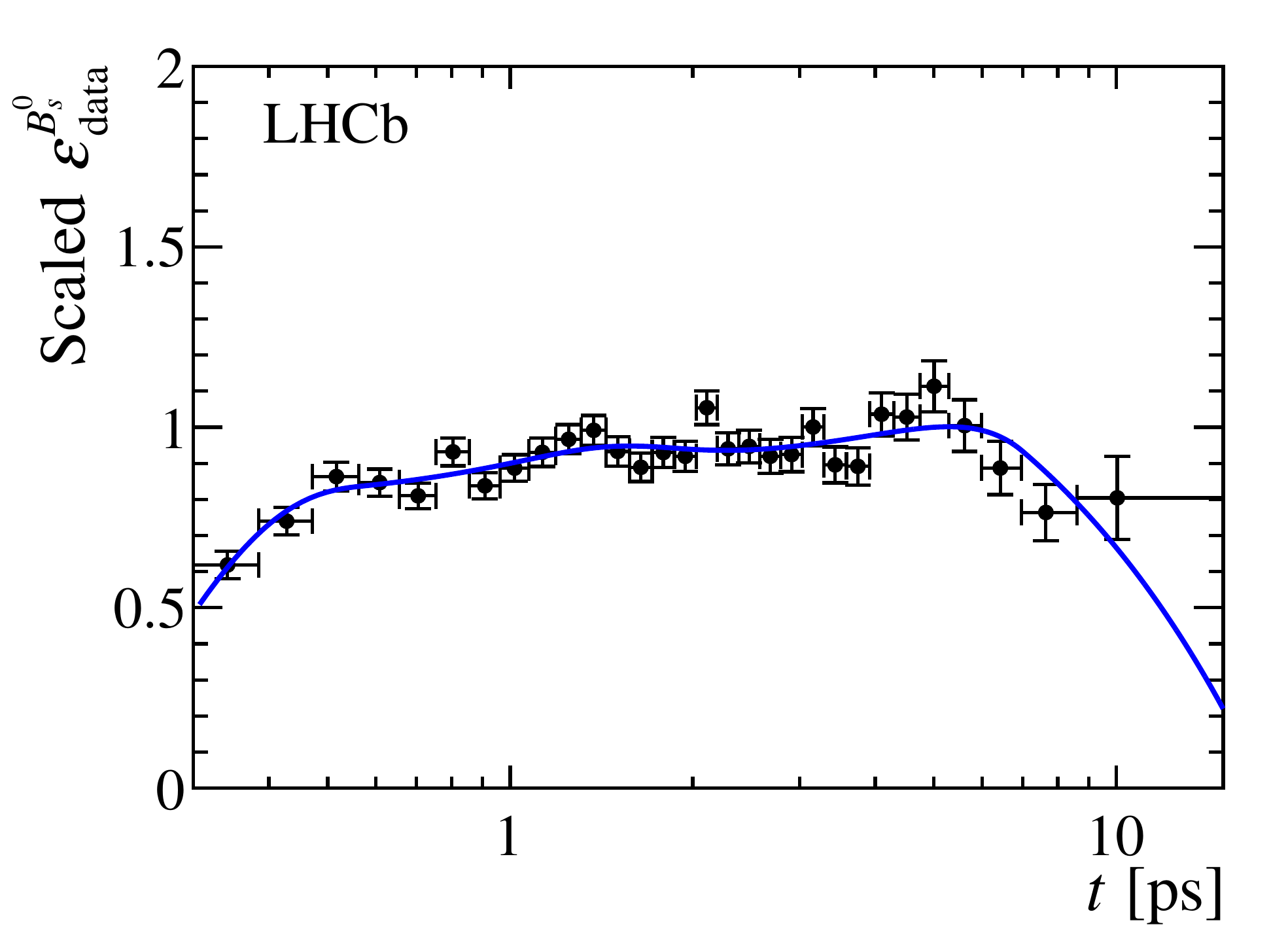}
        \put(-140,140){(b)}\\
    \includegraphics[width=0.49\textwidth]{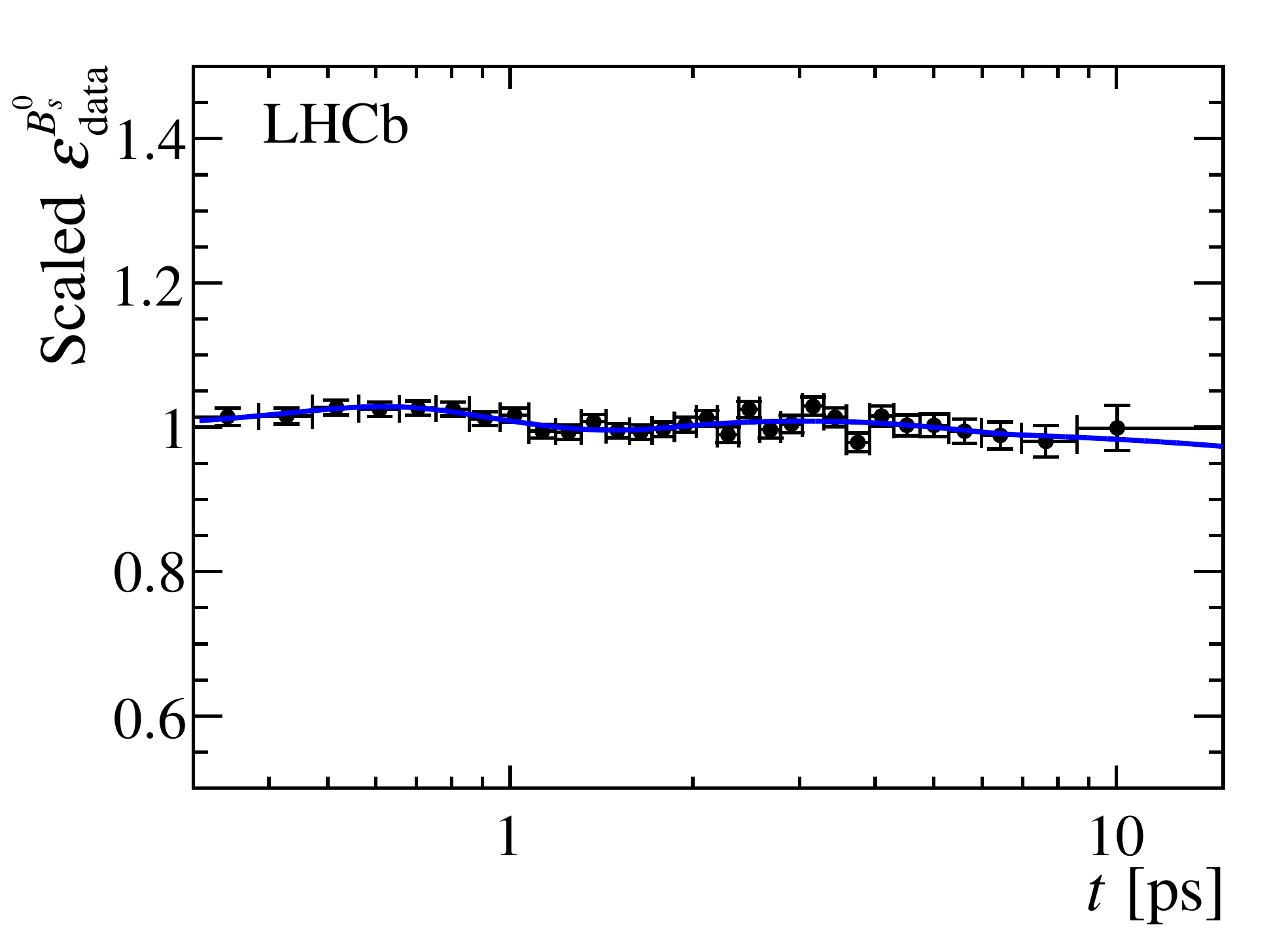}
        \put(-140,140){(c)}
    \includegraphics[width=0.49\textwidth]{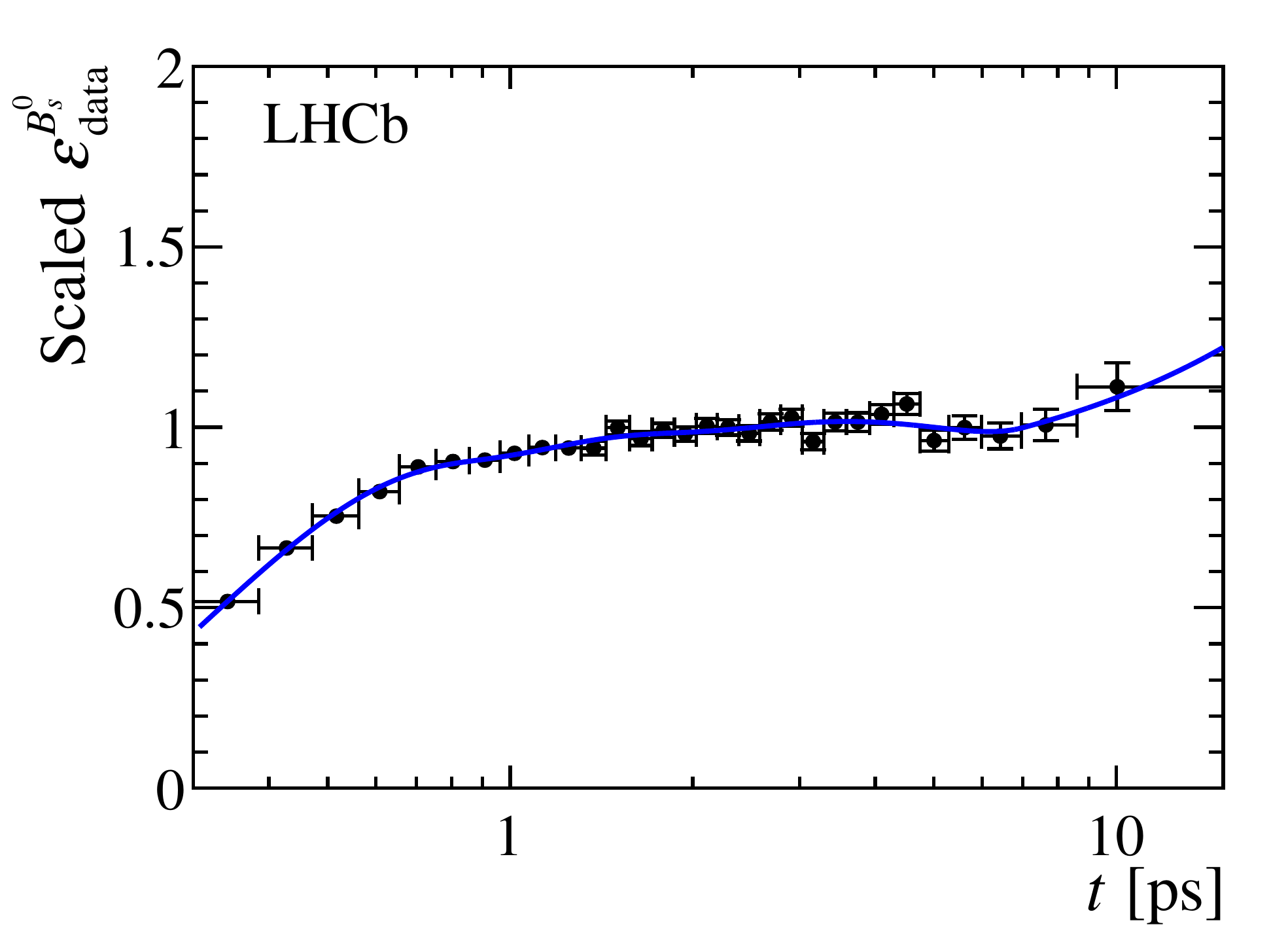}
        \put(-140,140){(d)}
    \caption{\label{fig:time_acc}\small Decay-time efficiency for the (a) 2015 unbiased, (b) 2015 biased, (c) 2016 unbiased and
    (d) 2016 biased \mbox{$\Bs\to\jpsi\phi$} sample. The cubic-spline
    function described in the text is shown by the blue line. For comparison, the black points show the efficiency when computed using
    histograms for each of the input component efficiencies.
    }
\end{figure}

The efficiency functions are parametrised using cubic splines with nodes at 0.3, 0.58, 0.91, 1.35,
1.96, 3.01, 7.00\ps and the first coefficient fixed to unity. 
The node positions are defined as to create six
uniformly populated bins in the interval $0.3$--$15\ps$, assuming an
exponential distribution with $\Gamma = 0.66 \ps^{-1}$.
The position of the last node
is chosen due to the lack of candidates at large decay times in the 2015 data control sample.
The final decay-time efficiencies, $\varepsilon^{\Bs}_{\mathrm{data}}(t)$,
are shown in Fig.~\ref{fig:time_acc}. The structure around 1\ps visible in Fig.~\ref{fig:time_acc}(a) and (c) is due to the different definition of the origin vertex used in the trigger and in the offline selection.

The full procedure is validated in data using two approaches where
the \Bs samples are replaced with alternative
\B meson samples of known lifetime. First, a sample of approximately 1.6~million $B^+ \to \jpsi(\to \mup \mu^-) K^+$ candidates
is reconstructed in the same data set as the $\Bs\to\jpsi\Kp\Km$ candidates and selected using similar selection requirements. 
The mass distribution of these candidates is shown in Fig.~\ref{fig:control_samples}(b). This sample is used to measure the difference 
of the $\Bu$ and $\Bd$ decay widths, $\Gamma_u-\Gamma_d$, with the same methods used for the measurement of $\Gamma_s-\Gamma_d$. 
A simulated sample of $\Bu$ decays is used in the calculation of the numerator of Eq.~\eqref{eq:timeacc} and this sample is corrected such that the particle-identification, event-multiplicity and
other kinematic and selection variables match those in data. 
The measured difference of decay widths is \mbox{$\Gamma_u-\Gamma_d = -0.0478 \pm 0.0013\invps$}, where the uncertainty is statistical only. This is in agreement with the 
world average value, \mbox{$-0.0474 \pm 0.0023\invps$}~\cite{PDG2018}, and validates the measurement of $\Gamma_s-\Gamma_d$ with a precision of 0.003\invps. 

A similar test is done using the $\Bd \to \jpsi \Kp \pim$ decays both as the signal and the reference to measure a null decay-width difference. 
The sample is split into two independent sets according to different selection criteria, where one is used to evaluate the decay-time efficiency 
with the procedure defined above, and the other is used as the signal sample. 
In all cases, the measured decay-width difference is found to be consistent with zero with a precision around 0.003\invps. 

\section{Angular efficiency}
\label{sec:AngAcc}

\begin{figure}
    \centering
    \includegraphics[width=\textwidth]{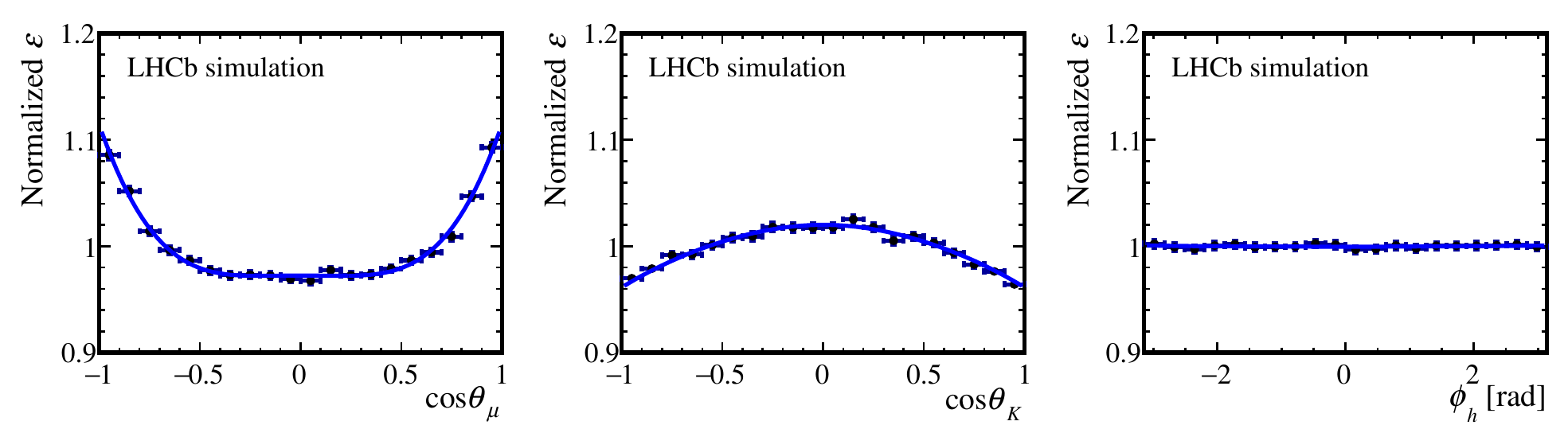}
    \put(-330,99){(a)}
    \put(-180,99){(b)}
    \put(-30,99){(c)}
    \caption{\label{fig:ang_eff}\small
    Normalised angular efficiency as a function of (a) $\cos\theta_K$, (b)
    $\cos\theta_\mu$ and (c) $\phi_h$, where in all cases the efficiency is
    integrated over the other two angles.  The efficiency is evaluated using
    simulated $\Bs\to\jpsi\phi$ decays that have been weighted to match the
    kinematics and physics of $\Bs\to\jpsi\Kp\Km$ decays in data, as described
    in the text. The points are obtained by dividing the angular distribution in the simulated sample 
    by the distribution expected without any efficiency effect and the curves represent an even fourth-order polynomial
    parameterisation of each one-dimensional efficiency. The figure is for illustration only as the angular efficiency is accounted for by normalisation weights in the signal PDF.}
\end{figure}

The LHCb detector geometry and the selection requirements give rise to efficiencies that vary as a function of the helicity angles $\theta_K$, $\theta_\mu$ and $\phi_h$. 
The three-dimensional angular-efficiency correction is
determined from simulated signal events to which the same trigger and selection
criteria as in the data are applied. 
The efficiency is evaluated separately for the different years of data taking and for the two trigger categories.
Two sets of corrections are applied to the simulated  events  such  that  they match the data.
First, the simulated samples are weighted, using a boosted decision tree method~\cite{Rogozhnikov:2016bdp}, to match the \pt, $p$ and $m(\Kp\Km)$ distributions of the 
\Bs signal.
A second procedure is performed to correct the differences observed in the kinematic distributions of the final-state particles and the fact that the 
simulated events do not include $\Kp\Km$ pairs in an S-wave configuration. 
This correction is implemented as an iterative procedure that gradually modifies the simulation such that the S-wave fraction matches
the value measured in the data. As a result, the agreement of the kaon momentum and
\pt distributions between the simulation and the data is improved. 
The efficiencies as a
function of the three helicity angles are shown for illustration in Fig.~\ref{fig:ang_eff}.
The angular efficiency correction is introduced in the analysis through normalisation weights 
in the PDF describing the signal decays in the fit of Sec.~\ref{sec:Fit}, 
following the procedure described in Ref.~\cite{TristansThesis}. 
The weights are calculated using simulated candidates and their statistical uncertainties are propagated to the parameters of interest as a systematic uncertainty.

A cross-check of the angular efficiency procedure is made using the $\Bd\to\jpsi \Kp \pim$ data
and $\Bd\to\jpsi K^*(892)^0(\to K^{+}\pi^{-})$ simulated samples. The simulation contains $K^{+}\pi^{-}$ systems in P-wave only and
is corrected to match the kinematic
distributions of the data using the iterative method defined above and the angular-efficiency
weights are determined. The P- and S-wave
$\Bd\to\jpsi K^{+}\pi^{-}$ polarisation amplitudes are measured by means of an unbinned fit to the distribution of helicity angles of the final-state particles and found to be consistent with those
in Ref.~\cite{LHCb-PAPER-2013-023}.

Another high-precision test of the angular-efficiency correction is made by using the large sample of 
$\Bu\to\jpsi\Kp$ decays presented in Sec.~\ref{sec:TimeAcc}. In $\Bu\to\jpsi\Kp$, the helicity angle $\theta_\mu$ distribution follows a $1-\cos^2\theta_\mu$ 
dependence. The $\Bu$ data sample is split into nine disjoint subsets according to the pseudorapidity of the $\Bu$ meson, 
to check the large efficiency variation as a function of this quantity. 
In each subset, background is subtracted with the \sPlot technique using the $\Bu$ candidate mass as a discriminating variable. 
Prior to any angular efficiency correction, the $\theta_\mu$ distribution presents up to a 30\% deviation from the expected shape, three times larger than in $\Bs\to\jpsi\Kp\Km$ decays. 
However, when the $\Bu\to\jpsi\Kp$ simulation is used to correct the data with the same method used for this analysis, a fit of the background-subtracted and efficiency-corrected data demonstrates that the expected distribution is fully recovered in each bin, 
with an overall precision of about 0.1$\%$. The test is stable against variation of the binning of the $\Bu$ sample and choice of different 
variables used to correct the simulation to
match the data with respect to the baseline strategy.

\section{Tagging the \boldmath\Bs meson flavour at production}
\label{sec:Tagging}

The determination of the initial flavour of the \Bs meson, called tagging, is a fundamental component for
measuring \CP asymmetries in the decays of \Bs mesons to \CP eigenstates. Two classes of algorithms are used. The opposite side (OS) tagger exploits the
fact that $b$ and $\overline{b}$ quarks are almost exclusively produced in pairs
in $pp$ collisions, allowing the
flavour of the signal \Bs candidate to be inferred from the flavour of the other
$b$ hadron in the event. The OS tagger combines information on the charge of the
muon or electron from semileptonic $b$ decays, the charge of the kaon from the
$b\to c\to s$ decay chain, the charge of a reconstructed secondary charm hadron and the charges of the tracks
that form the secondary vertex of the other $b$-hadron decay, combined into a weighted average, with weights depending on the transverse momenta of the tracks. The same-side kaon
(SSK) tagger exploits the additional correlated kaon that tends to be produced
during the hadronisation of the \bquarkbar (\bquark) quark that forms the signal
\Bs (\Bsb) candidate, with its initial flavour identified by the kaon
charge. These flavour tagging algorithms have been revisited and optimised using Run 2 data~\cite{Fazzini:2018dyq},
obtaining significantly higher combined tagging performances with respect to Run 1.
Further details on the OS and SSK taggers can be found in 
Refs.~\cite{LHCb-PAPER-2015-027,LHCb-PAPER-2011-027,1748-0221-11-05-P05010}.

The
tagging algorithms each provide a flavour-tagging decision, $\mathfrak{q}$,
and an estimate, $\eta$, of the probability that the decision is incorrect
(mistag) for each reconstructed \Bs candidate.
The tagging decision takes the value $+1$ ($-1$) for each tagged \Bs (\Bsb) candidate
and 0 if the taggers cannot make a decision (untagged). The
mistag probability is defined in the range from 0 to 0.5, since $\eta > 0.5$
corresponds to the opposite decision with a mistag of $(1-\eta)$. For 
untagged events $\eta$ is 0.5.

Each tagging algorithm is implemented as a BDT that is trained and
optimised using large samples of simulated $b$-hadron decays for the SSK tagger and a large data sample 
of $\Bu\to\jpsi\Kp$ decays for the OS tagger.
The mistag probability for each tagger is given by the
output of the BDT, which is calibrated using dedicated data control channels to
relate $\eta$ to the true mistag probability, $\omega$, as described
in the following sections. Each tagger has a
corresponding tagging power given by $\epsilon_{\rm tag}D^2$, where
$\epsilon_{\rm tag}$ is the fraction of tagged candidates and $D=1-2\omega$ is the dilution  induced on the amplitude of the \Bs oscillation. 
The tagging power represents
the effective reduction in statistical power due to imperfect tagging.

\subsection{Opposite-side tagging}

The OS tagging algorithm is calibrated using the
sample of $\Bu\to\jpsi\Kp$ decays (Sec.~\ref{sec:TimeAcc}), whose flavour is determined by the kaon charge.
This sample of $\Bu\to\jpsi\Kp$ decays is independent of that used to train and optimise the BDT of the tagging algorithm.
The result of the fit to the distribution of $m(\jpsi\Kp)$ shown
in Fig.~\ref{fig:control_samples}(b) is used to compute
{\it sWeights}, which are applied in
subsequent stages of the analysis to subtract the background. The
$\Bu\to\jpsi\Kp$ sample is further weighted to match the background-subtracted
$\Bs\to\jpsi\phi$ sample in the distributions of charged-track and PV
multiplicities and the \pt and rapidity of the $\B$ meson.

\begin{figure}[tb]
    \centering
    \includegraphics[width=0.49\textwidth]{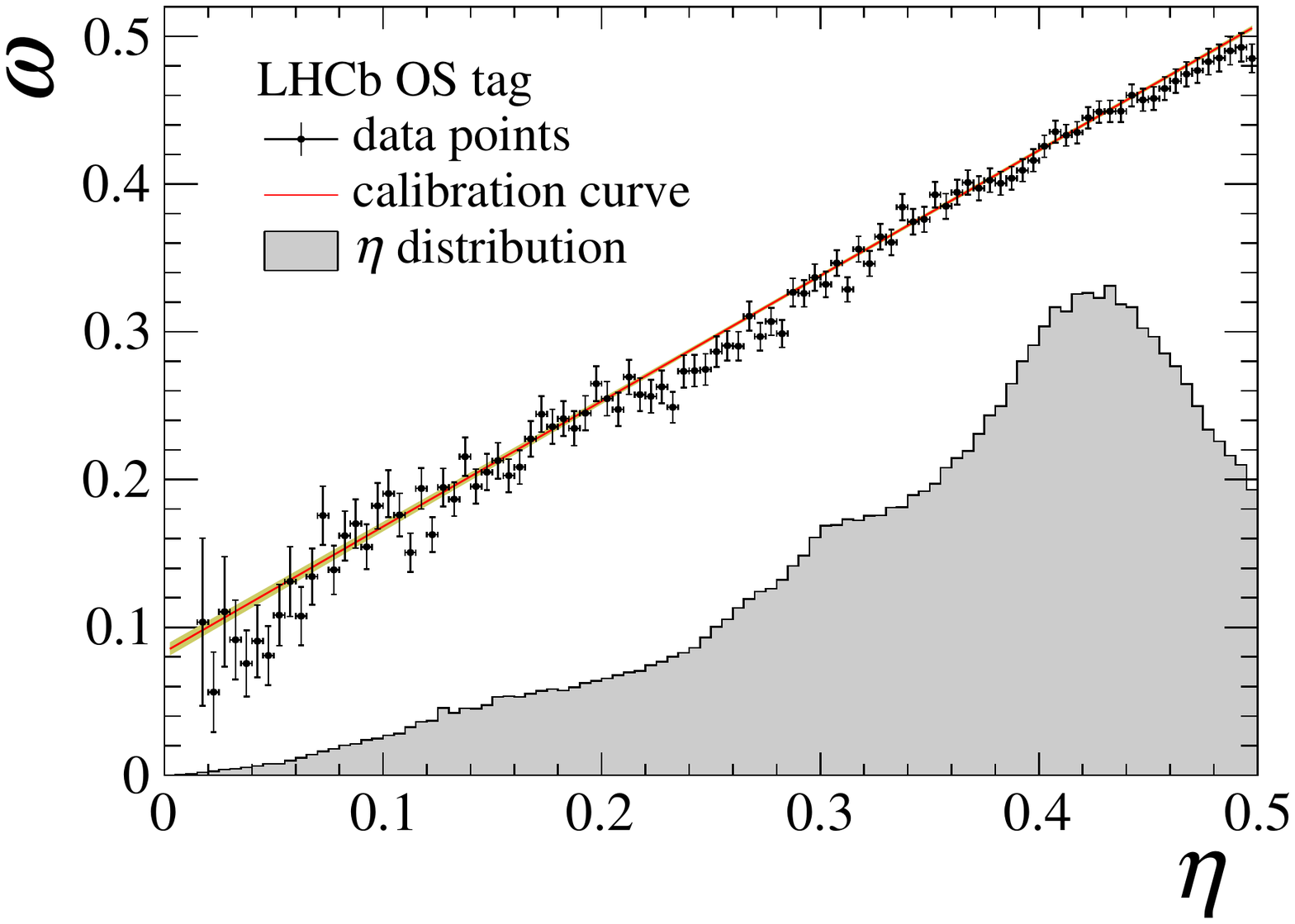}
    \caption{\label{fig:tagging_calib_OS}\small Calibration of the OS tagger
    using $\Bu\to\jpsi\Kp$ decays. The black points show the average measured mistag
    probability, $\omega$, in bins of predicted mistag, $\eta$, the red line shows the calibration as described in the text and the yellow area the calibration uncertainty within one standard deviation. The shaded histogram shows the
    distribution, with arbitrary normalisation, of $\eta$ in the background subtracted $\Bs\to\jpsi\phi$ sample,
    summing over candidates tagged as $\Bs$ or $\Bsb$.
    }
\end{figure}

The calibration between true and estimated mistag for each tagging algorithm
is given by an empirically determined linear relationship,
\begin{eqnarray}
\label{eqn:tagging_calib}
    \omega(\eta) &=& \left( p_0  + \frac{\Delta p_0}{2}\right) +
\left(p_1 + \frac{\Delta p_1 }{2}\right)
\left(\eta - \langle\eta\rangle\right)\\
\overline{\omega}(\eta) &=& \left( p_0  - \frac{\Delta p_0}{2}\right) +
\left(p_1 - \frac{\Delta p_1 }{2}\right)
\left(\eta - \langle\eta\rangle\right),
\end{eqnarray}
where $\omega(\eta)$ and $\overline{\omega}(\eta)$ are the calibrated mistag
probabilities for $B^+$ and $B^-$ mesons, respectively, $\Delta p_{0,1}$ are mistag
asymmetries and $\langle\eta\rangle$
is the average estimated mistag of the
$\Bu\to\jpsi\Kp$ sample. The calibration parameters are determined from
an unbinned maximum-likelihood fit to the $\eta$ distribution of the
probability 
\begin{equation}
    \mathcal{P}(a|\eta) = (1-a)\overset{\textbf{\fontsize{3pt}{3pt}\selectfont(---)}}{\omega}(\eta) +
    a(1-\overset{\textbf{\fontsize{3pt}{3pt}\selectfont(---)}}{\omega}(\eta)),
\end{equation}
for an initial flavour of the $B^+$ ($B^-$) meson. The discrete variable $a$ has the
value 0 or 1 for an incorrect or correct tagging decision, respectively, based
upon comparing the decision $\mathfrak{q}$ to the kaon charge.
Figure~\ref{fig:tagging_calib_OS} shows the relation between the flavour-averaged value of $\omega$ and
$\eta$ determined by the fit and the values of the measured mistag in bins of
estimated mistag, which supports the use of a linear calibration function.
The final calibration parameters are given in Table~\ref{tab:tagging} and the
overall tagging power for candidates with an OS tag only can be found in
Table~\ref{tab:tagging_summary}. Differences of the tagging efficiency are expected to be negligible as their effects are washed out by the fast 
$\Bs$-$\Bsb$ oscillations.
The applicability of the calibration from
$\Bu\to\jpsi\Kp$ to $\Bs\to\jpsi\phi$ decays is tested using simulated samples
and observed differences between the calibration parameters are treated
as a source of systematic uncertainty.  Variations in the parameters caused by the
use of a different model for the combinatorial background in the fit to the
$m(\jpsi\Kp)$ distribution are found to be negligible.

\renewcommand{\arraystretch}{1.1}
\begin{table}[tb]
\caption{\label{tab:tagging}\small Calibration parameters for the OS and SSK
taggers. Where given, the first uncertainty is statistical and the second is
systematic.}
\begin{center}
\begin{tabular}{c|cc}
    Tagger  & OS                                & SSK \\ \hline
    $p_0$   & $0.3890 \pm 0.0007 \pm 0.0028$    & $0.4325 \pm 0.0108 \pm 0.0030$ \\
    $p_1$   & $0.849 \pm 0.006 \pm 0.027$    & $0.92 \pm 0.13 \pm 0.02$ \\
    $\Delta p_0$& $0.0090 \pm 0.0014$           & $0.00   \pm 0.03$ \\
    $\Delta p_1$& $0.014 \pm 0.012$           & $0.00   \pm 0.03$ \\
    $\langle\eta\rangle $& $0.360$             & $0.417$\\
    \hline
    \end{tabular}
\end{center}
\end{table}

\begin{table}[tb]
  \begin{center}
     \caption{ \small Overall tagging performance for $\Bs\to\jpsi K^{+}K^{-}$. The uncertainty on $\epsilon_{\rm tag}D^{2}$ is obtained by varying the tagging calibration parameters within their statistical and systematic uncertainties summed in quadrature.}
  \label{tab:tagging_summary}
  \begin{tabular}{lccc}
      Category     & $\epsilon_{\rm tag}$(\%)  & $D^{2}$   & $\epsilon_{\rm tag}D^{2}(\%)$ \\
    \hline
    OS only  & 11.4& 0.078 & $0.88\pm0.04$ \\
    SSK only & 42.6& 0.032 & $1.38\pm0.30$ \\
    OS \& SSK  & 23.8& 0.104 & $2.47\pm0.15$ \\
    \hline
    Total    & 77.8& 0.061 & $4.73\pm0.34$ \\
  \end{tabular}
  \end{center}
\end{table}

\subsection{Same-side tagging}

\begin{figure}[t]
    \centering
    \includegraphics[width=0.51\textwidth]{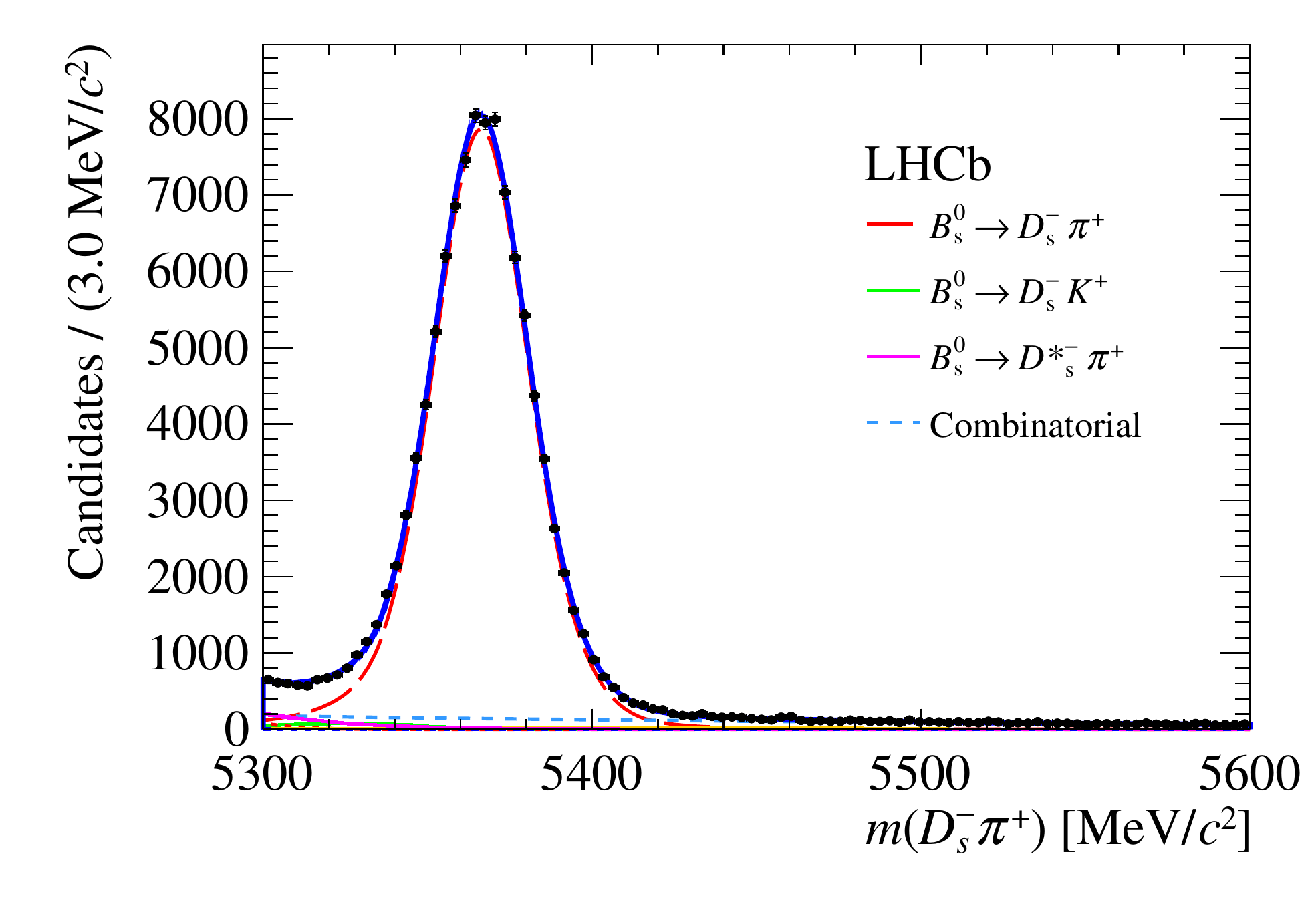}
    \caption{\label{fig:tagging_calib_SS_mass}\small Distribution of the invariant mass of selected
    $\Bs\to\Dsm\pi^+$ candidates (black points). The total fit function is shown as the solid blue line. The signal component is shown by the red long-dashed line, the
    combinatorial background by the light-blue short-dashed line and other small background
    components are also shown as specified in the legend. Only the dominant backgrounds are shown. 
    }
\end{figure}

The SSK tagger is calibrated by resolving the $\Bs$-$\Bsb$ flavour oscillations in
a sample of flavour-specfic $\Bs\to\Dsm\pip$ decays. The amplitude of this
oscillation is related to the averaged \Bs-\Bsb mistag probability, $\tilde{\omega}$, via the PDF of the decay-time
distribution of flavour-tagged $\Bs\to\Dsm\pip$ decays, given by 
\begin{align}
\label{eqn:dspi_time}
     \mathcal{P}(t)  & =  \epsilon(t) \left[ \Gamma(t) \otimes  R(t-t^\prime)\right], \\
 \Gamma(t) & =  \Gamma_s e^{-\Gamma_s t} \left[ \cosh (\Delta \Gamma_s t /2) + q^{\rm mix}
(1-2 \tilde{\omega} (\eta)) \cos ( \Delta m_s t) \right],\nonumber
\end{align}
where $t^\prime$ and $t$ are the true and reconstructed decay time of the $\Bs$ meson, respectively, and $\Gamma(t)$ is the $\Bs$ decay rate. 
The decay time and the decay-time
uncertainty are estimated from a kinematic fit~\cite{Hulsbergen:2005pu} in which
the $\Dsm\pi^+$ candidate is constrained to originate from the PV.
The decay-time efficiency is empirically
parameterised as \mbox{$\epsilon (t) =$ $1 - 1/(1+ (at)^n + b)$}, and $R(t-t^\prime)$
is the decay-time resolution model.
Here \mbox{$q^{\rm mix} = +1$ $(-1)$} if the $\Bs$ meson has (has not) changed flavour between its
production and decay, determined by comparing the flavour-tagging decision and
charge of the pion. A linear relationship between
the true and estimated mistag probabilities is assumed, as given in
Eq.~\eqref{eqn:tagging_calib}.

Approximately $70\,000$ same-side flavour-tagged \mbox{$\Bs\to\Dsm\pi^+$} decays, with \mbox{$\Dsm\to\Kp\Km\pi^-$}, are
selected with similar requirements as in Ref.~\cite{1748-0221-11-05-P05010}.  Due
to trigger requirements, only candidates with $\pt(\Bs)$ larger than $2\gevcc$
are used to perform the calibration.  Figure~\ref{fig:tagging_calib_SS_mass} shows
the distribution of $m(\Dsm\pi^+)$ for the selected sample.
Superimposed is the result of a fit with a model composed of a signal contribution
described by a Hypatia with tail
parameters fixed to those from simulation and a combinatorial background
component modelled by an exponential function. In addition, template shapes for
several peaking backgrounds ($\Bs\to\Dspm K^{\mp}$, $\Bd\to\Dsm\pi^+$,
$\Lbbar\to\Lcbar\pi^+$, $\Bs\to\Dssm\pi^+$ and $\Bs\to\Dsm\rho^+$) are evaluated
from simulation and included in the fit model. The yield of the peaking backgrounds is determined from a fit to \mbox{$m(\Dsm\pi^+)$} in the mass range \mbox{$5100$--$5600\mevcc$}. Using the fit results, the yield is extrapolated to the narrower region \mbox{$5300$--$5600\mevcc$} and fixed in the subsequent $m(\Dsm\pi^+)$ fit, which is used to compute {\it sWeights} for background subtraction as in the OS calibration. 
The $\Bs\to\Dsm\pi^+$ sample is also weighted to match the
background-subtracted $\Bs\to\jpsi\phi$ sample in the distributions of
charged-track and PV multiplicities and the \pt and rapidity of the $\Bs$ meson.

\begin{figure}[tb]
    \centering
    \includegraphics[width=0.49\textwidth]{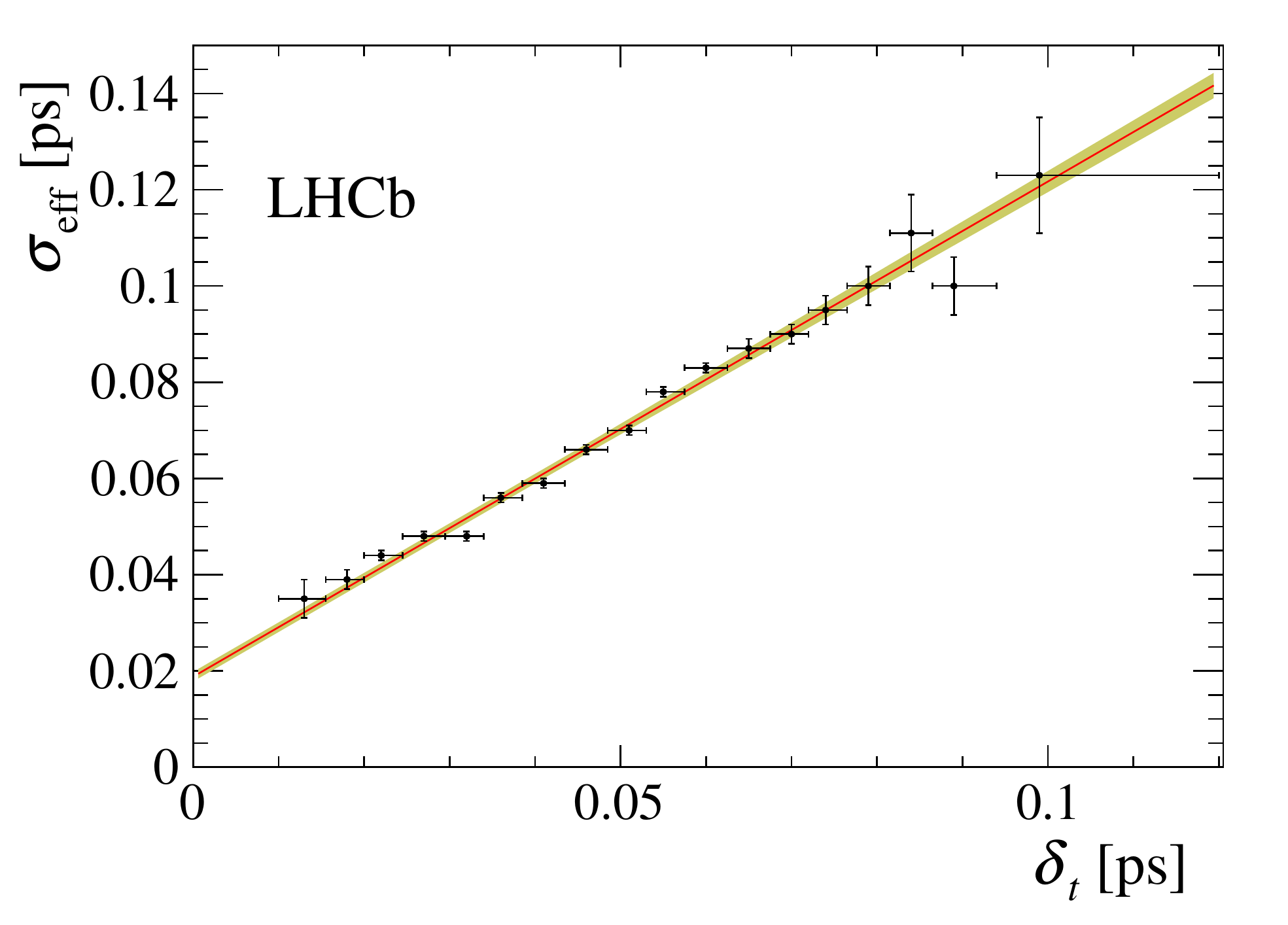}
    \caption{\label{fig:Dspi_time_res_calib}\small
    Variation of the effective single-Gaussian decay-time resolution,
    $\sigma_{\rm eff}$, as a
    function of the estimated per-event decay-time uncertainty, $\delta_t$,
    obtained from the prompt $\Dsm\pi^+$
    sample. The red line shows the result of a linear fit to the data and the yellow band its uncertainty within one standard deviation.  
    }
\end{figure}

To calibrate the decay-time resolution in Eq.~\eqref{eqn:dspi_time}, a sample
of promptly produced $\Dsm\pi^+$ candidates is selected following the
requirements defined in Ref.~\cite{LHCb-PAPER-2017-047}.  The procedure to
obtain the calibration for the decay-time resolution is similar to that
described in Sec.~\ref{sec:TimeResolution}. An unbinned maximum-likelihood fit is made to the $\Dsm$ candidate
invariant-mass distribution in 18 bins of
$\delta_t$.  The model consists of a single Gaussian component
for the signal and a second-order polynomial for the combinatorial background.
From this fit, {\it sWeights} are computed that are used to
subtract the background contribution in an unbinned fit to the decay-time
distribution in each $\delta_t$ bin. The model for this fit is composed of two
Gaussian functions with a common mean and different widths.
Only candidates with reconstructed decay time in the range from $-1.0$ to $0.1\ps$ are fitted. At such low values the longer-lived background components can be neglected.
The effective single-Gaussian resolution is calculated from the double-Gaussian
model using Eqs.~\eqref{eqn:time_dilution_tr}
and~\eqref{eqn:time_eff_res_tr}. The variation of the effective resolution with the
average value of $\delta_t$ in each bin is shown in
Fig.~\ref{fig:Dspi_time_res_calib}.  From a binned fit using a linear calibration
function, \mbox{$\sigma_{\rm eff}(\delta_t) = c_0 + c_1\delta_t$}, 
the calibration constants are determined to be \mbox{$c_0
= 18.8 \pm 1.0$\fs} and \mbox{$c_1 = 1.03 \pm 0.02$}, where the uncertainties are
statistical only.  Applying a similar procedure to a sample of simulated
$\Bs\to\Dsm\pi^+$ decays indicates a difference in the calibration parameters between prompt \mbox{$\Dsm\pi^+$} candidates and $\Bs\to\Dsm\pi^+$
decays.
A systematic uncertainty of $0.1$ is assigned to $c_1$ to account for this difference.

\begin{figure}[tb]
    \centering
    \includegraphics[width=0.505\textwidth]{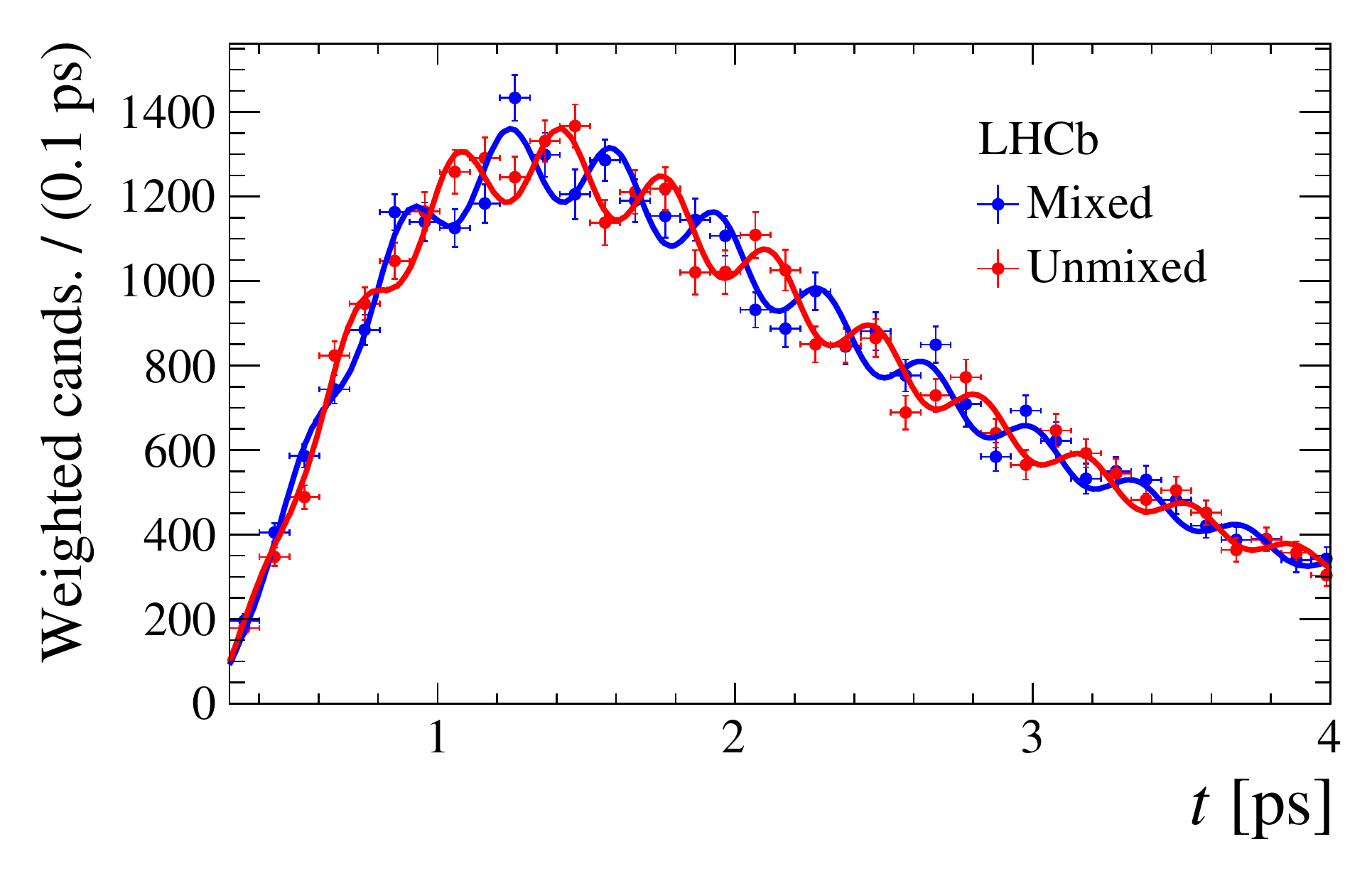}
    \put(-30,123){(a)}
    \includegraphics[width=0.47\textwidth]{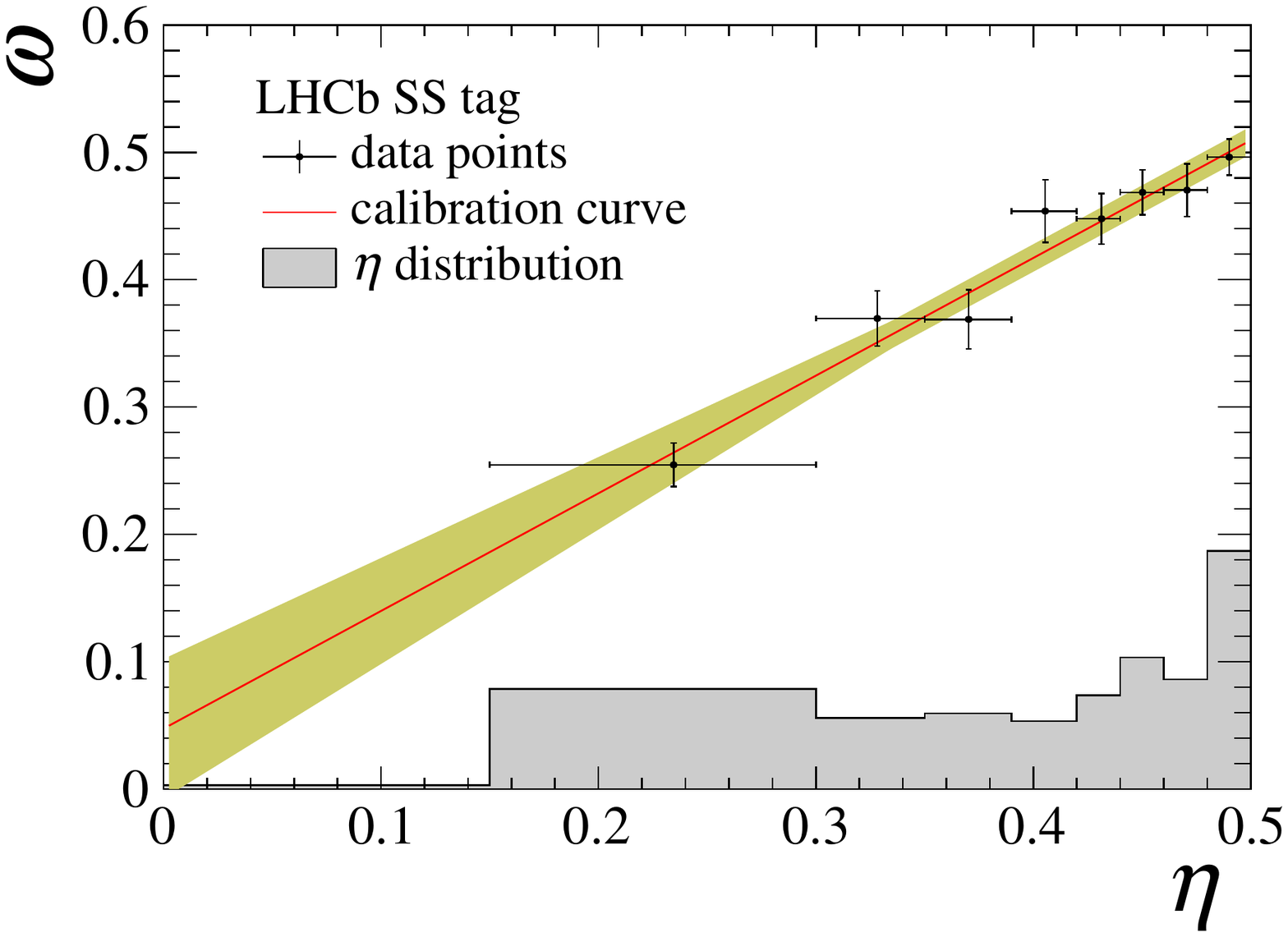}
    \put(-40,125){(b)}
    \caption{\label{fig:tagging_calib_SS}\small (a) Distribution of the decay time for
    \mbox{$\Bs\to\Dsm\pi^+$} candidates tagged as mixed and unmixed with the projection of the fit result, which is described in the text.
    (b) Calibration of the SSK tagger
    using $\Bs\to\Dsm\pi^{+}$ decays. The black points show the average measured mistag
    probability, $\omega$, in bins of predicted mistag, $\eta$, the red line shows the calibration
    obtained from the fit described in the text, and the yellow area the calibration uncertainty within one standard deviation. The shaded histogram shows the distribution of $\eta$ in the background subtracted $\Bs\to\jpsi\phi$ sample. 
    }
\end{figure}

To determine the SSK tagger calibration parameters from the $\Bs\to\Dsm\pi^+$
decay candidates, an unbinned maximum-likelihood fit, which uses the PDF of Eq.~\eqref{eqn:dspi_time}, is performed. 
Uncertainties due to the use of external
measurements~\cite{PDG2018} of $\Gamma_s$, $\Delta\Gamma_s$, $\Delta m_s$, and
the decay-time resolution parameters ($c_0$, $c_1$) are accounted for via
Gaussian constraints in the likelihood function.  The parameters of the SSK
tagger calibration and decay-time efficiency are free in the fit.
Figure~\ref{fig:tagging_calib_SS} shows the result of this fit, split by
decays that are tagged as being mixed or unmixed, and the obtained
relation between $\omega$ and $\eta$. Also shown are the values of the measured mistag
in bins of estimated mistag, which supports the use of a linear calibration function for
$\omega(\eta)$. The final calibration parameters are given in
Table~\ref{tab:tagging}. Two sources of systematic uncertainty are studied in addition to the knowledge of the time resolution, which is incorporated in the statistical uncertainty. The first and larger one is due to the applicability of the calibration from
$\Bs\to\Dsm\pi^+$ to $\Bs\to\jpsi\phi$ decays. It is tested using simulated
events and the observed difference
between the calibrations is assigned as a systematic uncertainty. Variations in the parameters through the use of a
different model for the combinatorial background in the fit to the
$\Bs\to\Dsm\pi^+$ invariant-mass distribution are treated as systematic
uncertainties. The tagging asymmetry parameters, $\Delta p_0$ and $\Delta p_1$,
are both assumed to be \mbox{$0.00\pm0.03$}. The uncertainty is estimated by studying the tagging
calibration using a sample of over 3.1~million promptly produced $\Dsm\to\Kp\Km\pi^-$
decays, with the method described in Ref.~\cite{1748-0221-11-05-P05010}. The overall tagging power for \mbox{$\Bs\to\jpsi K^{+}K^{-}$} candidates with only an SSK tag can be found
in Table~\ref{tab:tagging_summary}.

\subsection{Tagger combination}

Approximately 31\% of the tagged candidates in the \mbox{$\Bs\to\jpsi K^{+}K^{-}$} sample are tagged by both the OS and the SSK algorithms. Since the algorithms are uncorrelated, as they select mutually exclusive charged particles, the two tagging results are combined taking into
account both decisions and their corresponding estimate of $\eta$. The combined
estimated mistag probability and the corresponding uncertainties are obtained
by combining the individual calibrations for the OS and SSK tagging and
propagating their uncertainties.
The effective tagging power and efficiency for these both OS and SSK tagged
candidates is given in Table~\ref{tab:tagging_summary}.

\section{Maximum-likelihood fit}
\label{sec:Fit}

The maximum-likelihood fitting procedure is similar to that in Ref.~\cite{LHCb-PAPER-2014-059}, the only major differences being the treatment of the decay-time efficiency and 
that the quantity $\Gs-\Gd$
is measured instead of $\Gamma_s$.
It has been checked via pseudoexperiments that, given
that the decay-time efficiency is obtained using $\Gamma_d$ as an input parameter (see Sec.~\ref{sec:TimeAcc}), the fitted value of $\Gs-\Gd$ and its uncertainty
are independent of the value and uncertainty of $\Gamma_d$. This strategy has the advantage that the measured value of
$\Gs-\Gd$ can be combined with the most up-to-date value of $\Gamma_d$ to obtain $\Gamma_s$ or $\Gamma_s/\Gamma_d$.

Each candidate $i$ is given a signal weight $W_i$ using the \sPlot method with $m(\jpsi\Kp\Km)$  
as a discriminating variable and $\sigma_m$ as a conditional variable as explained in Sec.~\ref{sec:Selection}.
A weighted fit is then performed to the \Bs decay time and helicity-angle distributions using a PDF that describes only the signal. 
The log-likelihood in each of the 24 data subsamples is scaled by a per-sample factor $\alpha = \sum_i W_i  /  \sum_i W_i^2 $ to account for the effect
of the weights in the determination of the parameter uncertainties~\cite{sFit}.

The distribution of the decay time and angles for a \Bs\ meson produced at time $t=0$ is described by a sum of ten terms, corresponding to the four polarisation amplitudes squared
and their interference terms. Each of these is given by the product of a
decay-time-dependent function and an angular function
\begin{equation}
  \frac{\deriv^{4} \Gamma(\Bs\to J/\psi K^{+}K^{-}) }{\deriv t \;\deriv\Omega} \; \propto \;
  \sum^{10}_{k=1} \: N_{k} \: h_k(t) \: f_k( \Omega) \,,
  \label{eq:Eqbsrate}
\end{equation}
with
\begin{align}
    h_k(t|B^{0}_{s}) &= \frac{3}{4\pi}e^{-\Gamma t}\left( a_{k}\cosh\frac{\Delta\Gamma t}{2} + b_{k}\sinh\frac{\Delta\Gamma t}{2} + c_{k}\cos(\Delta m t) + d_{k}\sin(\Delta m t) \right),\\
    h_k(t|\bar{B}^{0}_{s}) &= \frac{3}{4\pi}e^{-\Gamma t}\left( a_{k}\cosh\frac{\Delta\Gamma t}{2} + b_{k}\sinh\frac{\Delta\Gamma t}{2} - c_{k}\cos(\Delta m t) - d_{k}\sin(\Delta m t) \right),
\end{align}
where the definition of the parameters $N_{k}$, $a_{k}$, $b_{k}$, $c_{k}$, $d_{k}$ and of the function $f_k( \Omega)$ can be found in Table~\ref{tab:abcd_rate}.
\begin{sidewaystable}[tp]
  
  \caption{\small Angular and time-dependent functions used in the fit to the data. Abbreviations used include $c_{K}=\cos\theta_K$, $s_{K}=\sin\theta_K$,
    $c_{l}=\cos\theta_l$, $s_{l}=\sin\theta_l$,
    $c_{\phi}=\cos\phi$ and $s_{\phi}=\sin\phi$.
  }

  \label{tab:abcd_rate}
\resizebox*{\textheight}{!}{\centering
    \begin{tabular}[t]{c|c|c|c |ccc}
      $f_k$ & $N_k$ &  $a_k$ & $b_k$ & $c_k$ & $d_k$ \\
      \hline
      $ c^2_K s^2_l $
      & $ |A_0|^2     $
      & $\frac{1}{2}(1+ |\lambda_{0}|^2)$
      & $-|\lambda_0| \cos(\phi_0) $
      & $\frac{1}{2}(1-|\lambda_{0}|^2)$
      & $ |\lambda_0| \sin(\phi_0)$ \\
      \hline
      $ \frac{1}{2}{s^2_K (1- c_\phi^2 s_l^2)} $
      & $ |A_{\parallel}|^2$
      & $\frac{1}{2}(1+ |\lambda_{{\parallel}}|^2)$
      & $-|\lambda_{{\parallel}}| \cos(\phi_{{\parallel}}) $
      & $\frac{1}{2}(1-|\lambda_{{\parallel}}|^2)$
      & $ |\lambda_{{\parallel}}| \sin(\phi_{{\parallel}})$ \\
      \hline
      $ \frac{1}{2}s^2_K (1- s_\phi^2 s_l^2) $
      & $|A_{\perp} |^2 $
      & $\frac{1}{2}(1+ |\lambda_{{\perp}}|^2)$
      & $|\lambda_{{\perp}}| \cos(\phi_{{\perp}}) $
      & $\frac{1}{2}(1-|\lambda_{{\perp}}|^2)$
      & $ -|\lambda_{{\perp}}| \sin(\phi_{{\perp}})$ \\
     \hline
      $  s^2_K   s_l^2  s_\phi c_\phi  $
      &  $  |A_{\perp} A_{\parallel}|$
      &  $\begin{array}{l}
        \frac{1}{2} \bigg[\sin (\delta_{\perp}-\delta_{\parallel})  - |\lambda_{\perp}  \lambda_{\parallel} |\\  \sin( \delta_{\perp}-\delta_{\parallel} -\phi_{\perp} +\phi_{\parallel})   \bigg]
      \end{array}$
      &  $\begin{array}{l}
        \frac{1}{2} \bigg[ |\lambda_{\perp}| \sin (\delta_{\perp}-\delta_{\parallel} -\phi_{\perp}) \\    + | \lambda_{\parallel} |\sin( \delta_{\parallel}-\delta_{\perp}  -\phi_{\parallel})   \bigg]
      \end{array}$
      & $\begin{array}{l}
        \frac{1}{2} \bigg[\sin (\delta_{\perp}-\delta_{\parallel})  + |\lambda_{\perp}  \lambda_{\parallel} |\\  \sin( \delta_{\perp}-\delta_{\parallel} -\phi_{\perp} +\phi_{\parallel})   \bigg]
      \end{array}$
      &  $\begin{array}{l}
        -\frac{1}{2} \bigg[ |\lambda_{\perp}| \cos (\delta_{\perp}-\delta_{\parallel} -\phi_{\perp}) \\    + | \lambda_{\parallel} |\cos( \delta_{\parallel}-\delta_{\perp}  -\phi_{\parallel})   \bigg]
      \end{array}$    \\
       \hline
      $\sqrt 2 s_K c_K s_l c_l c_\phi   $
      &  $|A_{0} A_{\parallel}|$
      &  $\begin{array}{l}
        \frac{1}{2} \bigg[\cos (\delta_{0}-\delta_{\parallel})  + |\lambda_{0}  \lambda_{\parallel} |\\  \cos( \delta_{0}-\delta_{\parallel} -\phi_{0} +\phi_{\parallel})   \bigg]
      \end{array}$
      &  $\begin{array}{l}
        -\frac{1}{2} \bigg[|\lambda_{0}|\cos (\delta_{0}-\delta_{\parallel}-\phi_0 )  \\  + | \lambda_{\parallel} |\cos( \delta_{\parallel}-\delta_{0} -\phi_{\parallel})   \bigg]
      \end{array}$
      &  $\begin{array}{l}
        \frac{1}{2} \bigg[\cos (\delta_{0}-\delta_{\parallel})  - |\lambda_{0}  \lambda_{\parallel} |\\  \cos( \delta_{0}-\delta_{\parallel} -\phi_{0} +\phi_{\parallel})   \bigg]
      \end{array}$
      &  $\begin{array}{l}
        -\frac{1}{2} \bigg[|\lambda_{0}|\sin (\delta_{0}-\delta_{\parallel}-\phi_0 )  \\  + | \lambda_{\parallel} |\sin( \delta_{\parallel}-\delta_{0} -\phi_{\parallel})   \bigg]
      \end{array}$ \\
      \hline
      $ -\sqrt 2 s_K c_K s_l c_l s_\phi  $
      &  $|A_{0} A_{\perp}|$
      &  $\begin{array}{l}
        -\frac{1}{2} \bigg[\sin (\delta_{0}-\delta_{\perp})  - |\lambda_{0}  \lambda_{\perp} |\\  \sin( \delta_{0}-\delta_{\perp} -\phi_{0} +\phi_{\perp})   \bigg]
      \end{array}$
      &  $\begin{array}{l}
        \frac{1}{2} \bigg[ |\lambda_{0}| \sin (\delta_{0}-\delta_{\perp} -\phi_{0}) \\    + | \lambda_{\perp} |\sin( \delta_{\perp}-\delta_{0}  -\phi_{\perp})   \bigg]
      \end{array}$
      & $\begin{array}{l}
        -\frac{1}{2} \bigg[\sin (\delta_{0}-\delta_{\perp})  + |\lambda_{0}  \lambda_{\perp} |\\  \sin( \delta_{0}-\delta_{\perp} -\phi_{0} +\phi_{\perp})   \bigg]
      \end{array}$
      &  $\begin{array}{l}
        -\frac{1}{2} \bigg[ |\lambda_{0}| \cos (\delta_{0}-\delta_{\perp} -\phi_{0}) \\    + | \lambda_{\perp} |\cos( \delta_{\perp}-\delta_{0}  -\phi_{\perp})   \bigg]
      \end{array}$    \\
      \hline
      $\frac{1}{3} s_l^2$
      & $|A_{\rm S}|^2 $
      & $\frac{1}{2}(1+ |\lambda_{\rm S}|^2)$
      & $  |\lambda_{\rm S}| \cos(\phi_{\rm S})$
      & $\frac{1}{2}(1 - |\lambda_{\rm S}|^2)$
      & $-|\lambda_{\rm S}| \sin(\phi_{\rm S})$\\
      \hline
      $ \frac{2}{\sqrt 6}s_K s_lc_l c_\phi  $
      &  $  |A_{\rm S} A_{\parallel}|$
      &  $\begin{array}{l}
        \frac{1}{2} \bigg[\cos (\delta_S-\delta_{\parallel})  - |\lambda_S  \lambda_{\parallel} |\\  \cos( \delta_S-\delta_{\parallel} -\phi_S +\phi_{\parallel})   \bigg]
      \end{array}$
      &  $\begin{array}{l}
        \frac{1}{2} \bigg[|\lambda_S|\cos (\delta_S-\delta_{\parallel}-\phi_S )  \\  - | \lambda_{\parallel} |\cos( \delta_{\parallel}-\delta_S -\phi_{\parallel})   \bigg]
      \end{array}$
      &  $\begin{array}{l}
        \frac{1}{2} \bigg[\cos (\delta_S-\delta_{\parallel})  + |\lambda_S \lambda_{\parallel} |\\  \cos( \delta_S-\delta_{\parallel} -\phi_S +\phi_{\parallel})   \bigg]
      \end{array}$
      &  $\begin{array}{l}
        \frac{1}{2} \bigg[|\lambda_S|\sin (\delta_S-\delta_{\parallel}-\phi_S )  \\  - | \lambda_{\parallel} |\sin( \delta_{\parallel}-\delta_S -\phi_{\parallel})   \bigg]
      \end{array}$  \\
      \hline
      $ -\frac{2}{\sqrt 6} s_K s_l c_l s_\phi  $
      &  $ |A_{\rm S} A_{\perp}|$
      &  $\begin{array}{l}
        -\frac{1}{2} \bigg[\sin (\delta_S-\delta_{\perp})  + |\lambda_S \lambda_{\perp} |\\  \sin( \delta_S-\delta_{\perp} -\phi_S +\phi_{\perp})   \bigg]
      \end{array}$
      &  $\begin{array}{l}
        -\frac{1}{2} \bigg[ |\lambda_S| \sin (\delta_S-\delta_{\perp} -\phi_S) \\    - | \lambda_{\perp} |\sin( \delta_{\perp}-\delta_S  -\phi_{\perp})   \bigg]
      \end{array}$
      & $\begin{array}{l}
        -\frac{1}{2} \bigg[\sin (\delta_S-\delta_{\perp})  - |\lambda_S \lambda_{\perp} |\\  \sin( \delta_S-\delta_{\perp} -\phi_S +\phi_{\perp})   \bigg]
      \end{array}$
      &  $\begin{array}{l}
        -\frac{1}{2} \bigg[- |\lambda_S| \cos (\delta_S-\delta_{\perp} -\phi_S) \\    + | \lambda_{\perp} |\cos( \delta_{\perp}-\delta_S  -\phi_{\perp})   \bigg]
      \end{array}$    \\
      \hline
      $ \frac{2}{\sqrt 3} c_K s^2_l  $
      &  $| A_{\rm S} A_{0}|$
      &  $\begin{array}{l}
        \frac{1}{2} \bigg[\cos (\delta_S-\delta_{0})  - |\lambda_S \lambda_{0} |\\  \cos( \delta_S-\delta_{0} -\phi_S +\phi_{0})   \bigg]
      \end{array}$
      &  $\begin{array}{l}
        \frac{1}{2} \bigg[|\lambda_S|\cos (\delta_S-\delta_{0}-\phi_S )  \\  - | \lambda_{0} |\cos( \delta_{0}-\delta_S -\phi_{0})   \bigg]
      \end{array}$
      &  $\begin{array}{l}
        \frac{1}{2} \bigg[\cos (\delta_S-\delta_{0})  + |\lambda_S  \lambda_{0} |\\  \cos( \delta_S-\delta_{0} -\phi_S +\phi_{0})   \bigg]
      \end{array}$
      &  $\begin{array}{l}
        \frac{1}{2} \bigg[|\lambda_S|\sin (\delta_S-\delta_{0}-\phi_S )  \\  - | \lambda_{0} |\sin( \delta_{0}-\delta_S -\phi_{0})   \bigg]
      \end{array}$ \\
      \hline
    \end{tabular}
}
\end{sidewaystable}
The interference between the different S- and P-wave contributions is accounted
for via an effective coupling factor, $C_{\rm SP}$. The $C_{\rm SP}$ factors are 
computed by integrating the interference between the S- and P-wave contributions in each of the six
$m(\Kp\Km)$ bins in which the analysis is performed, using the same strategy as in the previous analysis. They are applied by multiplication to the relevant terms in Eq.~\eqref{eq:Eqbsrate}. The $C_{\rm SP}$ factors are unity for terms involving P-wave and S-wave amplitudes only ($k<8$). 
In the determination of the $C_{\rm SP}$ factors, the $m(\Kp\Km)$ lineshape of the P-wave component is described by a relativistic Breit--Wigner distribution, while the S-wave is taken as an $f_0(980)$
resonance modelled as a Flatt\'{e} amplitude with parameters from Ref.~\cite{Aaij:2014emv}.
The $C_{\mathrm{SP}}$ correction factors are calculated
to be $0.8463,0.8756,0.8478,0.8833,0.9415$ and 0.9756 from the lowest to the highest $m(\Kp\Km)$ bin.
Their effect on the fit results is small and is discussed further in
Sec.~\ref{sec:Systematics}, where three different S-wave lineshapes are considered to assign a systematic uncertainty.
The PDF considers four disjoint tagging cases: only OS tagged candidates, 
only SSK-tagged, OS and SSK tagged, and untagged candidates. 
Taking into account all detector response effects, the full PDF is conditional upon the mistag probability and the estimated decay-time uncertainty.

A simultaneous fit is made to the different subsamples, divided by $m(\Kp\Km)$ bin,
year of data taking and trigger category.
The PDF for each subsample, up to a normalisation constant, is given by
\small{
\begin{align}
 &{\cal P}\left( t,\Omega| \mathfrak{q}^{\rm OS},  \mathfrak{q}^{\rm SSK}, \eta^{\rm OS}, \eta^{\rm SSK}, \delta_{t} \right) \propto \sum_{k=1}^{10} C^{k}_{\rm SP} N_{k} f_{k}(\Omega)\varepsilon^{\Bs}_{\rm data}(t) \\
 & \cdot \left\{\left[
 {\cal Q}\left(\mathfrak{q}^{\rm OS}, \mathfrak{q}^{\rm SSK}, \eta^{\rm OS}, \eta^{\rm SSK}\right)
 h_k\left(t | \Bs\right) +
 \bar{\cal Q}\left(\mathfrak{q}^{\rm OS}, \mathfrak{q}^{\rm SSK}, \eta^{\rm OS}, \eta^{\rm SSK}\right)\right. h_k\left(t | \Bsb\right) \right]\otimes \left. {\cal R}\left(t -t' | \delta_{t}\right) \right\}, \nonumber
 \label{eq:fullpdf}
 \end{align}}
 \normalsize
where ${\cal R}$ is the time resolution function defined in Eq.~\eqref{eqn:decay_time_res_model} and the terms
\small{
\begin{align}
    {\cal Q}\left(\mathfrak{q}^{\rm OS},  \mathfrak{q}^{\rm SSK}, \eta^{\rm OS}, \eta^{\rm SSK}\right) & = \left[1+\mathfrak{q}^{\rm OS}\left(1-2\omega\left(\eta^{\rm OS}\right)\right)\right]\left[1+\mathfrak{q}^{\rm SSK}\left(1-2\tilde{\omega}\left(\eta^{\rm SSK}\right)\right)\right],\\
    \bar{{\cal Q}}\left(\mathfrak{q}^{\rm OS},  \mathfrak{q}^{\rm SSK}, \eta^{\rm OS}, \eta^{\rm SSK}\right) & = \left[1-\mathfrak{q}^{\rm OS}\left(1-2\bar{\omega}\left(\eta^{\rm OS}\right)\right)\right]\left[1-\mathfrak{q}^{\rm SSK}\left(1-2\tilde{\omega}\left(\eta^{\rm SSK}\right)\right)\right].
\end{align}}\normalsize
account for the measured flavour of the \Bs candidate. All physics parameters are free in the fit and are common across the subsamples,
except for the S-wave fraction and the phase difference
$\delta_{\rm S}-\delta_{\perp}$, which are independent parameters for each $m(\Kp\Km)$ bin.

\section{Systematic uncertainties}
\label{sec:Systematics}

Systematic uncertainties on the measured physics parameters arise from a variety of
sources that are described in the following. They are summarised in Table~\ref{tab:tot_syst_1}.

\begin{sidewaystable}[p]
 \begin{center}
        \caption{\small Summary of the systematic uncertainties.}
\resizebox*{!}{0.5\textwidth}{
\renewcommand{\arraystretch}{1.2}
   \begin{tabular}{lccccccccc}
   Source  & $\phis$ & $|\lambda|$  & $\Gs-\Gamma_{d}$ 
   & $\Delta \Gamma_{s}$ & $\dms$ & $|A_{\perp}|^{2}$ & $|A_{0}|^{2}$ & $\delta_\perp-\delta_{0}$ & $\delta_\parallel-\delta_{0}$ \\
    & [$\rad$] &  & [$\invps$] 
   & [$\invps$] & [$\invps$]& & & [$\rad$] & [$\rad$]\\
   \hline
   Mass: width parametrisation                & -  & -   &   - 	& 0.0002 & 0.001 & 0.0004 & 0.0006  & - 	& 0.003   \\
   Mass: decay-time $\&$ angles dependence    & 0.004  & 0.0037  & 0.0007  	& 0.0022 & 0.016  & 0.0005	& 0.0002  & 0.05 	& 0.009  \\
   Multiple candidates                       & 0.0011 & 0.0011 & 0.0003 	& 0.0001 & 0.001  & 0.0001  & 0.0001 	& 0.01 & 0.002\\
   Fit bias                                   & 0.0010  &   -      &   -     	& 0.0003 & 0.001 & 0.0006& 0.0001  	& 0.02  & 0.033 \\
   $C_{\rm SP}$ factors                        & 0.0010  & 0.0010   &   -     	& 0.0001 & 0.002   & 0.0001&   -   	& 0.01  & 0.005 \\
   Time resolution: model applicability                       &   -    &   -      &   -     	&   -    & 0.001 &   -     &   -   	&  - 	& 0.001  \\
   Time resolution: $t$ bias               & 0.0032 & 0.0010    & 0.0002  	&  0.0003 &  0.005  &   -     &   -    	& 0.08  & 0.001 \\
   Time resolution: wrong PV                        &   -    &   -    	  &   -     	&   -     & 0.001&   -     &   -   & -  	& 0.001 \\
   Angular efficiency: simulated sample size                    & 0.0011 & 0.0018 	& - 		&   -     & 0.001 & 0.0004 & 0.0003 & - 	& 0.004 \\
   Angular efficiency: weighting                    & 0.0022 & 0.0043  & 0.0001 	& 0.0002 & 0.001 & 0.0011 & 0.0020	& 0.01 	& 0.008 \\
   Angular efficiency: clone candidates    	     & 0.0005 & 0.0014 	 & 0.0002 	& 0.0001 &   -   & 0.0001& 0.0002 &   -    & 0.002\\
   Angular efficiency: $t$ \& $\sigma_{t}$ dependence & 0.0012 & 0.0007  & 0.0002 	& 0.0010 & 0.003& 0.0012 & 0.0008 	& 0.03  & 0.006 \\
   Decay-time efficiency: statistical             &   -    &   -    	 & 0.0012 	& 0.0008 &   -     & 0.0003 & 0.0002 &   -    &   -   \\
   Decay-time efficiency: kinematic weighting     	     &   -    &   -    	  & 0.0002 	&   -     &   - &   -     &   -    &   -    &   -    \\
   Decay-time efficiency: PDF weighting           &   -    &   -    	  & 0.0001 	& 0.0001 &   -  &   -     &   -   &   -    &   -   \\
   Decay-time efficiency: $\Delta\Gamma_{s}=0$ simulation               &   -    &   -    & 0.0003 	& 0.0005 &   -    & 0.0002& 0.0001 	&   -    &   -    \\
   Length scale                              &   -    &   -     & -       	&   -     & 0.004 &   -     &   -      	&   -    &   -  \\
   \hline
   Quadratic sum of syst.                    & 0.0061  & 0.0064  & 0.0015  		& 0.0026 & 0.018 & 0.0019 & 0.0023    & 0.10 & 0.036 \\
   \hline
 \end{tabular}
}
 \label{tab:tot_syst_1}
\end{center}
\end{sidewaystable}

Three systematic effects due to the $m(\jpsi K^{+}K^{-})$ model and the {\it sWeights} computation are taken into account.
Firstly, the systematic effect due to statistical uncertainties in the $m(\jpsi K^{+}K^{-})$ fit model is estimated. For this the {\it sWeights}
are recomputed after varying the fit parameters within their statistical uncertainties. The systematic uncertainties are obtained from the difference in fit results and are found to be negligible.
Secondly, the average width of the double-sided CB distribution is parametrised as a linear function of the per-candidate mass uncertainty, 
instead of a quadratic one.
The differences to the baseline result are assigned as systematic uncertainties on the mass shape.
Thirdly, the assumption that the $m(\jpsi K^{+}K^{-})$ distribution is independent of the decay time and angles is tested
by re-evaluating the {\it sWeights} in bins of these observables, repeating the fit
and assigning the differences in fit results as systematic uncertainties. 

The main physics background contribution comes from misidentified $\Lb \to \jpsi p \Km$ decays. Possible effects due to the limited knowledge of
the size of this component are estimated by repeating the fit after varying
the amount of this background by one standard deviation of its measured
yield. The maximum difference is found to be negligible and thus no systematic uncertainties are assigned.
A further systematic effect due to the $\Bd \to \jpsi \Kp \Km$ background is evaluated by repeating the $m(\jpsi K^{+}K^{-})$ fit while
leaving the mass resolution for this component free. A new
set of {\it sWeights} are computed, leading to negligible systematic uncertainties.
Finally, approximately 0.5\% of $\Bs\to\jpsi\Kp\Km$ candidates
come from the decays of $\Bc$ mesons via the $\Bc\to\Bs\pip$ decay~\cite{LHCb-PAPER-2013-044,Kiselev:2003mp}.
The effect of ignoring this component in the fit is evaluated using simulated pseudoexperiments where
$0.5\%$ of the candidates are replaced with \Bs-from-$\Bc$ decays that are randomly
sampled from simulated $\Bc\to\Bs(\to \jpsi\phi)\pi^+$ decays. This is found to have a negligible effect on all parameters.

In the baseline strategy, all candidates are retained even if multiple candidates are present in a single event. A part of these multiple candidates
is found to peak in the $m(\jpsi K^{+}K^{-})$ distribution,
which introduces a bias in the physics parameters, mainly on $\Gamma_s$. The peaking component is due to so-called clone candidates originating from final
state tracks that are duplicated
in the reconstruction process. Candidates are considered to be clones if they belong to the same event and their final-state tracks are
separated by an angle smaller than 5\,mrad. To assign systematic uncertainties, a single random candidate is selected among all clone candidates in an
event and the fit is repeated.
Approximately 0.35\% (0.2\%) of all selected $\Bs$ ($\Bd$) candidates are removed. The maximum resulting variations 
of the fit values are assigned as systematic uncertainties.

Possible biases of the fitting procedure are studied by generating and fitting over eight thousand
pseudoexperiments of the same size as the data.
The biases are determined from the resulting pull distributions. The ones
that are significantly different from zero
are assigned as systematic uncertainties.

Different models of the S-wave lineshape based on the results in Ref.~\cite{LHCb-PAPER-2013-069} are used to evaluate the
coupling factors $C_{\rm SP}$ in each of the six $m(\Kp\Km)$ bins, according to Ref.~\cite{LHCb-PAPER-2017-008}. This includes an S-wave parametrisation with a cubic spline function determined from data,
a variation of the $f_{0}(980)$ pole and width parameters used in the baseline model within their uncertainties,
and the variation of the $f_{0}(980)$ parameters according to the second solution found in the analysis in Ref.~\cite{LHCb-PAPER-2013-069}. The maximum resulting variations of the fit values, mostly due to
the spline parametrisation, are assigned as systematic uncertainties.

The tagging parameters are Gaussian-constrained in the fit and therefore their uncertainties
contribute to the statistical uncertainty of each fit value. This mainly affects the parameter $\phis$,
with a contribution to the uncertainty of  15\mrad. In addition, the calibration of the OS tagging is re-evaluated using a quadratic function instead of a linear one. 
The observed differences when repeating the fit are found to be negligible.

The systematic uncertainties associated with
decay-time resolution originate from four different sources.
The first is due to the statistical uncertainties on the calibration parameters and is found to be negligible.
The second is related to the assumption that the resolution model obtained in the calibration sample applies also to the signal sample. The corresponding systematic uncertainty is determined by evaluating the ratio of the calibration effective resolutions, obtained
from the simulated samples of the calibration and signal decays, and using
it to scale the
effective resolutions in the prompt data sample. These scaled effective resolutions are then described by a quadratic function and used to
determine the physics parameters.
The differences with respect to the baseline result are assigned as systematic uncertainties. A third source of uncertainty is due to a possible bias
of the Gaussian resolution mean, which is assumed to be zero in the baseline model.
A quadratic dependence of the mean on the decay-time uncertainty is observed in the calibration sample, with a maximum deviation of about $5\fs$ from zero. It is modelled in the prompt data sample
after weighting it in order to match the signal data sample. Corresponding systematic uncertainties are
evaluated as the differences between the results obtained with this bias and the baseline model.
Finally, the fourth systematic
effect is estimated by varying the contribution in the fit of candidates with an associated wrong origin vertex. The fraction of these candidates is varied
between 0 and 1.5\%, corresponding to about three times the fraction that is measured in the calibration sample, the calibration updated and the fit to data repeated. The maximum deviations from the baseline fit are assigned as systematic uncertainties.

The angular efficiency is determined from simulated
signal, weighted such that the
kinematic distributions of the final-state particles match those in the data.
Systematic uncertainties are assigned to account for the limited
size of the simulated sample by varying the normalisation weights according to their uncertainties and their covariance matrix
and repeating the fit with a new varied set of weights. The resulting RMS of each fitted observable is taken as a systematic uncertainty.
In addition, the impact of the specific configuration of the gradient-boost tree method used in the reweighting of the simulation is studied by testing approximately one hundred alternative configurations.
The maximal deviations from the fit result obtained with the default angular efficiency are assigned as systematic uncertainties.
The differences between the fit results obtained using angular corrections
from the baseline or alternative weighting procedures of the simulated candidates are also considered as
systematic uncertainties. 
An imperfect removal of clone candidates, in simulation, that peak in the \Bs candidate mass is tested as follows.
The peaking component is separated from the underlying background via {\it sWeights} using all simulated events to determine its shape. As an alternative, it is modelled according to the distribution of the corresponding background classification that is available in simulations, which however is limited by the small sample size and
is therefore not used as the baseline strategy. In addition, the two components are separated by matching the reconstructed daughter particles to the simulated particles by comparing their track momentum magnitudes and directions. The angular efficiency is determined according to these two changes and the larger differences    
are assigned as systematic uncertainties. Finally, from a fit to several simulated samples of the same size as data, uncertainties are evaluated as the differences between fitted and generated values to account for correlations between the angular efficiency and the decay time as well as the decay-time uncertainty. Such correlations are neglected in the baseline fit.

Several sources of systematic uncertainties related to the determination
of the spline-based decay-time efficiency are studied and found to be small.
First, the effect due to the limited size of the data and simulated samples is estimated by
repeating the fit several times with the spline coefficients varied according to their covariance matrix and the RMS of the
fitted observable distributions is taken as systematic uncertainties. Two further contributions are evaluated
by taking the difference between the baseline fit and alternative fits where the time efficiency is determined
without applying either the kinematic or the PDF weighting procedures used to correct the physics parameters of $\Bs$ and $\Bz$ simulated samples.
Next, the number of spline nodes is doubled and found to have a negligible effect on the result.
Another systematic uncertainty source is due to the differences observed in decay-time
efficiency derived from the simulated samples with $\Delta \Gamma_s$ equal to or different from zero. It has been also
checked that varying the decay-time resolutions used in the determination of the decay-time efficiency by $10\%$ has negligible effects.

The uncertainty on the LHCb length scale is estimated to be 0.022\%~\cite{LHCb-PAPER-2013-006}, as determined
from metrology and track-based alignment. This translates directly into an uncertainty on $\Gs-\Gd$,
$\Delta\Gamma_s$ and $\Delta m_s$, which is non-negligible only in the case of $\Delta m_s$. Other parameters are unaffected.
The precision on the track momentum scale is $0.03\%$. Its effect largely cancels in the computation of the decay time,
leading to negligible uncertainties on all observables.

 Asymmetries between \Bs and \Bsb production rates are diluted by the fast oscillation between particle and antiparticle. They are found to have a negligible effect on the fit parameters. 
 
 No statistically significant systematic effect on the results is observed when repeating the analysis
on subsets of the data, splitting by magnet polarity, trigger conditions,
year of data taking, number of primary vertices, bins of \Bs \pt, pseudorapidity and decay-time
uncertainty.

\section{Results}
\label{sec:Results}

\begin{figure}[tbp]
  \centering
    \includegraphics[width=0.49\textwidth]{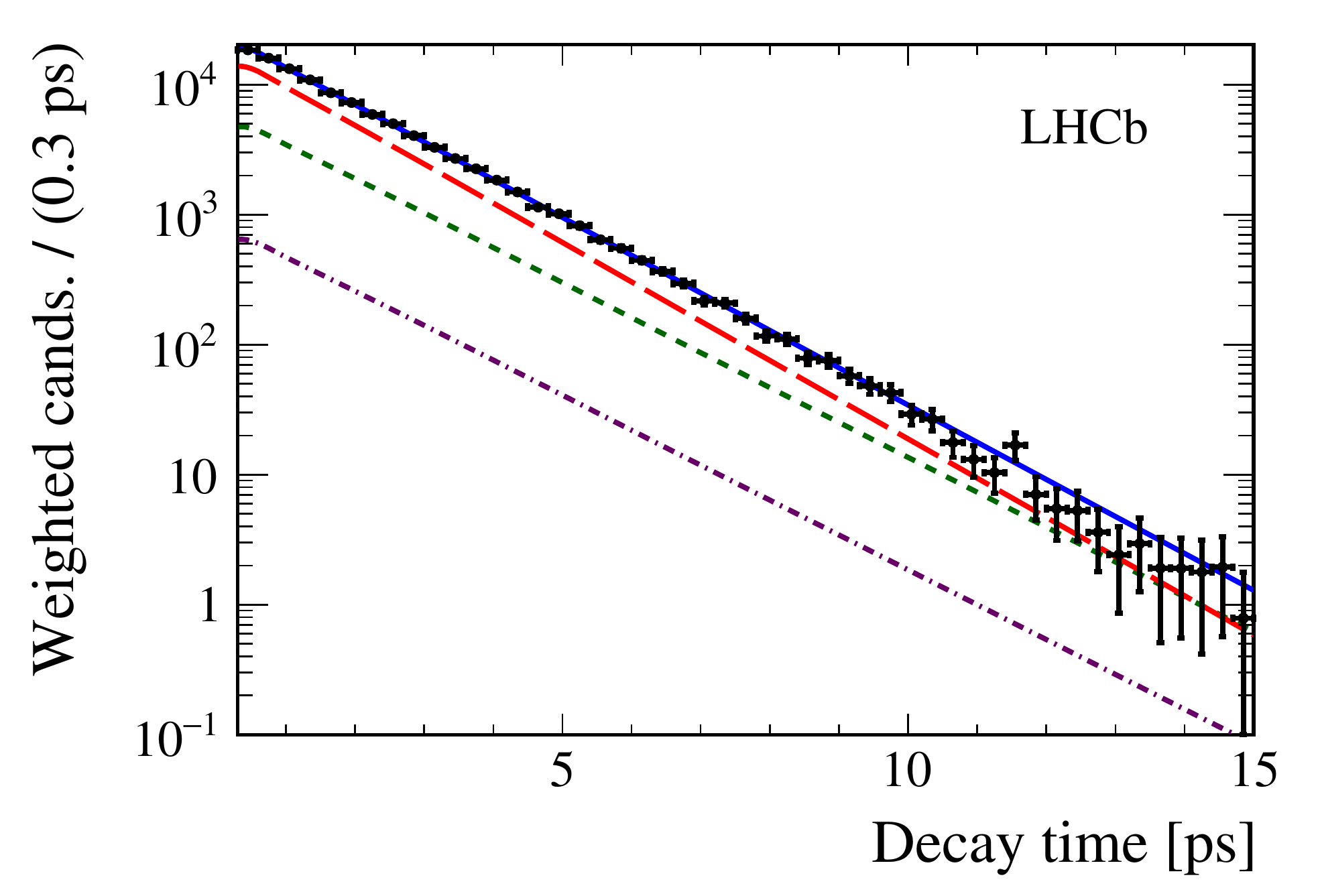}
    \includegraphics[width=0.49\textwidth]{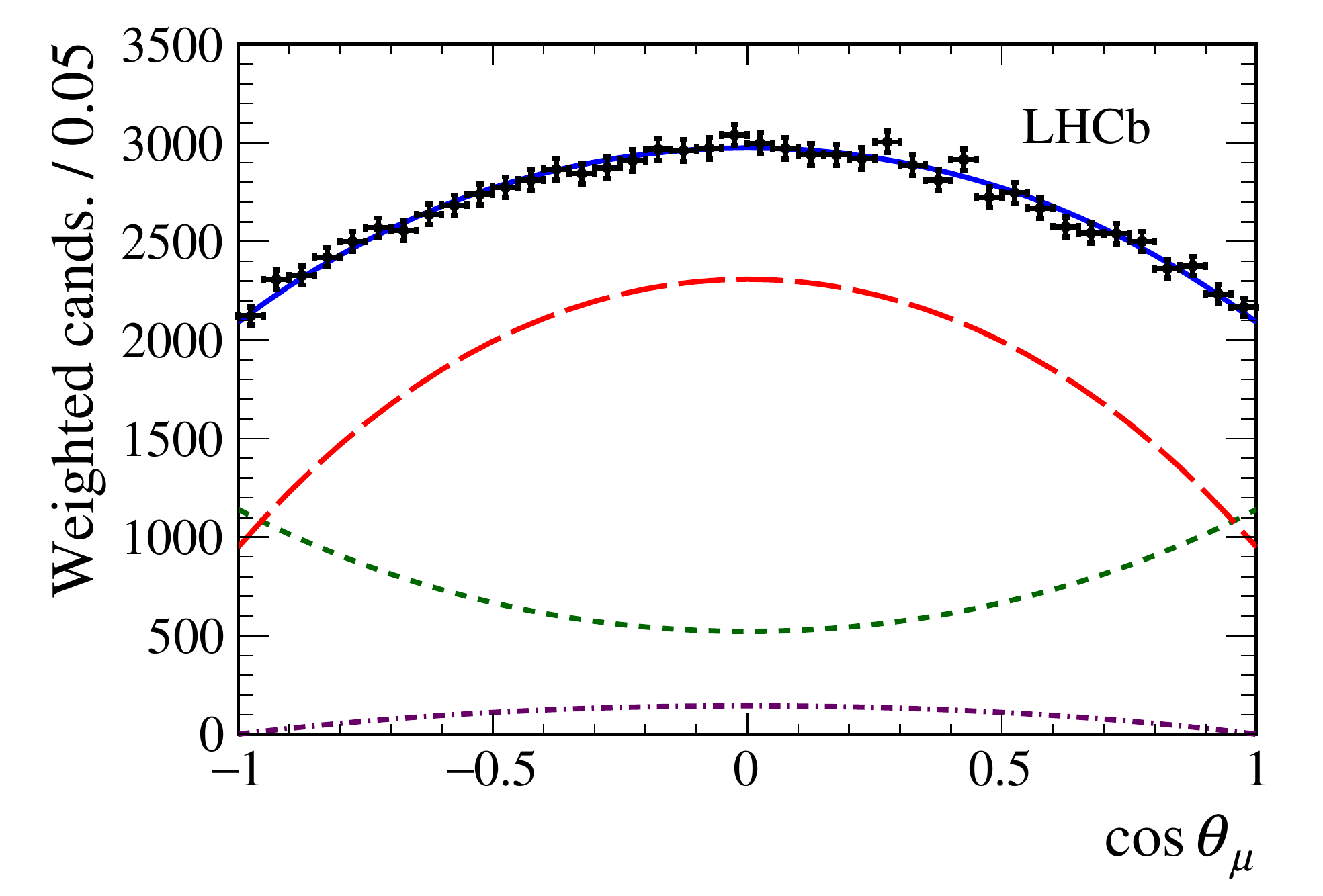}
    \includegraphics[width=0.49\textwidth]{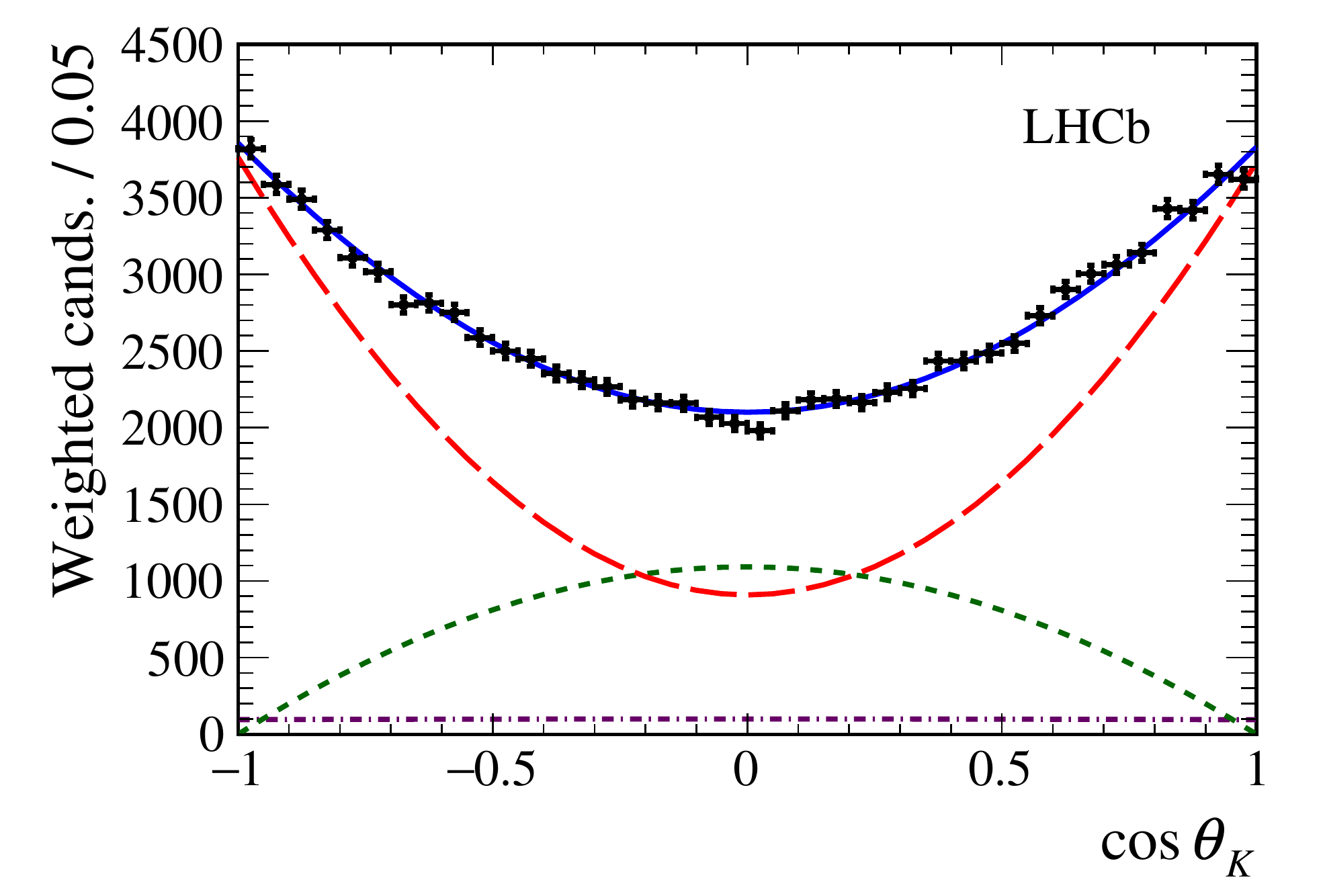}
    \includegraphics[width=0.49\textwidth]{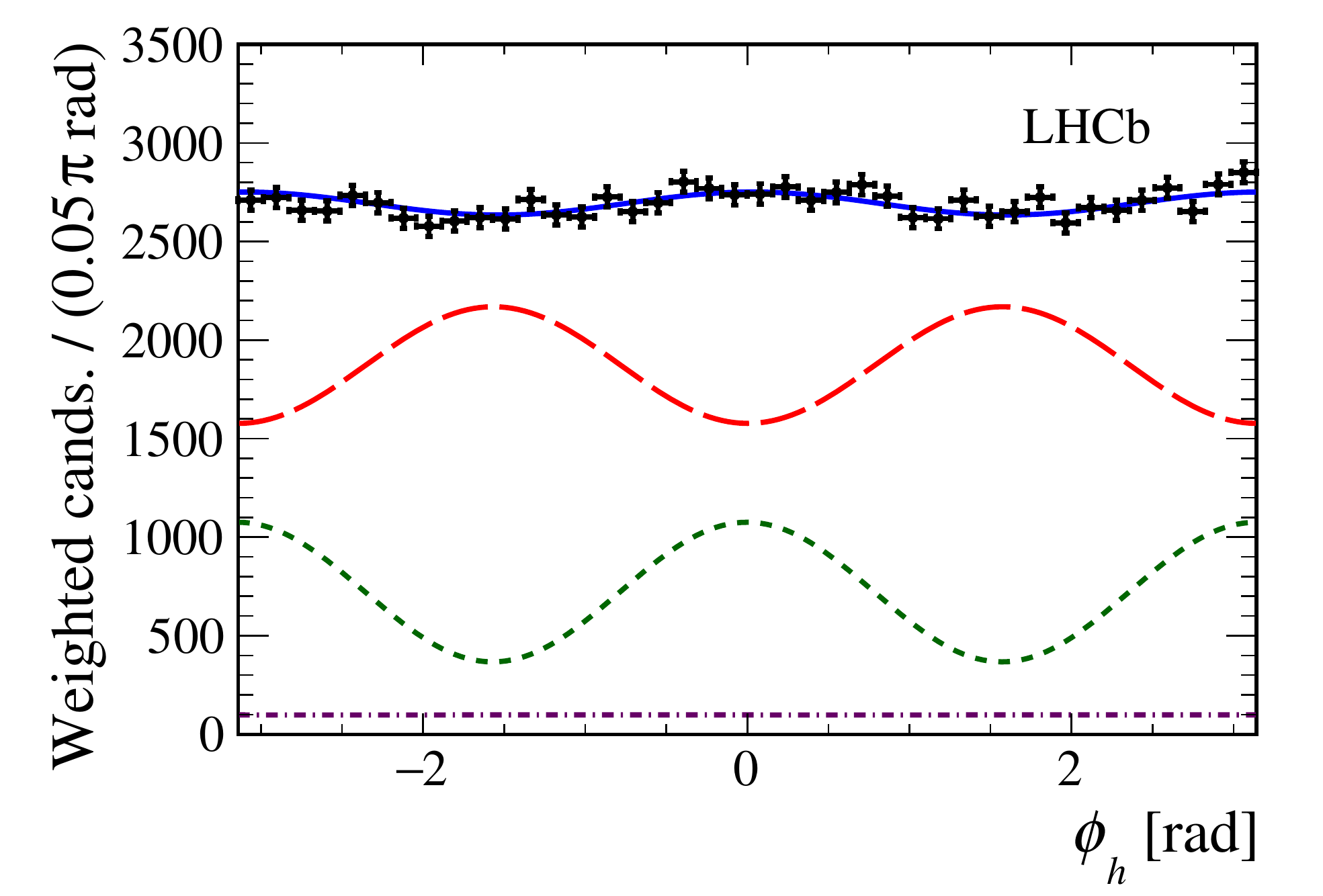}
  \caption{\small Decay-time and helicity-angle distributions for background subtracted \mbox{$\Bs\to\jpsi\Kp\Km$} decays
  (data points) with the one-dimensional projections of the PDF at
           the maximum-likelihood point.
           The solid blue line shows the total signal contribution, which contains  (long-dashed red) \CP-even,  (short-dashed
           green) \CP-odd and (dotted-dashed purple) S-wave contributions. Data and fit projections for the different samples considered (data-taking year, trigger and tagging categories, $m(K^{+}K^{-})$ bins) are combined.}
  \label{fig:results_projections}
\end{figure}

The results of the maximum-likelihood fit described in Sec.~\ref{sec:Fit} are
\begin{align}
     \phi_{s} &= -0.083\pm0.041\pm0.006\rad \nonumber \\
     |\lambda|           		&= 1.012\pm0.016\pm0.006   \nonumber \\
     \Gamma_{s}-\Gamma_{d} &= -0.0041\pm0.0024\pm0.0015\invps \nonumber\\
     \Delta\Gamma_{s} &= 0.077\pm0.008\pm0.003\invps \nonumber\\
     \Delta m_{s}&= 17.703\pm0.059\pm0.018 \invps \nonumber\\
     |A_{\perp}|^2	&= 0.2456\pm0.0040\pm0.0019\nonumber  \\
     |A_0|^2             		&= 0.5186\pm0.0029\pm0.0023  \nonumber  \\
     \delta_\perp-\delta_0      	&= 2.64\pm0.13\pm0.10 \rad\nonumber  \\
     \delta_\parallel-\delta_0    	&= 3.06^{\:+\:0.08}_{\:-\:0.07}\pm0.04 \rad.
 \end{align}
The S-wave fractions and phase differences with respect to $\delta_{\perp}$ in each $m(K^{+}K^{-})$ bin are summarized in Appendix~\ref{app:S-wave parameters}.
The background-subtracted data distributions with fit projections are shown in Fig.~\ref{fig:results_projections}.

The results are in good agreement with the previous LHCb measurement. 
The measurements of $\phi_s$, $\Delta\Gamma_s$ and $\Gs-\Gd$
are the most precise to date and agree with the SM expectations~\cite{UTfit-UT,CKMfitter2015,Artuso:2015swg,Kirk:2017juj}. The results also indicate 
no \CP violation in $\Bs \to J/\psi \Kp \Km$ decays. The value of $\Delta m_{s}$ is in a good agreement with the world average value~\cite{PDG2018}.
Relaxing the assumption that $\lambda_r$ is the same for all polarisation
states and repeating the fit shows no evidence for any polarisation dependence.
The correlation matrix including systematic uncertainties can be found in Table~\ref{tab:correlations_result}.

\begin{table}[htbp]
  \centering
  \caption{\label{tab:correlations_result}\small Correlation matrix including the statistical and systematic correlations between the parameters.}
  \begin{tabular}{l| c c c c c c c c c }
  & $\phi_{s}$ & $|\lambda|$ & $\Gamma_{s}-\Gamma_{d}$ & $\Delta\Gamma_s$ & $\dms$ & $|A_{\perp}|^2$ & $|A_0|^2$ & $\delta_\perp -\delta_{0}$ & $\delta_\parallel -\delta_{0}$ \\
  \hline
  $\phi_{s}$ & 1.00 & \phantom{+}0.16 & $-$0.05 & \phantom{+}0.01 & $-$0.02 & \phantom{+}0.01 & \phantom{+}0.00 & \phantom{+}0.03 & \phantom{+}0.00 \\
  $|\lambda|$ &  & \phantom{+}1.00 & \phantom{+}0.07 & $-$0.09 & \phantom{+}0.06 & \phantom{+}0.04 & $-$0.02 & \phantom{+}0.04 & \phantom{+}0.01 \\
  $\Gamma_{s}-\Gamma_{d}$ & & & \phantom{+}1.00 & $-$0.46 & \phantom{+}0.06 & \phantom{+}0.35 & $-$0.24 & $-$0.01 & \phantom{+}0.03 \\
  $\Delta\Gamma_s$ & & & & \phantom{+}1.00 & $-$0.05 & $-$0.64 & \phantom{+}0.46 & $-$0.02 & \phantom{+}0.00 \\
  $\dms$ & & & & & \phantom{+}1.00 & \phantom{+}0.01 & \phantom{+}0.01 & \phantom{+}0.55 & $-$0.01 \\
  $|A_{\perp}|^2$ & & & & & & \phantom{+}1.00 & $-$0.64 & \phantom{+}0.01 & \phantom{+}0.07 \\
  $|A_0|^2$ & & & & & & & \phantom{+}1.00 & \phantom{+}0.01 & $-$0.02 \\
  $\delta_\perp -\delta_{0}$ & & & & & & & & \phantom{+}1.00 & \phantom{+}0.25 \\
  $\delta_\parallel -\delta_{0}$ & & & & & & & & & \phantom{+}1.00 \\
  \end{tabular}
\end{table}

\section{Combination with other results}
\label{sec:Combination}
The results presented in this paper are combined with related Run 1 and Run 2 LHCb measurements, taking into account all statistical correlations, all systematic errors and their correlations, and correlations between different run periods.

\subsection{\boldmath Combination with Run 1 $\Bs\to\jpsi\Kp\Km$}
The measurements presented in this paper are consistent with those 
obtained from the analysis of the data collected by LHCb during the LHC Run 1~\cite{LHCb-PAPER-2014-059}. The \mbox{Run 1} measurements are combined with the results of this analysis taking into account a covariance matrix that includes the statistical uncertainties with their correlations,
and the systematic uncertainties with their correlations, both between the parameters in a single run period and between the two run periods. 

The sources of systematic uncertainty that are correlated between the analyses are the applicability of the time resolution obtained from the prompt control sample on the signal sample, the $C_{\rm SP}$ factors, the correction of simulation for the angular efficiency determination, and the length scale. In the case of the angular efficiency, a correlation matrix is determined from the RMS distributions of the parameters in Run 2 and the same matrix is taken to account for correlations between Run 1 and Run 2. For all other sources of systematic uncertainty no correlation is assumed.
For the parameters showing asymmetric uncertainties, the larger uncertainty has been used in the combination. 

It has been 
verified that using the average of the two asymmetric uncertainties does not change the combination, and 
that completely ignoring the systematic correlations has a negligible effect. 
In the Run 1 measurement, $\Gamma_{s}$ was measured instead of $\Gamma_{s}-\Gamma_{d}$,
hence a linear transformation is taken into account in the combination, constraining $\Gamma_d$ to the known value~\cite{HFLAV16}. The combined results are 
\begin{align}
\label{eq:comb1}
\phi_{s} &= -0.081\pm0.032\rad\,, \nonumber \\
|\lambda| &= 0.994\pm0.013\,, \nonumber \\
\Gamma_{s} &= 0.6572\pm0.0023\invps\,,\nonumber \\
\Delta\Gamma_{s} &= 0.0777\pm0.0062\invps\,,\nonumber \\
\Delta m_{s} &= 17.694 \pm 0.042\invps\,,\nonumber \\
|A_{\perp}|^2 &= 0.2489 \pm 0.0035\,, \nonumber \\
|A_0|^2 &= 0.5195 \pm 0.0034\,, \nonumber \\
\delta_\perp -\delta_{0}&= 2.87 \pm 0.11\rad\,,\nonumber \\
\delta_\parallel -\delta_{0} &= 3.153 \pm 0.079\rad.
\end{align}
The correlation matrix can be found in Table~\ref{tab:comb_correlations}. The correlation between $\Gamma_{s}$ and $\Gamma_d$ is 0.39. 
The combined value of $\phi_s$ is $2.5$ standard deviations from zero and agrees with expectations 
based on the SM~\cite{UTfit-UT,CKMfitter2015}.
\begin{table}[ht]
  \centering
  \caption{\label{tab:comb_correlations}\small Correlation matrix for the results in Eq.~\eqref{eq:comb1} taking into account correlated systematics between Run 1 and the 2015 and 2016 results.}
  \begin{tabular}{l| c c c c c c c c c }
 & $\phi_{s}$ & $|\lambda|$ & $\Gamma_{s}$ & $\Delta\Gamma_s$ & $\dms$ & $|A_{\perp}|^2$ & $|A_0|^2$ & $\delta_\perp -\delta_{0}$ & $\delta_\parallel -\delta_{0}$ \\
  \hline
  $\phi_{s}$                        & 1.00                          & \phantom{+}0.10   & $-$0.02 & $-$0.03 & \phantom{+}0.02 & \phantom{+}0.01 & $-$0.01 & \phantom{+}0.07 & \phantom{+}0.00 \\
  $|\lambda|$                       & & \phantom{+}1.00             & \phantom{+}0.04   & $-$0.04           & $-$0.05         & \phantom{+}0.03 & $-$0.02 & $-$0.04 & \phantom{+}0.03 \\
  $\Gamma_{s}$                      & & & \phantom{+}1.00           & $-$0.35           & \phantom{+}0.04   & \phantom{+}0.28 & $-$0.17 & \phantom{+}0.01 & \phantom{+}0.01 \\
  $\Delta\Gamma_s$                  & & & & \phantom{+}1.00         & $-$0.01           & $-$0.62           & \phantom{+}0.40 & $-$0.05 & $-$0.01 \\
  $\dms$                            & & & & & \phantom{+}1.00       & \phantom{+}0.01   & $-$0.01           & \phantom{+}0.62 & \phantom{+}0.02 \\
  $|A_{\perp}|^2$                   & & & & & & \phantom{+}1.00     & $-$0.67           & \phantom{+}0.03   & \phantom{+}0.01 \\
  $|A_0|^2$                         & & & & & & & \phantom{+}1.00   & $-$0.06           & $-$0.06 \\
  $\delta_\perp -\delta_{0}$        & & & & & & & & \phantom{+}1.00 & \phantom{+}0.28 \\
  $\delta_\parallel -\delta_{0}$    & & & & & & & & & \phantom{+}1.00 \\
  \end{tabular}
\end{table}

\subsection{\boldmath Combination with other LHCb $\phi_s$ results}
The results obtained in the previous section are further combined with the recent results from \mbox{$\Bs\to\jpsi\pip\pim$}~\cite{LHCb-PAPER-2019-003} decays, and the Run 1
results from \mbox{$\Bs\to\jpsi\pip\pim$}~\cite{LHCb-PAPER-2014-019}, \mbox{$\Bs \to \jpsi \Kp \Km$} for the $\Kp\Km$ invariant mass region above
1.05\gevcc~\cite{LHCb-PAPER-2017-008}, \mbox{$\Bs\to\psi(2S)\phi$}~\cite{LHCb-PAPER-2016-027} and \mbox{$\Bs\to D^+_sD^-_s$}~\cite{LHCb-PAPER-2014-051} decays. 

The Run 1 analysis of $\Bs\to\jpsi\pip\pim$ decays measured $|\lambda|$ and $\phi_s$ assuming a value of $\dms$ fixed to $17.768\pm0.024\invps$. 
Before the combination, this value is updated to $17.711\pm0.059\invps$~\cite{LHCb-PAPER-2014-059}, and the analysis is repeated to obtain updated values of $|\lambda|$ and $\phi_s$. 
The analysis of 2015 and 2016 data, instead, measured $\phi_s$, $|\lambda|$ and $\Gamma_{\rm H} - \Gamma_d$, assuming the $\Delta m_s$ value determined in this analysis. In the combination, $\Gamma_{\rm H} - \Gamma_d$ is parametrised as $\Gamma_{s}-\Gamma_{d}-\Delta \Gamma_s/2$, and the value of $\Gamma_d$ is constrained to the known value.
The combined values are
\begin{align}
\label{eq:comb2}
\phi_{s} &= -0.042 \pm 0.025\rad\,, \nonumber \\
|\lambda| &= 0.993 \pm 0.010\,, \nonumber \\
\Gamma_{s} &= 0.6563 \pm 0.0021\invps\,,\nonumber \\
\Delta\Gamma_{s} &= 0.0813 \pm 0.0048\invps\,.
\end{align}
The correlation matrix can be found in Table~\ref{tab:comb_correlations_pipi}. The correlation between $\Gamma_{s}$ and $\Gamma_{d}$ is 0.48.
\begin{table}[t]
  \centering
  \caption{\label{tab:comb_correlations_pipi}\small Correlation matrix for the results in Eq.~\eqref{eq:comb2} obtained taking into account correlated systematics between the considered analyses.}
  \begin{tabular}{l| c c c c }
  & $\phi_{s}$ & $|\lambda|$ & $\Gamma_{s}$ & $\Delta\Gamma_s$  \\
  \hline
  $\phi_{s}$        & 1.00  & \phantom{+}0.06 & $-$0.01 & $-$0.03 \\
  $|\lambda|$       &       & \phantom{+}1.00 & \phantom{+}0.03 & $-$0.02  \\
  $\Gamma_{s}$      &       &                 & \phantom{+}1.00 & $-$0.17 \\
  $\Delta\Gamma_s$  &       &                 & & \phantom{+}1.00  \\
  \end{tabular}
\end{table}
\begin{figure}
    \centering
    \includegraphics[width=0.6\textwidth]{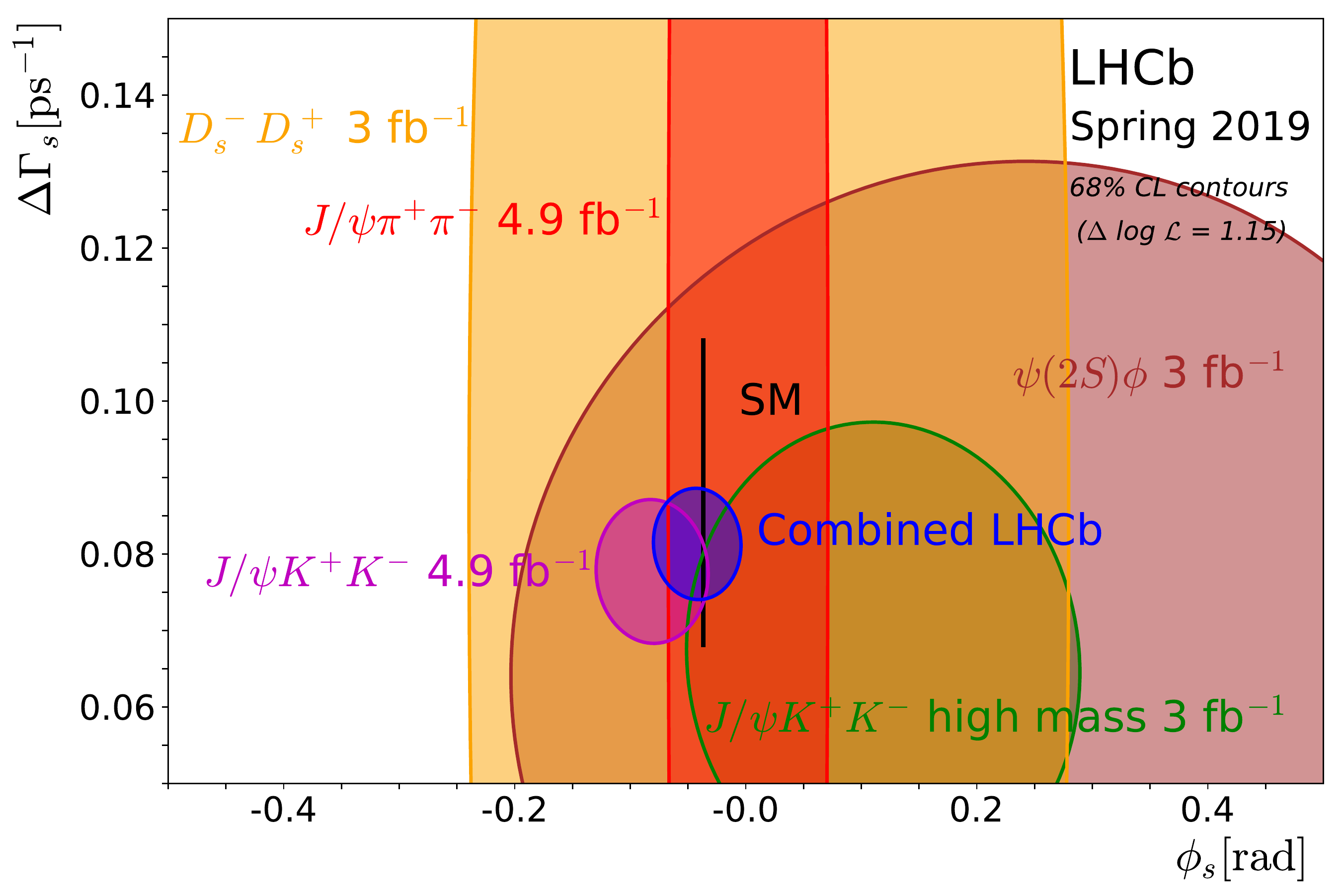}
    \caption{\label{fig:lhcb_combination}\small
    Regions of 68\% confidence level in the $\phi_s$-$\Delta\Gamma_s$ plane for the individual \lhcb measurements and a combined contour (in blue).
        The $\Bs \to \jpsi \Kp \Km$ (magenta) and \mbox{$\Bs\to\jpsi\pip\pim$}~\cite{LHCb-PAPER-2019-003} (red) contours show the Run 1 and Run 2 combined numbers.
        The $\phi_s$~\cite{CKMfitter2015} and $\Delta\Gamma_{s}$~\cite{Artuso:2015swg} predictions are indicated by the thin black rectangle.}
\end{figure}
The values of these parameters are the most precise to date. Figure~\ref{fig:lhcb_combination} shows the 68\% confidence level regions in the $\phi_s$ vs.
$\Delta\Gamma_s$ plane for the considered analyses and the \lhcb combination. The combined value of $\phi_s$ is consistent with 
global fits to data.  The parameter $|\lambda|$ agrees with the hypothesis of no \CP violation in the decay. The values of $\Gamma_s$ and $\Delta \Gamma_s$
are consistent with expectations from HQE models.

\section{Conclusions}
\label{sec:Conclusions}

In summary, a flavour-tagged decay-time-dependent angular analysis of $\Bs\to\jpsi\Kp\Km$ decays
has been performed, using 1.9\invfb of $pp$ collision data recorded by the \lhcb experiment during
the 2015 and 2016 runs of the LHC. Approximately $117\,000$ signal decays are selected,
with a decay-time resolution of about 45\fs and a tagging power of 4.7\%. The \CP-violating phase $\phis$ is measured to be $-0.083\pm0.041\pm0.006$\rad, 
the decay width difference of the $\Bs$ mass eigenstates, $\Delta\Gamma_s = 0.077\pm0.008\pm0.003$\invps, and the difference of the average decay widths of the \Bs and \Bd
mesons, \mbox{$\Gamma_s-\Gamma_d = -0.0041\pm0.0024\pm0.0015$\invps.}
Using the known value for the \Bd
meson lifetime $1.520\pm0.004\ps$~\cite{HFLAV16}, the ratio of \Bs and \Bd meson decay widths is
measured to be $\Gamma_s/\Gamma_d = 0.9938 \pm 0.0036 \pm 0.0023$.
All results are shown with first the statistical and second the
systematic uncertainty. These are the single most precise measurements of these quantities to date. In addition, the mass difference between the \Bs mass eigenstates is measured to be $\Delta m_{s}=17.703\pm0.059\pm0.018\invps$. All results are consistent with
theoretical predictions based on the SM~\cite{UTfit-UT,CKMfitter2015}. The \CP-violating parameters are also determined assuming that they are not the same
for all $\Bs\to\jpsi\Kp\Km$ polarisation states and no polarisation dependence is observed.

The measurements presented here for the parameters $\phis$, $|\lambda|$, $\Gs-\Gamma_{d}$, $\DGs$, $\dms$, $|A_{\perp}|^{2}$, $|A_{0}|^{2}$, $\delta_\perp-\delta_{0}$ and $\delta_\parallel-\delta_{0}$
are consistent with those from $\Bs\to\jpsi\Kp\Km$
decays obtained using data collected by the \lhcb experiment during Run 1 of the
LHC~\cite{LHCb-PAPER-2014-059}. The two sets of measurements are combined
accounting for the statistical and systematic correlations between parameters
in each and the systematic correlations between the two run periods. The combined
values are 
$\phi_s = -0.081 \pm 0.032$\rad, $|\lambda| = 0.994 \pm 0.013$, $\Gamma_s = 0.6572 \pm 0.0023\invps$,
$\Delta\Gamma_s = 0.0777\pm0.0062\invps$ and $\Delta m_{s}= 17.694\pm 0.042\invps$. The value of $\phi_s$ is
2.5 standard deviations from zero and consistent with theoretical predictions based on the SM~\cite{UTfit-UT,CKMfitter2015}.

The results are further combined with the recent results from $\Bs\to\jpsi\pip\pim$~\cite{LHCb-PAPER-2019-003}, and the Run 1
results from $\Bs\to\jpsi\pip\pim$~\cite{LHCb-PAPER-2014-019}, $\Bs \to \jpsi \Kp \Km$ for the $\Kp\Km$ invariant mass region above
1.05\gevcc~\cite{LHCb-PAPER-2017-008}, $\Bs\to\psi(2S)\phi$~\cite{LHCb-PAPER-2016-027} and $\Bs\to D^+_sD^-_s$~\cite{LHCb-PAPER-2014-051}. The combined values are 
$\phi_s = -0.042 \pm 0.025$\rad, $|\lambda| = 0.993 \pm 0.010$, $\Gs = 0.6563 \pm 0.0021$\invps and $\DGs = 0.0813 \pm 0.0048$\invps.
These values are consistent with
theoretical predictions based on the SM~\cite{UTfit-UT,CKMfitter2015}. In particular, the value
of $\phis$ is consistent with a non-zero \CP-violation predicted within the SM and with no \CP-violation in the interference of \Bs meson mixing and decay.
The parameter $|\lambda|$ is consistent with unity, implying no evidence for direct
\CP-violation in $\Bs\to\jpsi\Kp\Km$ decays.

\section*{Acknowledgements}
%
%
\noindent We express our gratitude to our colleagues in the CERN
accelerator departments for the excellent performance of the LHC. We
thank the technical and administrative staff at the LHCb
institutes.
We acknowledge support from CERN and from the national agencies:
CAPES, CNPq, FAPERJ and FINEP (Brazil); 
MOST and NSFC (China); 
CNRS/IN2P3 (France); 
BMBF, DFG and MPG (Germany); 
INFN (Italy); 
NWO (Netherlands); 
MNiSW and NCN (Poland); 
MEN/IFA (Romania); 
MSHE (Russia); 
MinECo (Spain); 
SNSF and SER (Switzerland); 
NASU (Ukraine); 
STFC (United Kingdom); 
DOE NP and NSF (USA).
We acknowledge the computing resources that are provided by CERN, IN2P3
(France), KIT and DESY (Germany), INFN (Italy), SURF (Netherlands),
PIC (Spain), GridPP (United Kingdom), RRCKI and Yandex
LLC (Russia), CSCS (Switzerland), IFIN-HH (Romania), CBPF (Brazil),
PL-GRID (Poland) and OSC (USA).
We are indebted to the communities behind the multiple open-source
software packages on which we depend.
Individual groups or members have received support from
AvH Foundation (Germany);
EPLANET, Marie Sk\l{}odowska-Curie Actions and ERC (European Union);
ANR, Labex P2IO and OCEVU, and R\'{e}gion Auvergne-Rh\^{o}ne-Alpes (France);
Key Research Program of Frontier Sciences of CAS, CAS PIFI, and the Thousand Talents Program (China);
RFBR, RSF and Yandex LLC (Russia);
GVA, XuntaGal and GENCAT (Spain);
the Royal Society
and the Leverhulme Trust (United Kingdom).
\textcolor{white}{\tiny   Little Buddy!\normalsize}


\section*{Appendix}

\appendix


\section{S-wave parameters}
\label{app:S-wave parameters}
The results for the S-wave parameters in each of the six $m(K^{+}K^{-})$ bins are given in Table~\ref{tab:Swave_pars}. The main sources of systematic uncertainties are the $C_{\rm SP}$ factors, the mass factorization and biases of the fitting procedure.

\begin{table}[ht]
    \centering
    \caption{Values of the S-wave parameters in each $m(K^{+}K^{-})$ bin. The first uncertainty is statistical and the second systematic.}
    \label{tab:Swave_pars}
    \begin{tabular}{ll}
                                 Parameter  &   Value \\ 
\hline
$                             F_{S1}$ & $0.492\pm 0.043\pm0.010$ \\ 
$                             F_{S2}$ & $0.041\pm0.008\pm0.006$\\ 
$                             F_{S3}$ & $0.0044^{+0.0030}_{-0.0017}\pm0.0014$ \\ 
$                             F_{S4}$ & $0.0069^{+0.0062}_{-0.0045}\pm0.0016$ \\ 
$                             F_{S5}$ & $0.073\pm0.013\pm0.004$ \\ 
$                             F_{S6}$ & $0.152^{+0.019}_{-0.018}\pm0.009$ \\ 
\hline 
$  \delta_{S1}-\delta_{\perp} [\rad]$ & $+2.21^{+0.17}_{-0.20}\pm0.20$\\ 
$  \delta_{S2}-\delta_{\perp} [\rad]$ & $+1.56\pm 0.29\pm0.05$ \\ 
$  \delta_{S3}-\delta_{\perp} [\rad]$ & $+1.09^{+0.47}_{-0.36}\pm0.10$ \\ 
$  \delta_{S4}-\delta_{\perp} [\rad]$ & $-0.28^{+0.16}_{-0.26}\pm0.12$ \\ 
$  \delta_{S5}-\delta_{\perp} [\rad]$ & $-0.54^{+0.09}_{-0.10}\pm0.02$ \\ 
$  \delta_{S6}-\delta_{\perp} [\rad]$ & $-1.10^{+0.13}_{-0.16}\pm0.11$\\ 
\hline 
    \end{tabular}
\end{table}

\newpage

\addcontentsline{toc}{section}{References}
\setboolean{inbibliography}{true}
\bibliographystyle{LHCb}
\bibliography{standard,main,LHCb-PAPER,phis,LHCb-CONF,LHCb-DP,LHCb-TDR}

\newpage



\newpage
\centerline{\large\bf LHCb collaboration}
\begin{flushleft}
\small
R.~Aaij$^{28}$,
C.~Abell{\'a}n~Beteta$^{46}$,
T.~Ackernley$^{56}$,
B.~Adeva$^{43}$,
M.~Adinolfi$^{50}$,
H.~Afsharnia$^{6}$,
C.A.~Aidala$^{78}$,
S.~Aiola$^{22}$,
Z.~Ajaltouni$^{6}$,
S.~Akar$^{61}$
% ...

{\footnotesize \it
% ...
$ ^{6}$Universit{\'e} Clermont Auvergne, CNRS/IN2P3, LPC, Clermont-Ferrand, France\\
% ...
$ ^{22}$INFN Sezione di Milano, Milano, Italy\\
% ...
$ ^{28}$Nikhef National Institute for Subatomic Physics, Amsterdam, Netherlands\\
% ...
$ ^{43}$Instituto Galego de F{\'\i}sica de Altas Enerx{\'\i}as (IGFAE), Universidade de Santiago de Compostela, Santiago de Compostela, Spain\\
% ...
$ ^{46}$Physik-Institut, Universit{\"a}t Z{\"u}rich, Z{\"u}rich, Switzerland\\
% ...
$ ^{50}$H.H. Wills Physics Laboratory, University of Bristol, Bristol, United Kingdom\\
% ...
$ ^{56}$Oliver Lodge Laboratory, University of Liverpool, Liverpool, United Kingdom\\
% ...
$ ^{61}$University of Cincinnati, Cincinnati, OH, United States\\
% ...
$ ^{63}$Syracuse University, Syracuse, NY, United States\\
% ...
$ ^{78}$University of Michigan, Ann Arbor, United States, associated to $^{63}$\\
% ...
}
\end{flushleft}

\end{document}